\documentclass[aps,pra,10pt, twocolumn, superscriptaddress, 
]{revtex4-1}

\usepackage[utf8]{inputenc}
\usepackage{amsmath,amsfonts,amssymb,amsthm,bbm,bm, braket}
\usepackage{scalerel}
\usepackage{wasysym}
\usepackage{comment}
\usepackage{enumerate}
\usepackage{graphicx}
\usepackage{mathtools}
\usepackage{xparse}
\usepackage{ifthen}
\usepackage{tensor}
\usepackage{tikz}
\usepackage{bbold}
\usepackage{tikz-network}
\usetikzlibrary{patterns,decorations.pathreplacing}

\usepackage{color}
\definecolor{myred}{RGB}{232,102,102}
\definecolor{myblue}{RGB}{187,187,255}
\definecolor{myviolet}{RGB}{210,145,178}
\definecolor{myvioletc}{RGB}{45,130,60}
\definecolor{mygreen}{RGB}{34,139,34}
\definecolor{myorange}{RGB}{255,165,0}
\definecolor{OliveGreen}{RGB}{85,107,47}
\definecolor{NavyBlue}{RGB}{0,0,128}

\usepackage{tensor}

\newcommand{\be}{\begin{equation}}
\newcommand{\ee}{\end{equation}}
\newcommand{\ba}{\begin{aligned}}
\newcommand{\ea}{\end{aligned}}

\newcommand{\1}{\mathbbm{1}}

\theoremstyle{plain}
\newtheorem{property}{Property}
\theoremstyle{plain}

\usepackage[colorlinks,bookmarks=false,citecolor=NavyBlue,linkcolor=OliveGreen,urlcolor=blue]{hyperref}

\newcommand{\mcirc}{\mathbin{\scalerel*{\fullmoon}{G}}}
\newcommand{\msquare}{\mathord{\scalerel*{\square}{G}}}
\newcommand{\mcircf}{\mathbin{\scalerel*{\newmoon}{G}}}
\newcommand{\msquaref}{\mathord{\scalerel*{\blacksquare}{G}}}

\newcommand{\Wgate}[2]{

\draw[very thick] (#1-0.5, #2 +0.5) -- (#1+0.5,#2-0.5);
\draw[very thick] (#1-0.5,#2-0.5) -- (#1+0.5,#2+0.5);
\draw[ thick, fill=myvioletc, rounded corners=2pt] (#1-0.25,#2+0.25) rectangle (#1+0.25,#2-0.25);
\draw[thick] (#1,#2+0.15) -- (#1+0.15,#2+0.15) -- (#1+0.15,#2);
}

\newcommand\encircle[1]{%
  \tikz[baseline=(X.base)] 
   \node (X) [draw, shape=circle, inner sep=0.05] {\strut #1};}

\renewcommand{\boxed}[1]{%
  \framebox{\raisebox{0pt}[0.4\baselineskip][0.025\baselineskip]{\hbox to 0.25cm{\hss#1\hss}}}}

\begin{document}

\title{Scrambling in Random Unitary Circuits: Exact Results}

\author{Bruno Bertini}
\affiliation{Department of Physics, Faculty of Mathematics and Physics, University of Ljubljana, Jadranska 19, SI-1000 Ljubljana, Slovenia}
\author{Lorenzo Piroli}
\affiliation{Max-Planck-Institut f\"ur Quantenoptik, Hans-Kopfermann-Str.~1, 85748 Garching, Germany}
\affiliation{Munich Center for Quantum Science and Technology, Schellingstra\ss e 4, 80799 M\"unchen, Germany}

\date{\today}

\begin{abstract}
	We study the scrambling of quantum information in local random unitary circuits by focusing on  the tripartite information proposed by Hosur \emph{et al}. We provide exact results for the averaged R\'enyi-$2$ tripartite information in two cases: (i) the local gates are Haar random and (ii) the local gates are dual-unitary and randomly sampled from a single-site Haar-invariant measure. We show that the  latter case defines a one-parameter family of circuits, and prove that for a ``maximally chaotic'' subset of this family quantum information is scrambled faster than in the Haar-random case. Our approach is based on a standard mapping onto an averaged folded tensor network, that can be studied by means of appropriate recurrence relations. By means of the same method, we also revisit the computation of out-of-time-ordered correlation functions, re-deriving known formulae  for Haar-random unitary circuits, and presenting an exact result for maximally chaotic random dual-unitary gates. 
\end{abstract}

\maketitle

\section{Introduction}

Characterising chaos and dynamical complexity in quantum many-body physics is a long standing problem. While these concepts can be defined very precisely in classical mechanics~\cite{arnold, cornfeld, gaspard, ott}, the extension to the quantum theory is far from straightforward. In the past decades, a large body of the literature has focused on quantum Hamiltonians with a well-defined semi-classical limit, aiming at understanding the features that give rise to classical chaotic behaviour when  $\hbar\to 0$~\cite{Berry1991}. Unfortunately, this approach is inadequate to analyse many-body systems such as quantum spin-$1/2$ chains, where there is no obvious semi-classical limit and the problem appears even harder~\cite{JPA07}. 

Recently, a renewed interest in quantum chaos came from the study of black hole physics, also due to the discovery of interesting connections to the phenomenon of \emph{scrambling} of quantum information~\cite{HaPr07, SeSu08}. In this context, an important realisation was that several aspects of quantum chaos and scrambling are encoded in how the support of local observables increases during the quantum dynamics, as computed by the out-of-time-ordered correlation (OTOC) functions~\cite{LaOv69, LSHO13, ShSt14, Kitaev_talk, ShSt14_bh,RoSS15,MaSS16,RoSS18}. The latter are expected to be  complementary to other measures quantifying fingerprints of quantum chaos, such as the operator-space entanglement entropy (OSEE)~\cite{PL:OE, Zana01, PZ07,  PP:OEIsing, PP:OEXY,AlDM19,BKP:OEsolitons,BKP:OEergodicandmixing, JHN:OperatorEntanglement} of local operators.

The appeal to characterise chaos in terms of the dynamics of local operators lies in the fact that one does not need to work in a semi-classical regime, and that quantities such as OTOCs can be computed (and in principle measured) directly for any many-body quantum system~\cite{TsWU17, AlFI16,SwCh17,Yung17,PaSa17,KuGP17,DoMo17,LiMo18,LiMo18,SKMK18,RoLa18,LFSL19,YoYa19,NIDS19,McNK19, HuBZ19,CLBS19, XuSw20}. In practice, however, the calculation of OTOCs represents a major challenge, which has provided an  increasing motivation to find simplified models for the chaotic dynamics where these objects could be studied in some detail~\cite{RandomCircuitsEnt,Nahum:operatorspreadingRU,Keyserlingk,RPK:conservationlaws1,Kemani,Chalker,Chalker2,SPHS18,BKP:kickedIsing,KoLP18,Chalker3, BrVi10,ShSt15,SPQS19,GHST18,ZhCh19,SaSS18,PiSQ20}. In particular, a lot of attention has been devoted to different kinds of \emph{random} unitary quantum circuits, both with local~\cite{RandomCircuitsEnt,Nahum:operatorspreadingRU,Keyserlingk,RPK:conservationlaws1,Kemani,Chalker,Chalker2,SPHS18, Chalker3, JHN:OperatorEntanglement, LiCF18,CBQA19,GuHu19,LiCF19,SkRN19,VPYL19,ZGWG19,BaCA20, ZN:statmech, RPK:conservationlaws, H:conservationlaws, RPK:conservationlaws2,LaAB20,Hunt19} and non-local gates~\cite{BrVi10,ShSt15,SPQS19,GHST18,ZhCh19,SaSS18,PiSQ20}. At the same time, interesting classes of non-disordered, \emph{dual unitary} quantum circuits have been introduced~\cite{BKP:dualunitary, BKP:kickedIsing} and studied~\cite{BKP:entropy, BKP:OEsolitons, BKP:OEergodicandmixing,GoLa19,PBCP20,RaAL19,ClLa20,GBAW20,CL:OTOCsDU}, displaying the unique property of being both chaotic and analytically tractable. In fact, apart from representing a convenient idealisation of the generic quantum many-body dynamics, these systems can also be realised in practice using arrays of superconducting qubits~\cite{Google}.

We note that, despite being very useful probes for the operator growth, some aspects of OTOCs are still not completely understood. For instance, in lattice models with a finite local Hilbert-space dimension,  the semi-classical justification of OTOCs appears problematic~\cite{KuGP17}. Furthermore, it is presently debated which of their features, if any, would be able to distinguish a typical local Hamiltonian dynamics from an \emph{interacting} integrable one~\cite{GHKV18}. For these reasons, it is still of great interest to consider other possible measures of quantum chaos and scrambling, and study their behaviour in different models.

A relevant example in this direction is given by the (negative) tripartite information (TI) of the evolution operator introduced in Ref.~\cite{HQRY:tripartiteinfo}. This was suggested as a valuable tool to quantify the scrambling power of a quantum channel, namely its ability to delocalise information provided as an input. Differently from OTOCs, this quantity is not defined in terms of the dynamics of local operators, but is rather a property of the time-evolution operator itself. Although very appealing, the TI turned out to be very difficult to compute. So far, it has only been obtained  numerically in non-local random circuits~\cite{SPQS19} and for small systems without disorder~\cite{SBPM19,HQRY:tripartiteinfo}, while analytic results could be derived exclusively for ``perfect tensor'' circuits~\cite{HQRY:tripartiteinfo}, defined by gates that remain unitary under arbitrary permutation of their indices (see also Refs.~\cite{IySa18, PRZI18, SeML18} , where the tripartite information of given states, and not of the channel, was considered). Clearly, it would be highly desirable to study the TI in other models, to further explore its general features.

In this work we address this issue by computing exactly the TI in local random unitary circuits. In particular, we consider its R\'enyi-$2$ version, and focus on two relevant cases: (i) the local gates are Haar random and (ii) the local gates are dual-unitary and randomly sampled from a single-site Haar-invariant probability distribution. We also formulate a conjecture for the behaviour of the TI in completely chaotic~\cite{BKP:OEergodicandmixing} clean dual-unitary circuits. 

Our approach is based on a standard mapping of averages of relevant physical quantities to ``folded'' tensor networks, which are studied by deriving and solving appropriate recurrence equations. We note that similar recurrence relations appeared recently in the literature of quantum circuits (not necessary random)~\cite{ZN:statmech, ZN:nonrandommembrane}, resulting from a mapping to a classical spin model. 

As a further application of this method, we also revisit the computation of OTOCs. We will show that our recurrence relations can be solved exactly for Haar-random unitary circuits, re-deriving the formulae first obtained in Refs.~\cite{Nahum:operatorspreadingRU,Keyserlingk}, by means of a the aforementioned spin-model mapping~\cite{ZN:nonrandommembrane,ZN:statmech}. Finally, we also present an exact result for random dual-unitary gates.

The rest of this article is organised as follows. In Sec.~\ref{sec:unitary_circuits} we introduce the unitary circuits studied in this work and lay out the formalism that we will employ for our calculations. Sec.~\ref{sec:TI} is devoted to the computation of the tripartite information, while in Sec.~\ref{sec:OTOC} we tackle OTOCs. Finally, our conclusions are reported in Sec.~\ref{sec:conclusions}, while the most technical parts of our work are consigned to several appendices.

\section{Formalism}
\label{sec:unitary_circuits}

We consider a chain of $d$-level systems (qudits), described by a local Hilbert space $\mathcal{H}_j\simeq \mathbb{C}^d$ with basis vectors $\ket{j}, j = 0 ,\ldots, d-1$. We are interested in local unitary quantum circuits where the dynamics is implemented by subsequent discrete applications of the evolution operator   
\be
\mathbb U = \bigotimes_{x \,\in \mathbb Z} U_{x,x+1/2}\bigotimes_{x \in  \mathbb Z+1/2} U_{x,x+1/2}\,.
\ee
Here $U_{x,y}\in {\rm End}(\mathbb C^d \otimes \mathbb C^d)$ are two-qudit gates acting on the local spaces labeled by $x$ and $y$. The quantum circuit dynamics is conveniently represented in terms of ``brick-wall'' diagrams which are reminiscent of the traditional notation of tensor-network theory. In particular, in this language matrix elements of local operators are denoted by boxes with a number of incoming and outgoing legs. To each leg corresponds an index associated with one of the local spaces on which the local operator acts on. For instance, the two-qudit unitary gates $U$ and $U^{\dagger}$ are written as
\begin{equation}
\label{eq:U}
U_{i,j}^{k,l}=
\begin{tikzpicture}[baseline=(current  bounding  box.center), scale=1]
\def\eps{0.5};
\draw[thick] (-2.25,0.5)node[left]{$k$}-- (-1.25,-0.5)node[right]{$j$};
\draw[thick] (-2.25,-0.5)node[left]{$i$} -- (-1.25,0.5)node[right]{$l$};
\draw[ thick, fill=myred, rounded corners=2pt] (-2,0.25) rectangle (-1.5,-0.25);
\draw[thick] (-1.75,0.15)-- (-1.6,0.15) -- (-1.6,0);
\end{tikzpicture}\,,
\qquad
\left(U^{\dagger}\right)_{i,j}^{k,l}=
\begin{tikzpicture}[baseline=(current  bounding  box.center), scale=1]
\def\eps{0.5};
\draw[thick] (-2.25,0.5)node[left]{$k$}-- (-1.25,-0.5)node[right]{$j$};
\draw[thick] (-2.25,-0.5)node[left]{$i$} -- (-1.25,0.5)node[right]{$l$};
\draw[ thick, fill=myblue, rounded corners=2pt] (-2,0.25) rectangle (-1.5,-0.25);
\draw[thick] (-1.75,0.15) -- (-1.6,0.15) -- (-1.6,0);
\end{tikzpicture}\,.
\end{equation}
When legs of different operators are joined together a sum over the index of the corresponding local space is understood. Note that we added a mark to stress that $U$ and $U^{\dagger}$ are generically not symmetric under space reflection (left to right flip) and time reversal (up to down flip, transposition of $U$). The time direction runs from bottom to top, hence lower legs correspond to incoming indices (matrix row) and upper legs to outgoing indices (matrix column). Finally, an explicit label for the legs can omitted when it does not generate confusion.

In this work we will be interested in two classes of models, where the two-site gates~\eqref{eq:U} are drawn out of an appropriate ensemble.

\subsection{Haar-random quantum circuits}

The first class we consider is that of Haar-random local  quantum circuits~\cite{RandomCircuitsEnt}, where the two-qudit unitaries are chosen by sampling the unitary group $U(d^2)$ from a uniform Haar probability distribution. 

Haar-random local quantum circuits have been extensively studied over the past few years. One of the main reasons lies in the fact that Haar-averages make the computation of several physical quantities within the reach of analytic inspection, providing tractable models for the chaotic dynamics and allowing, for instance, for the derivation of remarkable results on entanglement spreading~\cite{RandomCircuitsEnt, LiCF18,CBQA19,GuHu19,LiCF19,SkRN19,VPYL19,ZGWG19,BaCA20,LaAB20, ZN:statmech} and the growth of local operators ~\cite{Nahum:operatorspreadingRU,Keyserlingk,RPK:conservationlaws1,Kemani}.  

In the following, we will denote Haar averages of physical quantities by $\mathbb{E}_{\rm Haar}[\ldots]$, where the subscript will be omitted when it does not generate confusion.

\subsection{Dual-unitary quantum circuits}
\label{sec:du_random}

As a second class of models, we will consider dual-unitary quantum circuits, focusing in particular on two-site qubit gates ($d=2$) sampled from a single-site Haar-invariant measure, as we explain below.

Dual-unitary quantum circuits have been recently introduced in Ref.~\cite{BKP:dualunitary}, as analytically tractable \emph{non-disordered} models for the chaotic quantum dynamics. The defining property of a dual unitary gate $U$  is that it remains unitary after a given reshuffling of indices. More precisely, defining $\tilde{U}$ by
\be
\langle k|\langle l| \tilde{U}|i\rangle|j\rangle = \langle j|\langle l| U|i\rangle|k\rangle\,,
\ee
we say that $U$ is dual-unitary if  the operator $\tilde{U}$ is also unitary.  

We note that  both the unitarity and dual-unitarity conditions can be expressed in a simple way by means of the graphical notation introduced in Eq.~\eqref{eq:U}. In particular unitarity is represented as 
\be
\begin{tikzpicture}[baseline=(current  bounding  box.center), scale=0.8]
\def\ep{.5};
\def\eps{0.8};
\draw[thick] (-2.25,1) -- (-1.75,0.5);
\draw[thick] (-1.75,0.5) -- (-1.25,1);
\draw[thick] (-2.25,-1) -- (-1.75,-0.5);
\draw[thick] (-1.75,-0.5) -- (-1.25,-1);
\draw[thick] (-1.9,0.35) to[out=170, in=-170] (-1.9,-0.35);
\draw[thick] (-1.6,0.35) to[out=10, in=-10] (-1.6,-0.35);
\draw[thick, fill=myred, rounded corners=2pt] (-2,0.25) rectangle (-1.5,0.75);
\draw[thick, fill=myblue, rounded corners=2pt] (-2,-0.25) rectangle (-1.5,-0.75);
\draw[thick] (-1.75,0.65) -- (-1.6,0.65) -- (-1.6,0.5);
\draw[thick] (-1.75,-0.35) -- (-1.6,-0.35) -- (-1.6,-0.5);
\Text[x=-1.25,y=-0.5-\ep]{}
\Text[x=-0.4,y=0]{$=$}
\draw[ thick] (.7,-.75) -- (.7,.75) (1.3,-0.75) -- (1.3,0.75) (.7,-.75) -- (.45,.-1) (1.3,-.75) -- (1.55,.-1) (.7,.75) -- (.45,1) (1.3,.75) -- (1.55,1);
\Text[x=1.6,y=0]{,}
\end{tikzpicture}
\qquad
\begin{tikzpicture}[baseline=(current  bounding  box.center), scale=0.8]
\def\ep{.5};
\def\eps{0.8};
\draw[thick] (-2.25,1) -- (-1.75,0.5);
\draw[thick] (-1.75,0.5) -- (-1.25,1);
\draw[thick] (-2.25,-1) -- (-1.75,-0.5);
\draw[thick] (-1.75,-0.5) -- (-1.25,-1);
\draw[thick] (-1.9,0.35) to[out=170, in=-170] (-1.9,-0.35);
\draw[thick] (-1.6,0.35) to[out=10, in=-10] (-1.6,-0.35);
\draw[thick, fill=myblue, rounded corners=2pt] (-2,0.25) rectangle (-1.5,0.75);
\draw[thick, fill=myred, rounded corners=2pt] (-2,-0.25) rectangle (-1.5,-0.75);
\draw[thick] (-1.75,0.65) -- (-1.6,0.65) -- (-1.6,0.5);
\draw[thick] (-1.75,-0.35) -- (-1.6,-0.35) -- (-1.6,-0.5);
\Text[x=-1.25,y=-0.5-\ep]{}
\Text[x=-0.4,y=0]{$=$}
\draw[thick] (.7,-.75) -- (.7,.75) (1.3,-0.75) -- (1.3,0.75) (.7,-.75) -- (.45,.-1) (1.3,-.75) -- (1.55,.-1) (.7,.75) -- (.45,1) (1.3,.75) -- (1.55,1);
\Text[x=1.6,y=0]{,}
\end{tikzpicture}
\label{eq:unitarity_pic}
\ee
while dual-unitarity reads
\be
\begin{tikzpicture}[baseline=(current  bounding  box.center), scale=0.8]
\def\eps{0.2};
\def\ep{1.3}
\Text[x=-0.125,y=-\ep]{}
\draw[ thick] (-1.75+2,0.5+\eps) -- (-1.25+2,1+\eps);
\draw[ thick] (-1.25+2,0+\eps) -- (-1.75+2,0.5+\eps);
\draw[ thick] (-1.25+2,0-\eps) -- (-1.75+2,-0.5-\eps);
\draw[ thick] (-1.75+2,-0.5-\eps) -- (-1.25+2,-1-\eps);
\draw[ thick] (-1.9+2,0.7+\eps) to[out=170, in=-170] (-1.9+2,-0.7-\eps);
\draw[ thick] (-1.9+2,0.3+\eps) to[out=170, in=-170] (-1.9+2,-0.3-\eps);
\draw[thick, fill=myblue, rounded corners=2pt] (-2+2,0.25+\eps) rectangle (-1.5+2,0.75+\eps);
\draw[thick, fill=myred, rounded corners=2pt] (-2+2,-0.25-\eps) rectangle (-1.5+2,-0.75-\eps);
\draw[thick] (.25,0.65+\eps) -- (.4,0.65+\eps) -- (.4,0.5+\eps);
\draw[thick] (.25,-0.35-\eps) -- (.4,-0.35-\eps) -- (.4,-0.5-\eps);
\draw[ thick] (-1.45+4,0.7+\eps) to[out=170, in=-170] (-1.45+4,-0.7-\eps);
\draw[ thick] (-1.45+4,0.3+\eps) to[out=170, in=-170] (-1.45+4,-0.3-\eps);
\draw[ thick] (-1.1+4,0.7+\eps) -- (-1.45+4,0.7+\eps)(-1.1+4,0.7+\eps) -- (-.75+4,1+\eps);
\draw[ thick] (-1.1+4,0.3+\eps) -- (-1.45+4,0.3+\eps)(-1.1+4,0.3+\eps) -- (-.75+4,+\eps);
\draw[ thick] (-1.1+4,-0.3-\eps) -- (-1.45+4,-0.3-\eps)(-1.1+4,-0.3-\eps) -- (-.75+4,-\eps);
\draw[ thick] (-1.1+4,-0.7-\eps) -- (-1.45+4,-0.7-\eps)(-1.1+4,-0.7-\eps) -- (-.75+4,-1-\eps);
\Text[x=1.45,y=0]{$=$}
\Text[x=3.45,y=0]{,}
\end{tikzpicture}
\qquad 
\begin{tikzpicture}[baseline=(current  bounding  box.center), scale=0.8]
\def\eps{0.2};
\def\ep{1.3}
\Text[x=-0.125,y=-\ep]{}
\draw[ thick] (-2.25+2,1+\eps) -- (-1.75+2,0.5+\eps);
\draw[ thick] (-1.75+2,0.5+\eps) -- (-2.25+2,0+\eps);
\draw[ thick] (-1.75+2,-0.5-\eps) -- (-2.25+2,0-\eps);
\draw[ thick] (-2.25+2,-1-\eps) -- (-1.75+2,-0.5-\eps);
\draw[ thick] (-1.6+2,0.7+\eps) to[out=10, in=-10] (-1.6+2,-0.7-\eps);
\draw[ thick] (-1.6+2,0.3+\eps) to[out=10, in=-10] (-1.6+2,-0.3-\eps);
\draw[thick, fill=myblue, rounded corners=2pt] (-2+2,0.25+\eps) rectangle (-1.5+2,0.75+\eps);
\draw[thick, fill=myred, rounded corners=2pt] (-2+2,-0.25-\eps) rectangle (-1.5+2,-0.75-\eps);
\draw[thick] (.25,0.65+\eps) -- (.4,0.65+\eps) -- (.4,0.5+\eps);
\draw[thick] (.25,-0.35-\eps) -- (.4,-0.35-\eps) -- (.4,-0.5-\eps);
\draw[ thick] (-.7+3.2,0.7+\eps) to[out=10, in=-10] (-.7+3.2,-0.7-\eps);
\draw[ thick] (-.7+3.2,0.3+\eps) to[out=10, in=-10] (-.7+3.2,-0.3-\eps);
\draw[ thick] (-.7+3.2,0.7+\eps) -- (-1.05+3.2,0.7+\eps)(-1.35+3.2,1+\eps) -- (-1.05+3.2,0.7+\eps);
\draw[ thick] (-.7+3.2,0.3+\eps) -- (-1.05+3.2,0.3+\eps)(-1.35+3.2,+\eps) -- (-1.05+3.2,0.3+\eps);
\draw[ thick] (-.7+3.2,-0.3-\eps) -- (-1.05+3.2,-0.3-\eps)(-1.35+3.2,-\eps) -- (-1.05+3.2,-0.3-\eps);
\draw[ thick] (-.7+3.2,-0.7-\eps) -- (-1.05+3.2,-0.7-\eps)(-1.35+3.2,-1-\eps) -- (-1.05+3.2,-0.7-\eps);
\Text[x=1.5,y=0]{$=$}
\end{tikzpicture}.
\label{eq:dual_unitarity_pic}
\ee
Here continuous solid lines represent the identity operator.

In the case of qubits, \emph{i.e.} circuits with local dimension $d=2$, dual-unitary gates can be completely classified~\cite{BKP:dualunitary}. In particular, an arbitrary member of this family can be parameterised as
\be
U=e^{i \phi} (u_+ \otimes u_-)\cdot V[J]\cdot (v_-\otimes v_+)\,,
\label{eq:dualunitaryU}
\ee
where $\phi, J \in \mathbb R$, $u_\pm,v_\pm\in {\rm SU}(2)$ and 
\be
\!\!\!V[J]\!=\! \exp\!\!\left[-i\!\left(\frac{\pi}{4} \sigma^x\otimes\sigma^x +\frac{\pi}{4} \sigma^y\otimes\sigma^y+J \sigma^z\otimes\sigma^z\right)\!\right]\!.
\label{eq:v_mat_dual}
\ee
Some examples of dual unitarity gates for local dimension $d> 2$ have been constructed in Refs.~\cite{RaAL19,GBAW20}, although no complete parametrisation is known in these cases.

Several studies have now shown that the dual-unitarity condition allows for the derivation of exact results for interesting physical quantities such as dynamical correlation functions~\cite{BKP:dualunitary}, the dynamics of entanglement and correlations after a quench~\cite{BKP:entropy,PBCP20}, and the spectral statistics~\cite{BKP:kickedIsing}.  Up to now, however, the computation of dynamical chaos indicators (such as Local-Operator Entanglement or OTOCs) in the chaotic regime has been achieved only by formulating suitable  conjectures~\cite{BKP:OEergodicandmixing,ClLa20}. Here we follow a different route, showing that exact statements can be made by sampling dual-unitary gates from an appropriate probability distribution and considering the averaged results. From the  structure of the two-qubit gates~\eqref{eq:dualunitaryU}, it is natural to consider an ensemble where the matrix $V[J]$ is fixed, while the operator $u_{\pm}$, $v_{\pm}$ are drawn from a Haar-invariant distribution over the group $U(2)$, and independently at different time steps. This defines a single-site Haar-invariant measure for the two-site gates. This means that the average of physical quantities is not affected by transformations of the form
\be
U_{i} \leftrightarrow\left(w_{1} \otimes w_{2}\right) U_{i}\left(w_{3} \otimes w_{4}\right)
\ee
for arbitrary choices of $w_i\in U(2)$ (a similar random ensemble --- not involving dual-unitary matrices --- was considered in Ref.~\cite{SPHS18}). We stress that the family of random  dual-unitary circuits defined above depends on one free parameter $J$. In other words, this single-site Haar-invariant measure is more constrained that the one adopted in Haar-random circuits and, as a consequence, the averaged gate (see Sec.~\ref{sec:folding} for its definition and Eq.~\eqref{eq:averagedgateDU} for its explicit expression) has more structure in this case. 

In the following, we will denote averages of physical quantities over the above random ensemble by $\mathbb{E}_{\rm d.u.}[\ldots]$, where the subscript will be omitted when it does not generate confusion. As in the case of Haar-random quantum circuits, the averaged results are expected to be representative of the individual dual-unitary realizations in the ensemble.

{\subsection{The folded picture}
\label{sec:folding}

\begin{figure}
\begin{tikzpicture}[baseline=(current  bounding  box.center), scale=0.4]
\def\eps{0};
\def\shift{11}
\def\shifty{-2.5}
\foreach \i in{0,...,3}
{
\draw[thick] (-9.5+2*\i,6) arc (45:180:0.15);
\draw[thick] (-9.5+2*\i+1,6) arc (135:0:0.15);
\draw[thick] (-9.5+2*\i,-6) arc (-45:-180:0.15);
\draw[thick] (-9.5+2*\i+1,-6) arc (-135:0:0.15);
}
\foreach \i in {1,...,4}
{
\draw[thick] (-.5-2*\i,1) -- (0.5-2*\i,0);
\draw[thick] (-0.5-2*\i,0) -- (0.5-2*\i,1);
\draw[thick] (-.5-2*\i,-1+\eps) -- (0.5-2*\i,\eps);
\draw[thick] (-0.5-2*\i,0) -- (-0.5-2*\i,\eps);
\draw[thick] (-0.5-2*\i,\eps) -- (0.5-2*\i,-1+\eps);
\draw[thick] (-0.5-2*\i+1,0) -- (-0.5-2*\i+1,\eps);
\draw[thick, fill=myblue, rounded corners=1pt] (-0.25-2*\i,0.25) rectangle (.25-2*\i,0.75);
\draw[thick] (-2*\i,0.65) -- (.15-2*\i,.65) -- (.15-2*\i,0.5);
\draw[thick, fill=myred, rounded corners=1pt] (-0.25-2*\i,-0.25+\eps) rectangle (.25-2*\i,-0.75+\eps);
\draw[thick] (-2*\i,-0.35+\eps) -- (.15-2*\i,-.35+\eps) -- (.15-2*\i,-0.5+\eps);
}
\foreach \i in {2,...,5}
{
\draw[thick] (.5-2*\i,6) -- (1-2*\i,5.5);
\draw[thick] (1.5-2*\i,6) -- (1-2*\i,5.5);
\draw[thick] (.5-2*\i,-6+\eps) -- (1-2*\i,-5.5+\eps);
\draw[thick] (1.5-2*\i,-6+\eps) -- (1-2*\i,-5.5+\eps);
}
\foreach \jj[evaluate=\jj as \j using -2*(ceil(\jj/2)-\jj/2)] in {0,...,3}
\foreach \i in {2,...,5}
{
\draw[thick] (.5-2*\i-1*\j,2+1*\jj) -- (1-2*\i-1*\j,1.5+\jj);
\draw[thick] (1-2*\i-1*\j,1.5+1*\jj) -- (1.5-2*\i-1*\j,2+\jj);
}
\foreach \jj[evaluate=\jj as \j using -2*(ceil(\jj/2)-\jj/2)] in {0,...,4}
\foreach \i in {2,...,5}
{
\draw[thick] (.5-2*\i-1*\j,1+1*\jj) -- (1-2*\i-1*\j,1.5+\jj);
\draw[thick] (1-2*\i-1*\j,1.5+1*\jj) -- (1.5-2*\i-1*\j,1+\jj);
\draw[thick, fill=myblue, rounded corners=1pt] (0.75-2*\i-1*\j,1.75+\jj) rectangle (1.25-2*\i-1*\j,1.25+\jj);
\draw[thick] (1-2*\i-1*\j,1.65+1*\jj) -- (1.15-2*\i-1*\j,1.65+1*\jj) -- (1.15-2*\i-1*\j,1.5+1*\jj);
}
\foreach \jj[evaluate=\jj as \j using -2*(ceil(\jj/2)-\jj/2)] in {0,...,3}
\foreach \i in {2,...,5}
{
\draw[thick] (.5-2*\i-1*\j,-2-1*\jj+\eps) -- (1-2*\i-1*\j,-1.5-\jj+\eps);
\draw[thick] (1-2*\i-1*\j,-1.5-1*\jj+\eps) -- (1.5-2*\i-1*\j,-2-\jj+\eps);
}
\foreach \jj[evaluate=\jj as \j using -2*(ceil(\jj/2)-\jj/2)] in {0,...,4}
\foreach \i in {2,...,5}
{
\draw[thick] (.5-2*\i-1*\j,-1-1*\jj+\eps) -- (1-2*\i-1*\j,-1.5-\jj+\eps);
\draw[thick] (1-2*\i-1*\j,-1.5-1*\jj+\eps) -- (1.5-2*\i-1*\j,-1-\jj+\eps);
\draw[thick, fill=myred, rounded corners=1pt] (0.75-2*\i-1*\j,-1.75-\jj+\eps) rectangle (1.25-2*\i-1*\j,-1.25-\jj+\eps);
\draw[thick] (1-2*\i-1*\j,-1.35-1*\jj+\eps) -- (1.15-2*\i-1*\j,-1.35-1*\jj+\eps) -- (1.15-2*\i-1*\j,-1.5-1*\jj+\eps);
}
\draw[thick, fill=black] (-6.5,0) circle (0.1cm); 
\draw[thick, fill=black] (-3.5,6) circle (0.1cm); 
\Text[x=-6.9,y=-0.3]{$a_j$}
\Text[x=-3.9,y=6.5]{$a_k$}
\Text[x=1.2, y=-6.8]{}
\foreach \i in {1,...,4}
{
\draw[very thick] (\shift-.5-2*\i,1+\shifty) -- (\shift+0.5-2*\i,0+\shifty);
\draw[very thick] (\shift-0.5-2*\i,0+\shifty) -- (\shift+0.5-2*\i,1+\shifty);
\draw[ thick, fill=myviolet, rounded corners=1pt] (\shift-0.25-2*\i,0.25+\shifty) rectangle (\shift+.25-2*\i,0.75+\shifty);
\draw[thick] (\shift-2*\i,0.65+\shifty) -- (\shift+.15-2*\i,.65+\shifty) -- (\shift+.15-2*\i,0.5+\shifty);
}
\foreach \i in {2,...,5}
{
\draw[very thick] (\shift+.5-2*\i,6+\shifty) -- (\shift+1-2*\i,5.5+\shifty);
\draw[very thick] (\shift+1.5-2*\i,6+\shifty) -- (\shift+1-2*\i,5.5+\shifty);
}
\foreach \jj[evaluate=\jj as \j using -2*(ceil(\jj/2)-\jj/2)] in {0,...,3}
\foreach \i in {2,...,5}
{
\draw[very thick] (\shift+.5-2*\i-1*\j,2+1*\jj+\shifty) -- (\shift+1-2*\i-1*\j,1.5+\jj+\shifty);
\draw[very thick] (\shift+1-2*\i-1*\j,1.5+1*\jj+\shifty) -- (\shift+1.5-2*\i-1*\j,2+\jj+\shifty);
}
\foreach \i in {1,...,4}
{
\draw[very thick] (\shift-.5-2*\i,1+\shifty) -- (\shift+0.5-2*\i,0+\shifty);
\draw[very thick] (\shift-0.5-2*\i,0+\shifty) -- (\shift+0.5-2*\i,1+\shifty);
\draw[ thick, fill=myviolet, rounded corners=1pt] (\shift-0.25-2*\i,0.25+\shifty) rectangle (\shift+.25-2*\i,0.75+\shifty);
\draw[thick] (\shift-2*\i,0.65+\shifty) -- (\shift+.15-2*\i,.65+\shifty) -- (\shift+.15-2*\i,0.5+\shifty);
}
\foreach \jj[evaluate=\jj as \j using -2*(ceil(\jj/2)-\jj/2)] in {0,...,4}
\foreach \i in {2,...,5}
{
\draw[very thick] (\shift+.5-2*\i-1*\j,1+1*\jj+\shifty) -- (\shift+1-2*\i-1*\j,1.5+\jj+\shifty);
\draw[very thick] (\shift+1-2*\i-1*\j,1.5+1*\jj+\shifty) -- (\shift+1.5-2*\i-1*\j,1+\jj+\shifty);
\draw[ thick, fill=myviolet, rounded corners=1pt] (\shift+0.75-2*\i-1*\j,1.75+\jj+\shifty) rectangle (\shift+1.25-2*\i-1*\j,1.25+\jj+\shifty);
\draw[thick] (\shift+1-2*\i-1*\j,1.65+1*\jj+\shifty) -- (\shift+1.15-2*\i-1*\j,1.65+1*\jj+\shifty) -- (\shift+1.15-2*\i-1*\j,1.5+1*\jj+\shifty);
}
\draw[ thick, fill=white] (\shift-8.5,\shifty) circle (0.1cm); 
\draw[ thick, fill=white] (\shift-7.5,\shifty) circle (0.1cm);
\draw[ thick, fill=white] (\shift-6.5,\shifty) circle (0.1cm); 
\draw[ thick, fill=white] (\shift-5.5,\shifty) circle (0.1cm); 
\draw[ thick, fill=white] (\shift-4.5,\shifty) circle (0.1cm); 
\draw[ thick, fill=white] (\shift-3.5,\shifty) circle (0.1cm); 
\draw[ thick, fill=white] (\shift-2.5,\shifty) circle (0.1cm); 
\draw[ thick, fill=white] (\shift-1.5,\shifty) circle (0.1cm);
\draw[ thick, fill=white] (\shift-8.5,\shifty+6) circle (0.1cm); 
\draw[ thick, fill=white] (\shift-7.5,\shifty+6) circle (0.1cm);
\draw[ thick, fill=white] (\shift-6.5,\shifty+6) circle (0.1cm); 
\draw[ thick, fill=white] (\shift-5.5,\shifty+6) circle (0.1cm); 
\draw[ thick, fill=white] (\shift-4.5,\shifty+6) circle (0.1cm); 
\draw[ thick, fill=white] (\shift-3.5,\shifty+6) circle (0.1cm); 
\draw[ thick, fill=white] (\shift-2.5,\shifty+6) circle (0.1cm); 
\draw[ thick, fill=white] (\shift-9.5,\shifty+6) circle (0.1cm); 
\draw[thick, fill=black] (\shift-6.5,\shifty) circle (0.1cm); 
\draw[thick, fill=black] (\shift-3.5,6+\shifty) circle (0.1cm); 
\Text[x=-6.8+\shift,y=-0.4+\shifty]{$a_j$}
\Text[x=-3.9+\shift,y=6.5+\shifty]{$a_k$}
\Text[x=0,y=0]{$\longmapsto$}
\end{tikzpicture}
\caption{Pictorial representation of the folding procedure for the two-point function defined in Eq.~\eqref{eq:two_point_function} by means of the graphical notation introduced in Eq.~\eqref{eq:U} (left diagram). It is understood that lower and upper open lines are joined together. By ``folding'' the picture, each operator $U^\dag$ ends up lying on top of the corresponding gate $U^{T}$, leading  to an evolution dictated by the doubled gates $U^\dag\otimes U^{T}$. Note that white circles correspond to the boundary conditions induced by the trace in Eq.~\eqref{eq:two_point_function}, cf. Sec.~\ref{sec:folding}.}
	\label{fig:folded_picture}
\end{figure}
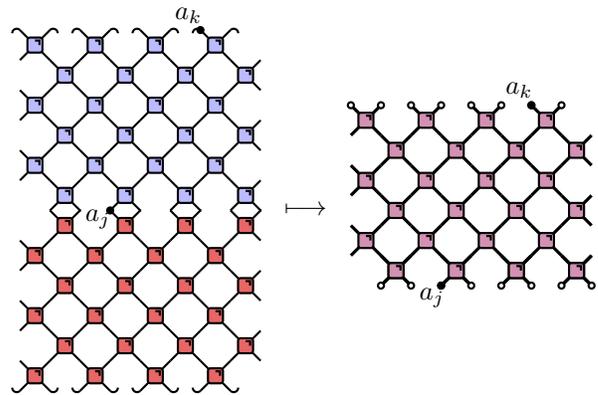

In this work we will make use of the so-called folded picture, which is by now a standard tool in tensor-network theory~\cite{BHVC09,MuCB12}, and have been used extensively in the recent literature of both clean~\cite{BKP:OEsolitons, BKP:OEergodicandmixing, ZN:nonrandommembrane} and random quantum circuits~\cite{Nahum:operatorspreadingRU, JHN:OperatorEntanglement,Keyserlingk,ZN:statmech}.

In essence, the folded picture consists in a way to represent diagrammatically quantities involving products of operators $\mathbb{U}$ and $\mathbb{U}^{\dagger}$. The basic idea is easily explained by considering the two-point correlation function on the infinite temperature state 
\be
\mathcal{C}_{j,k}(t)=\frac{1}{d^{2L}}{\rm tr}\left[a_j(t)a_k\right]\,,
\label{eq:two_point_function}
\ee
where $a_j(t)=\left[\mathbb{U}^\dagger \right]^t a_j \mathbb{U}^t$, with $a_j$ an operator acting non-trivially only on the local space $j$ and $2L$ is the length of the qudit chain. Eq.~\eqref{eq:two_point_function} can be represented by means of the graphical notation introduced in Eq.~\eqref{eq:U}, as reported in Fig.~\ref{fig:folded_picture}. We can now imagine to ``fold'' the picture, so that, after folding, each operator $U^\dag$ ends up lying on top of the corresponding gate $U^{T}$ ($(\cdot)^T$ denotes the transposition). This leads to a representation where the evolution is dictated by the ``doubled'' gates $U^\dag \otimes U^{T}$, acting on a pair of  ``doubled'' qudits, each one associated with a local space $\mathcal{H}_j\otimes \mathcal{H}_j $.

The advantage of the folded picture lies in the fact that the building blocks $U^\dag \otimes U^{T}$ can be averaged independently from one another. Furthermore, depending  on  the probability  distribution chosen, the disorder-averaged gates $\mathbb{E}\left[U^\dag \otimes U^{T}\right]$ might display a relatively simple structure. 

In this work we will be interested in quantities that involve two copies of the time evolution operator $\mathbb{U}$ and two of its conjugate $\mathbb{U}^{\dagger}$, so that we will need to repeat the folding procedure two times. The building block associated with this picture is  $U^\dag \otimes U^{T}\otimes U^\dag \otimes U^{T}$, for which we introduce the graphical notation
\begin{equation}
\begin{tikzpicture}[baseline=(current  bounding  box.center), scale=1]
\def\eps{0.5};
\Wgate{-3.75}{0.2};
\Text[x=-2.75,y=0.2, anchor = center]{$=$}
\draw[thick] (-.45,0.95) -- (.55,-0.05);
\draw[thick] (-.45,-0.05) -- (.55,0.95);
\draw[ thick, fill=myred, rounded corners=2pt] (-0.2,0.7) rectangle (0.3,0.2);
\draw[thick] (.05,0.3) -- (.2,0.3) -- (.2,0.45);

\draw[thick] (-1.05,0.8) -- (-0.05,-0.2);
\draw[thick] (-1.05,-0.2) -- (-0.05,0.8);
\draw[ thick, fill=myblue, rounded corners=2pt] (-0.8,0.55) rectangle (-.3,0.05);
\draw[thick] (-.55,0.45) -- (-.4,0.45) -- (-.4,0.3);

\draw[thick] (-1.65,0.65) -- (-0.65,-0.35);
\draw[thick] (-1.65,-0.35) -- (-0.65,0.65);
\draw[ thick, fill=myred, rounded corners=2pt] (-1.4,0.4) rectangle (-.9,-0.1);
\draw[thick] (-1.15,0) -- (-1,0) -- (-1,0.15);

\draw[thick] (-2.25,0.5) -- (-1.25,-0.5);
\draw[thick] (-2.25,-0.5) -- (-1.25,0.5);
\draw[ thick, fill=myblue, rounded corners=2pt] (-2,0.25) rectangle (-1.5,-0.25);
\draw[thick] (-1.75,0.15) -- (-1.6,0.15) -- (-1.6,0);

\Text[x=-4.8,y=0.2]{$W=$}
\end{tikzpicture}\,.
\label{eq:W}
\end{equation}
The folded local gate $W$ acts on ${\rm End}(\mathbb C^{d^4}\otimes \mathbb C^{d^4})$, namely the thick legs are now $d^4$-dimensional. For both Haar-random~\cite{Nahum:operatorspreadingRU} and random dual-unitary quantum circuits, $W$ can be computed exactly. In particular, as we will explicitly show later, in both cases $W$ acts non-trivially on a proper subspace of $\mathbb C^{d^4}\otimes \mathbb C^{d^4}$, leading to a lower effective local dimension for the dynamics in the folded picture.

Before leaving this section, we introduce a last piece of notation which we will use extensively in the next sections. In particular, we define the following two states in the $4$-fold local Hilbert space $\mathcal{H}_j^{\otimes 4}$
\begin{align}
\ket{\mcirc}&=\frac{1}{d}\left(\sum_{j=0}^{d-1}\ket{j}_1\ket{j}_2\right)\otimes \left(\sum_{j=0}^{d-1}\ket{j}_3\ket{j}_4\right)\,,\label{eq:square}\\
\ket{\msquare}&=\frac{1}{d}\left(\sum_{j=0}^{d-1}\ket{j}_1\ket{j}_4\right)\otimes \left(\sum_{j=0}^{d-1}\ket{j}_2\ket{j}_3\right)\,,\label{eq:circle}
\end{align}
 which are products of maximally entangled qudits between different pairs of the four local copies of $\mathcal{H}_j$. Note that $\braket{\mcirc|\mcirc}=\braket{\msquare|\msquare}=1$, and
 \be
 \braket{\mcirc|\msquare}=\frac{1}{d}\,.
 \ee
We chose to label these states by $\ket{\mcirc}$ and $\ket{\msquare}$ to make direct contact with the graphical notation that we use to denote them, i.e. a circle and a square respectively. Using this notation, we can rewrite Eqs.~\eqref{eq:square} and \eqref{eq:circle} as
\be
\begin{tikzpicture}[baseline={([yshift=-.5ex]current bounding box.center)}, scale=0.85]\draw[very thick] (-1.5,0) -- (-1.25,0);
\draw[thick, fill=white] (-1.5,0) circle (0.15cm); 
\Text[x=-0.8,y=0]{$= \frac{1}{d}$}
\draw[thick, fill=white] (-0.1,0.5) arc (90:270:0.2cm); 
\draw[thick, fill=white] (-0.1,-0.1) arc (90:270:0.2cm); 
\end{tikzpicture}\,,
\qquad\qquad
\begin{tikzpicture}[baseline={([yshift=-.5ex]current bounding box.center)}, scale=0.85]\draw[very thick] (-1.5,-0.2) -- (-1.25,-0.2);
\draw[thick, fill=white] (-1.65,-0.35) rectangle (-1.35,-0.05); 
\Text[x=-0.8,y=-0.2]{$= \frac{1}{d}$}
\draw[thick] (0.15,-0.1) arc (90:270:0.15cm); 
\draw[thick] (0.15,0.2) arc (90:270:0.45cm); 
\end{tikzpicture}\,.
\label{eq:states}
\ee
The notation introduced in Eq.~\eqref{eq:states} is particularly convenient, because it allows us to write down in a simple way several identities stemming from unitarity and dual unitarity. For example, it is a simple exercise to verify that for unitary gates
\begin{align}
&\begin{tikzpicture}[baseline=(current  bounding  box.center), scale=1]
\def\eps{0.5};
\draw[very thick] (-4.25,0.5) -- (-3.25,-0.5);
\draw[very thick] (-4.25,-0.5) -- (-3.25,0.5);
\draw[ thick, fill=myvioletc, rounded corners=2pt] (-4,0.25) rectangle (-3.5,-0.25);
\draw[thick] (-3.75,0.15) -- (-3.6,0.15) -- (-3.6,0);
\Text[x=-2.75,y=0.0, anchor = center]{$=$}
\draw[thick, fill=white] (-4.25,-0.5) circle (0.1cm); 
\draw[thick, fill=white] (-3.25,-0.5) circle (0.1cm); 
\draw[very thick] (-2.25,0.5) -- (-2.25,-0.5);
\draw[very thick] (-1.25,-0.5) -- (-1.25,0.5);
\draw[thick, fill=white] (-2.25,-0.5) circle (0.1cm); 
\draw[thick, fill=white] (-1.25,-0.5) circle (0.1cm); 
\Text[x=-1,y=0.0, anchor = center]{,}
\end{tikzpicture}
&&\begin{tikzpicture}[baseline=(current  bounding  box.center), scale=1]
\def\eps{0.5};
\draw[very thick] (-4.25,0.5) -- (-3.25,-0.5);
\draw[very thick] (-4.25,-0.5) -- (-3.25,0.5);
\draw[ thick, fill=myvioletc, rounded corners=2pt] (-4,0.25) rectangle (-3.5,-0.25);
\draw[thick] (-3.75,0.15) -- (-3.6,0.15) -- (-3.6,0);
\Text[x=-2.75,y=0.0, anchor = center]{$=$}
\draw[thick, fill=white] (-4.35,-0.6) rectangle (-4.15,-0.4); 
\draw[thick, fill=white] (-3.35,-0.6) rectangle (-3.15,-0.4); 
\draw[very thick] (-2.25,0.5) -- (-2.25,-0.5);
\draw[very thick] (-1.25,-0.5) -- (-1.25,0.5);
\draw[thick, fill=white] (-2.35,-0.6) rectangle (-2.15,-0.4); 
\draw[thick, fill=white] (-1.35,-0.6) rectangle (-1.15,-0.4); 
\Text[x=-1,y=0.0, anchor = center]{,}
\end{tikzpicture}\label{eq:unitary1}\\
&\notag\\
&\begin{tikzpicture}[baseline=(current  bounding  box.center), scale=1]
\def\eps{0.5};
\draw[very thick] (-4.25,0.5) -- (-3.25,-0.5);
\draw[very thick] (-4.25,-0.5) -- (-3.25,0.5);
\draw[ thick, fill=myvioletc, rounded corners=2pt] (-4,0.25) rectangle (-3.5,-0.25);
\draw[thick] (-3.75,0.15) -- (-3.6,0.15) -- (-3.6,0);
\Text[x=-2.75,y=0.0, anchor = center]{$=$}
\draw[thick, fill=white] (-4.25,0.5) circle (0.1cm); 
\draw[thick, fill=white] (-3.25,0.5) circle (0.1cm); 
\draw[very thick] (-2.25,0.5) -- (-2.25,-0.5);
\draw[very thick] (-1.25,-0.5) -- (-1.25,0.5);
\draw[thick, fill=white] (-2.25,0.5) circle (0.1cm); 
\draw[thick, fill=white] (-1.25,0.5) circle (0.1cm); 
\Text[x=-1,y=0.0, anchor = center]{,}
\end{tikzpicture}
&&\begin{tikzpicture}[baseline=(current  bounding  box.center), scale=1]
\def\eps{0.5};
\draw[very thick] (-4.25,0.5) -- (-3.25,-0.5);
\draw[very thick] (-4.25,-0.5) -- (-3.25,0.5);
\draw[ thick, fill=myvioletc, rounded corners=2pt] (-4,0.25) rectangle (-3.5,-0.25);
\draw[thick] (-3.75,0.15) -- (-3.6,0.15) -- (-3.6,0);
\Text[x=-2.75,y=0.0, anchor = center]{$=$}
\draw[thick, fill=white] (-4.35,0.4) rectangle (-4.15,0.6); 
\draw[thick, fill=white] (-3.35,0.4) rectangle (-3.15,0.6); 
\draw[very thick] (-2.25,0.5) -- (-2.25,-0.5);
\draw[very thick] (-1.25,-0.5) -- (-1.25,0.5);
\draw[thick, fill=white] (-2.35,0.4) rectangle (-2.15,0.6); 
\draw[thick, fill=white] (-1.35,0.4) rectangle (-1.15,0.6); 
\Text[x=-1,y=0.0, anchor = center]{.}
\end{tikzpicture}
\label{eq:unitary2}
\end{align}
We stress that similar notations and identities were also used, e.g., in Ref.~\cite{Nahum:operatorspreadingRU, ZN:statmech, ZN:nonrandommembrane}.

Finally, we introduce the two bases for the $4$-fold local Hilbert space $\mathcal{H}_j^{\otimes 4}$
\begin{align}
&\mathcal B_1 = \{ \ket{ \text{\small\encircle{$\alpha\beta$}}}, \alpha,\beta=0,\dots,d^2-1\}\,,\label{eq:B1basis}\\
&\mathcal B_2 = \{ \ket{ \text{\small \boxed{$\alpha\beta$}}} , \alpha,\beta=0,\dots,d^2-1\}\,,
\label{eq:B2basis}
\end{align}
where
\be
\label{eq:squciralphabeta}
\begin{tikzpicture}[baseline={([yshift=-.5ex]current bounding box.center)}, scale=0.85]\draw[very thick] (-1.5,0) -- (-1.25,0);
\draw[thick, fill=white] (-1.8,0) circle (0.32cm); 
\Text[x=-0.8,y=0]{$=$}
\draw[thick, fill=white] (0.1,0.5) arc (90:270:0.2cm); 
\draw[thick, fill=white] (0.1,-0.1) arc (90:270:0.2cm); 
\draw[thick, fill=black] (-0.1,0.3) circle (0.05cm); 
\draw[thick, fill=black] (-0.1,-0.3) circle (0.05cm);
\Text[x=-0.35,y=0.3]{$\alpha$}
\Text[x=-0.35,y=-0.3]{$\beta$}
\Text[x=-1.8,y=0]{$\alpha\beta$}
\end{tikzpicture}\,,
\qquad\qquad
\begin{tikzpicture}[baseline={([yshift=-.5ex]current bounding box.center)}, scale=0.85]
\draw[very thick] (-1.5,-0.2) -- (-1,-0.2);
\draw[thick, fill=white] (-1.8,-0.5) rectangle (-1.2,0.1); 
\Text[x=-0.5,y=-0.2]{$=$}
\draw[thick] (0.85,-0.1) arc (90:270:0.15cm); 
\draw[thick] (0.65,0.2) arc (90:270:0.45cm); 
\draw[thick, fill=black] (0.7,-0.25) circle (0.05cm); 
\draw[thick, fill=black] (0.2,-0.25) circle (0.05cm);
\Text[x=-1.5,y=-0.2]{$\alpha\beta$}
\Text[x=0,y=-0.2]{$\alpha$}
\Text[x=0.5,y=-0.2]{$\beta$}
\end{tikzpicture}\,,
\ee
and 
$
\begin{tikzpicture}[baseline={([yshift=-.5ex]current bounding box.center)}, scale=0.45]
\draw[thick] (0.65,0.2) arc (90:270:0.45cm);  
\draw[thick, fill=black] (0.2,-0.25) circle (0.05cm);
\Text[x=-0.2,y=-0.2]{$\alpha$}
\end{tikzpicture}
=
\begin{tikzpicture}[baseline={([yshift=-.5ex]current bounding box.center)}, scale=0.45]
\draw[thick] (0.2,0.25) -- (0.2,-0.75);  
\draw[thick, fill=black] (0.2,-0.25) circle (0.05cm);
\Text[x=-0.2,y=-0.2]{$\alpha$}
\end{tikzpicture}
= a^\alpha
$.
Here $\{a^{\alpha}\}_{\alpha=0}^{d^2-1}$ is a Hilbert-Schmidt-orthonormal basis of local operators on $\mathbb C^{d}$, namely ${\rm tr}[(a^\alpha)^\dag a^\beta]=\delta_{\alpha,\beta}$, with $a^0=\1/\sqrt{d}$.

\section{R\'enyi-$2$ tripartite Information}
\label{sec:TI}

\subsection{General definition}
\label{sec:TIdef}

\begin{figure}
	\includegraphics[width=8.cm]{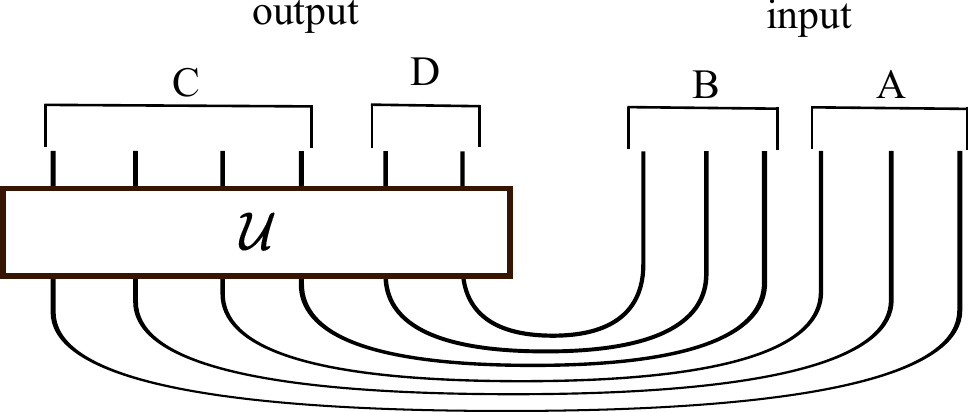}
	\caption{Pictorial representation of the state $\ket{\mathcal{U}}\rangle$ defined in Eq.~\eqref{eq:state_U}. The operator $\mathcal{U}$ is denoted by a box, while its legs correspond to the local Hilbert spaces $\mathcal{H}_j$, with the input legs being ``bent''.  Both the input and output sets of qudits have been partitioned into two regions: $A$, $B$ , and $C$, $D$ respectively.}
	\label{fig:entropy_definition}
\end{figure}

As anticipated in the introduction, the main object of study of this work is the tripartite information introduced in Ref.~\cite{HQRY:tripartiteinfo}, which represents a very intuitive measure to quantify the notion of scrambling of quantum information. We briefly define it here, following the discussion of Ref.~\cite{SPQS19}.

For the moment, we consider a finite system of $2L$ qudits (we will take the limit $L\to\infty$ at the end), associated with a Hilbert space $\mathcal{H}_{1,\ldots ,2L}$. In order to study the scrambling properties of the unitary operator $\mathcal{U}$ (which can be, for instance, a quantum circuit), we first promote it to be a state in a suitable space. To this end it is convenient to follow a different procedure with respect to the ``folding" discussed in the previous section. We introduce a copy of the original Hilbert space $\mathcal{H}^\prime_{1,\ldots ,2L}$, and define the maximally entangled state $\ket{I}\in \mathcal{H}_{1,\ldots ,2L} \otimes \mathcal{H}^\prime_{1,\ldots ,2L}$ as
\be
\ket{I}=\frac{1}{d^{L}}\sum_{\{j\}}\ket{\{j\}}\otimes \ket{\{j\}}\,,
\ee
were $\ket{\{j\}}$ are a set of basis vectors for $\mathcal{H}_{1,\ldots ,2L}$. Now, the operator $\mathcal{U}$ can be interpreted as a state in $\mathcal{H}_{1,\ldots ,2L} \otimes \mathcal{H}^\prime_{1,\ldots ,2L}$ through the Choi-Jamiolkowski mapping
\be
\mathcal{U}\mapsto \ket{\mathcal{U}}\rangle=\left(\mathbb{1}\otimes \mathcal{U}\right) \ket{I}\,,
\label{eq:state_U}
\ee
as pictorially reported in Fig. \ref{fig:entropy_definition}. 

Given $\ket{\mathcal{U}}\rangle$, we can compute the entanglement entropy between different spatial regions in the input and output qudits. We consider in particular
bipartitions into the complementary subsystems $A$, $B$ and $C$,$D$ respectively, as reported in Fig.~\ref{fig:entropy_definition}. Finally, given a pair of bipartitions $(A,B)$ and $(C,D)$, we define the tripartite information as~\cite{HQRY:tripartiteinfo}
\be
I_{3}(A : C : D) = I(A : C)+I(A : D)-I(A : C D)\,,
\label{eq:def_tripartite}
\ee
where $CD$ denotes the union of the regions $C$ and $D$. Here $I(X:Y)$ is the mutual information between the regions $X$ and $Y$ 
\be
I(X : Y)=S_{X}+S_{Y}-S_{X Y}\,,
\ee
where $S_{X}$ is the von Neumann entropy of the reduced density matrix $\rho_{X}$, namely $S_X=-{\rm tr }\rho_X\log \rho_X$. Note that, even if this is not apparent from the definition~\eqref{eq:def_tripartite}, the tripartite information is symmetric under all permutations of its arguments~\cite{HQRY:tripartiteinfo}.

From Eq.~\eqref{eq:def_tripartite}, we can appreciate that  $-I_3(A:C:D)$ quantifies the amount of information on the input region $A$ that can be recovered by global measurements in $C\cup D$, but can not be obtained by probing $C$ and $D$ individually. Thus, if $-I_3(A:C:D)$ is large the information localised in a subsystem $A$ of the input state can be recovered only by global measurements in the output state, signalling efficient scrambling of quantum information. Accordingly, we define $\mathcal{U}$ to be a good scrambler if for any bipartition of the sets of input and output qudits, $I_{3}(A:C:D)$ is negative with an absolute value close to the maximum possible value allowed by the geometry~\cite{HQRY:tripartiteinfo}. 

At this point we note that the computation of the tripartite information \eqref{eq:def_tripartite} is a very difficult task in random quantum circuits. Indeed, computing the von Neumann entanglement entropies is notoriously hard. For this reason, we study a simpler but closely related quantity, which is obtained from $I_3(A\! :\! C\! :\! D)$ by considering R\'enyi entropies
\be
\!\!\!\! I^{(2)}_{3}(A\! :\! C\! :\! D) \!=\! I^{(2)}(A\! :\! C)\!+\!I^{(2)}(A \!:\! D)\!-\!I^{(2)}(A \!:\! C D),
\label{eq:def_tripartite_renyi}
\ee
where
\be
I^{(2)}(X : Y)=S^{(2)}_{X}+S^{(2)}_{Y}-S^{(2)}_{X Y}\,,
\ee
and
\be
S^{(2)}_{X}=-\ln\,\mathbb{E}\left[\operatorname{tr}\left(\rho^2_{X}\right)\right]\,.
\label{eq:Renyi_2}
\ee
We stress that, strictly speaking, $S^{(2)}_X$ is not the averaged R\'enyi entropy of order $2$, as the disorder average is taken inside the logarithm. However, for large subsystems one expects the effect of fluctuations in the disorder to be small, so that $S^{(2)}_X$ can be considered a good approximation for the R\'enyi-$2$ entropy. This has been tested by studying quench problems in Haar-random circuits: it was shown that the two quantities show the same qualitative behaviour up to small corrections to the slope~\cite{Keyserlingk, ZN:statmech}.

\subsection{The folded tensor network}

In this section we show how the R\'enyi-$2$ tripartite information~\eqref{eq:def_tripartite_renyi} can be written in  terms of folded tensor networks. As a starting point, we note that Eq.~\eqref{eq:def_tripartite_renyi} can be simply rewritten as
\be
I^{(2)}_3(A,C,D)= 2 L \log d -S^{(2)}_{AC}-S^{(2)}_{AD}\,,
\label{eq:I3}
\ee
where $S^{(2)}_{AC}$ and $S^{(2)}_{AD}$ are the R\'enyi-$2$ operator entanglement entropies of the time evolving operator referring to the partitions depicted in Fig.~\ref{fig:tripartite}. 
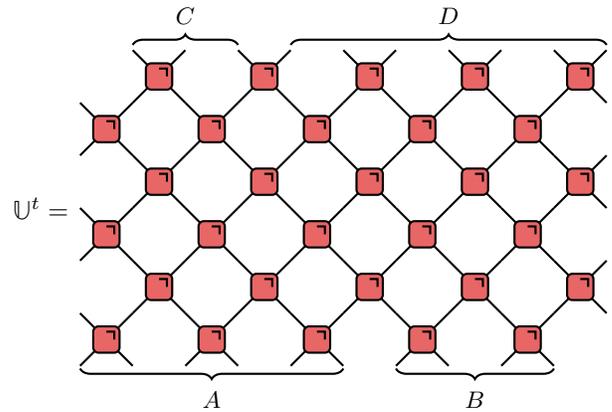
\begin{figure}
\centering
\begin{tikzpicture}[baseline=(current  bounding  box.center), scale=0.7]
\def\eps{-0.5};
\foreach \i in {0,...,4}
{
\draw[thick] (-.5-2*\i,-1+\eps) -- (0.5-2*\i,\eps);
\draw[thick] (-0.5-2*\i,\eps) -- (0.5-2*\i,-1+\eps);
\draw[thick, fill=myred, rounded corners=2pt] (-0.25-2*\i,-0.25+\eps) rectangle (.25-2*\i,-0.75+\eps);
\draw[thick] (-2*\i,-0.35+\eps) -- (.15-2*\i,-.35+\eps) -- (.15-2*\i,-0.5+\eps);
}
\foreach \i in {1,...,5}
{
\draw[thick] (.5-2*\i,-6+\eps) -- (1-2*\i,-5.5+\eps);
\draw[thick] (1.5-2*\i,-6+\eps) -- (1-2*\i,-5.5+\eps);
}
\foreach \jj[evaluate=\jj as \j using -2*(ceil(\jj/2)-\jj/2)] in {0,...,3}
\foreach \i in {1,...,5}
{
\draw[thick] (.5-2*\i-1*\j,-2-1*\jj+\eps) -- (1-2*\i-1*\j,-1.5-\jj+\eps);
\draw[thick] (1-2*\i-1*\j,-1.5-1*\jj+\eps) -- (1.5-2*\i-1*\j,-2-\jj+\eps);
}
\foreach \jj[evaluate=\jj as \j using -2*(ceil(\jj/2)-\jj/2)] in {0,...,4}
\foreach \i in {1,...,5}
{
\draw[thick] (.5-2*\i-1*\j,-1-1*\jj+\eps) -- (1-2*\i-1*\j,-1.5-\jj+\eps);
\draw[thick] (1-2*\i-1*\j,-1.5-1*\jj+\eps) -- (1.5-2*\i-1*\j,-1-\jj+\eps);
\draw[thick, fill=myred, rounded corners=2pt] (0.75-2*\i-1*\j,-1.75-\jj+\eps) rectangle (1.25-2*\i-1*\j,-1.25-\jj+\eps);
\draw[thick] (1-2*\i-1*\j,-1.35-1*\jj+\eps) -- (1.15-2*\i-1*\j,-1.35-1*\jj+\eps) -- (1.15-2*\i-1*\j,-1.5-1*\jj+\eps);
}
\draw [thick, decorate, decoration={brace,amplitude=4pt, mirror, raise=4pt},yshift=0pt]
(-9.5,-6.35) -- (-4.5,-6.35) node [black,midway,yshift=-0.55cm] {$A$};
\draw [thick, decorate, decoration={brace,amplitude=4pt, mirror, raise=4pt},yshift=0pt]
(-3.5,-6.35) -- (-.5,-6.35) node [black,midway,yshift=-0.55cm] {$B$};
\draw [thick, decorate, decoration={brace,amplitude=4pt, raise=-8pt},yshift=0pt]
(-8.5,0) -- (-6.5,0) node [black,midway,yshift=0.1cm] {$C$};
\draw [thick, decorate, decoration={brace,amplitude=4pt, raise=-8pt},yshift=0pt]
(-5.5,0) -- (.5,0) node [black,midway,yshift=0.1cm] {$D$};
\Text[x=-10.25,y=-3.5]{$\mathbb U^t =$}
\end{tikzpicture}
\caption{Pictorial representation of the bipartition considered in this work. The input qudits are divided into the sets $A =\! (-\infty,0], B= (0,+\infty)$, while for the bipartition of output qudits we choose $C=\! (-\infty, x], D=(x,+\infty)$.}
\label{fig:tripartite}
\end{figure}
More precisely, we have 
\be
S^{(2)}_I = - {\rm tr}\,\mathbb{E}\left[\rho_I^2\right],\qquad\qquad I=\{AC,AD\}\,,
\ee
where the disorder average is taken as in Eq.~\eqref{eq:Renyi_2}, while the density matrices are defined by their matrix elements 
\begin{align}
& [\rho_{AC}]_{\boldsymbol s'_A \boldsymbol s'_C}^{\boldsymbol s_A \boldsymbol s_C} = \frac{1}{d^{2L}} \sum_{\substack{\ket{\boldsymbol r} \in \mathcal C_B \\ \ket{\boldsymbol r'} \in \mathcal C_D}} [\mathbb U^t]_{\boldsymbol s_C \boldsymbol r'}^{\boldsymbol s_A \boldsymbol r}  [\mathbb U^{-t}]_{\boldsymbol s'_A \boldsymbol r}^{\boldsymbol s'_C \boldsymbol r'}\,,\label{eq:rhoAC}\\
&[\rho_{AD}]_{\boldsymbol s'_A \boldsymbol s'_D}^{\boldsymbol s_A \boldsymbol s_D} = \frac{1}{d^{2L}} \sum_{\substack{\ket{\boldsymbol r} \in \mathcal C_B \\ \ket{\boldsymbol r'} \in \mathcal C_C}} [\mathbb U^t]_{\boldsymbol r' \boldsymbol s_D}^{\boldsymbol s_A \boldsymbol r}  [\mathbb U^{-t}]_{\boldsymbol s'_A \boldsymbol r}^{\boldsymbol r' \boldsymbol s'_D}\,,
\label{eq:rhoAD}
\end{align}
where $\mathcal C_X$ is an orthonormal basis of $\mathbb C^{|X|}$. 

Equation~\eqref{eq:I3} can be further simplified, by taking the ``thermodynamic limit'' $L\to\infty$, and by restricting to a special class of bipartitions of the input and output qudits. In particular, in the rest of this work we choose
\begin{subequations}
\begin{align}
&A =\! (-\infty,0], & &B= (0,+\infty),\\
&C=\! (-\infty, x], & &D=(x,+\infty),
\end{align}
\label{eq:relevantpartition1}
\end{subequations}
where $|x|\leq t$ ($I^{(2)}_3$ vanishes if this condition is violated). We note that, while this choice is not the most general (it only involves connected intervals), it is the most natural one to consider in the thermodynamic limit $L\to\infty$~\cite{HQRY:tripartiteinfo}.

We now represent \eqref{eq:rhoAC} and \eqref{eq:rhoAD} diagrammatically and fold the circuit two times as described in Sec.~\ref{sec:folding} (see Appendix~\ref{sec:map_2d_pf} for the detailed procedure) to find 
\begin{align}
&S_{AC}^{(2)}= - \log \mathbb{E}\left[ Z_1(x_+,x_-)\right]\,,\label{eq:SAD}\\
&S_{AD}^{(2)}= (2L-x_+-x_-)\log d - \log \mathbb{E}\left[Z_2(x_+,x_-)\right]\,,\label{eq:SAC}
\end{align}
where we introduced the light-cone coordinates 
\be
x_+ \equiv t+ \lfloor x \rfloor\,,\qquad x_- \equiv t-\lceil x \rceil\,,
\label{eq:lightconecoord}
\ee
with $\lfloor x \rfloor$, $\lceil x \rceil$ denoting the floor and ceiling functions respectively~\cite{ceilnote}.
Here $\{Z_j(x,y)\}_{j=1}^2$ are partition functions admitting the graphical representation
\be
Z_1(x,y)=
\begin{tikzpicture}[baseline=(current  bounding  box.center), scale=0.55]
\def\eps{-0.5};
\foreach \i in {0,...,3}
\foreach \j in {2,...,-3}{
\draw[very thick] (-4.5+1.5*\j,1.5*\i) -- (-3+1.5*\j,1.5*\i);
\draw[very thick] (-3.75+1.5*\j,1.5*\i-.75) -- (-3.75+1.5*\j,1.5*\i+.75);
\draw[ thick, fill=myvioletc, rounded corners=2pt, rotate around ={45: (-3.75+1.5*\j,1.5*\i)}] (-4+1.5*\j,0.25+1.5*\i) rectangle (-3.5+1.5*\j,-0.25+1.5*\i);
\draw[thick, rotate around ={45: (-3.75+1.5*\j,1.5*\i)}] (-3.75+1.5*\j,.15+1.5*\i) -- (-3.6+1.5*\j,.15+1.5*\i) -- (-3.6+1.5*\j,1.5*\i);
}
\foreach \i in {0,...,3}{
\draw[thick, fill=white] (0.1,1.5*\i) circle (0.1cm);
\draw[thick, fill=white] (-9-0.1,1.5*\i-0.1) rectangle (-9+0.1,1.5*\i+0.1);
}
\foreach \i in {-3,...,2}{
\draw[thick, fill=white] (-3.75+1.5*\i-0.1,-0.75-0.1) rectangle (-3.75+1.5*\i+0.1,-0.75+0.1);
\draw[thick, fill=white] (-3.75+1.5*\i,5.25) circle (.1cm);
}
\draw [thick, decorate, decoration={brace,amplitude=4pt,mirror,raise=4pt},yshift=0pt]
(0.2,0) -- (0.2,4.5) node [black,midway,xshift=.45cm] {$y$};
\draw [thick, decorate, decoration={brace,amplitude=4pt, raise=4pt},yshift=0pt]
(-8.25,5.3) -- (-.75,5.3) node [black,midway,yshift=0.55cm] {$x$};
\Text[x=-8.25,y=-3]{}
\end{tikzpicture}\!,
\label{eq:SAC'}
\ee
and
\be
Z_2(x,y) = \begin{tikzpicture}[baseline=(current  bounding  box.center), scale=0.55]
\def\eps{-0.5};
\foreach \i in {0,...,3}
\foreach \j in {2,...,-3}{
\draw[very thick] (-4.5+1.5*\j,1.5*\i) -- (-3+1.5*\j,1.5*\i);
\draw[very thick] (-3.75+1.5*\j,1.5*\i-.75) -- (-3.75+1.5*\j,1.5*\i+.75);
\draw[ thick, fill=myvioletc, rounded corners=2pt, rotate around ={45: (-3.75+1.5*\j,1.5*\i)}] (-4+1.5*\j,0.25+1.5*\i) rectangle (-3.5+1.5*\j,-0.25+1.5*\i);
\draw[thick, rotate around ={45: (-3.75+1.5*\j,1.5*\i)}] (-3.75+1.5*\j,.15+1.5*\i) -- (-3.6+1.5*\j,.15+1.5*\i) -- (-3.6+1.5*\j,1.5*\i);}
\foreach \i in {0,...,3}{
\draw[thick, fill=white] (0,1.5*\i-0.1) rectangle (0.2,1.5*\i+0.1);
\draw[thick, fill=white] (-9,1.5*\i-0.1) rectangle (-8.8,1.5*\i+0.1);
}
\foreach \i in {-3,...,2}{
\draw[thick, fill=white] (-3.75+1.5*\i,-0.75) circle (0.1cm);
\draw[thick, fill=white] (-3.75+1.5*\i,5.25) circle (0.1cm);
}
\draw[thick, fill=white] (-8.25,5.25) circle (0.1cm);
\draw[thick, fill=white] (-8.25,-.75) circle (0.1cm);
\draw [thick, decorate, decoration={brace,amplitude=4pt,mirror,raise=4pt},yshift=0pt]
(0.2,0) -- (0.2,4.5) node [black,midway,xshift=.5cm] {$y$};
\draw [thick, decorate, decoration={brace,amplitude=4pt, raise=4pt},yshift=0pt]
(-8.25,5.3) -- (-.75,5.3) node [black,midway,yshift=0.55cm] {$x$};
\Text[x=-8.25,y=-3]{}
\end{tikzpicture}\,.
\label{eq:SAD'}
\ee
The green rectangles denote the folded gates introduced in Eq.~\eqref{eq:W}, while we made use of the notation in Eq.~\eqref{eq:states} for the boundary states. Next,  plugging Eqs.~\eqref{eq:SAD}, \eqref{eq:SAC}  into \eqref{eq:I3} we arrive at~\cite{note1} 
\begin{align}
\lim_{L\to\infty} I^{(2)}_3(x,t) = \log \left(d^{(x_++x_-)} \mathbb{E}\left[Z_1(x_+,x_-)\right]\right) \nonumber\\
+\log \mathbb{E} \left[Z_2(x_+,x_-)\right]\,.
\label{eq:I3TL}
\end{align}

We see that the tensor networks \eqref{eq:SAC'} and \eqref{eq:SAD'} differ only for the boundary conditions. In Sec.~\ref{sec:OTOC} we will see that also the OTOCs can be expressed in terms of very similar partition functions (again, with different boundary conditions). As discussed in Ref.~\cite{ZN:nonrandommembrane}, this is a generic feature of all scrambling measures. More precisely, considering a measure involving $n$ copies of the time evolution operator and $n$ copies of its Hermitian conjugate (for example the Tripartite Information~\eqref{eq:I3} defined with R\'enyi entropies of order $n$ or the Local Operator Entanglement measured with R\'enyi entropies of order $n/2$~\cite{BKP:OEergodicandmixing}) one can again write it in terms of partition functions like $Z_1(x,y)$ or $Z_2(x,y)$ with two main differences. First, the local gate is obtained by ``piling up'' $2n$ gates ($n$ copies of $U^T$ and $n$ copies of $U^\dag$), second the legs correspond to $d^{2n}$ dimensional local spaces. In this respect, \eqref{eq:I3} represents a minimal case: it is a measure of scrambling obtained by piling up the lowest number of local gates.

The discussion carried out so far is general, and applies for arbitrary ensembles of random quantum circuits. In the two following subsections we will show how \eqref{eq:I3TL} can be computed explicitly in the minimal settings discussed in the introduction, namely Haar-random unitary circuits and random dual-unitary circuits. Our approach is based on writing down suitable recurrence relations fulfilled by the tensor networks~\eqref{eq:SAC'} and \eqref{eq:SAD'}. These equations will be solved exactly for Haar random circuits while for random dual-unitary circuits they will be truncated to provide strict bounds. In the Haar-random case, this approach can be thought of as an alternative point of view with respect to the one put forward in Refs.~\cite{ZN:statmech,ZN:nonrandommembrane} (see also~\cite{RandomCircuitsEnt, Nahum:operatorspreadingRU, Keyserlingk}). While the latter references exploit a mapping between averaged tensor networks like~\eqref{eq:SAC'} and \eqref{eq:SAD'} and classical spin models in $2d$ here we will work directly with the tensor network.

\subsection{Tripartite Information in Haar Random Unitary Circuits}
\label{sec:TIHaar}

Our goal is to evaluate the averages in  Eqs.~\eqref{eq:SAC'}, \eqref{eq:SAD'}, when the gate $U$ [cf. Eqs.~\eqref{eq:U}, \eqref{eq:W}] is Haar-random and independently distributed at each site and half-time step. In this case the average of $W$ can be evaluated using simple group theoretical arguments, see, e.g., Refs.~\cite{Nahum:operatorspreadingRU, ZN:nonrandommembrane, ZN:statmech}. In our notation the result reads as  
\be
\label{eq:Haaraveragedgate}
 \mathbb{E}_{\rm Haar}\left[W\right]=
\begin{tikzpicture}[baseline=(current  bounding  box.center), scale=1]
\def\eps{0.5};
\draw[thick] (-2.25,0.5) -- (-1.25,-0.5);
\draw[thick] (-2.25,-0.5) -- (-1.25,0.5);
\draw[ thick, fill=gray, rounded corners=2pt] (-2,0.25) rectangle (-1.5,-0.25);
\draw[thick] (-1.75,0.15) -- (-1.6,0.15) -- (-1.6,0);
\Text[x=-2,y=-0.7]{}
\end{tikzpicture} = \frac{d^4}{d^4-1} \!\!\!\!\sum_{s,r\in\{\mcirc,\msquare\}} \!\! w(r,s) \ket{s s}\!\!\bra{r r}\,,
\ee
where $w(\mcirc,\mcirc)=w(\msquare,\msquare)=1$ and $w(\mcirc,\msquare)=w(\msquare,\mcirc)=-1/d^2$, and were we denoted the averaged gate by a gray rectangle. The operator~\eqref{eq:Haaraveragedgate} projects onto a two-dimensional Hilbert space spanned by the basis
\be
\mathcal B = \{\ket{\mcirc\mcirc}, \ket{\msquare\mcirc}, \ket{\mcirc\msquare}, \ket{\msquare\msquare}\}\,. 
\label{eq:nonorthogonalbasis}
\ee 
Note that this basis is \emph{not} orthogonal as $\braket{\mcirc|\msquare}=1/d$ [cf. Eq.~\eqref{eq:states}]. The action of the averaged gate on this non-orthogonal basis is defined by the relations \eqref{eq:unitary1}, coming from unitarity, together with   
\begin{align}
&\begin{tikzpicture}[baseline=(current  bounding  box.center), scale=1]
\def\eps{0.5};
\draw[very thick] (-4.25,0.5) -- (-3.25,-0.5);
\draw[very thick] (-4.25,-0.5) -- (-3.25,0.5);
\draw[ thick, fill=gray, rounded corners=2pt] (-4,0.25) rectangle (-3.5,-0.25);
\draw[thick] (-3.75,0.15) -- (-3.6,0.15) -- (-3.6,0);
\draw[thick, fill=white] (-4.25,-0.5) circle (0.1cm); 
\draw[thick, fill=white] (-3.35,-0.6) rectangle (-3.15,-0.4); 
\end{tikzpicture}
=
\frac{d}{d^2+1}\,\,
\begin{tikzpicture}[baseline=(current  bounding  box.center), scale=1]
\draw[very thick] (-2.25,0.5) -- (-2.25,-0.5);
\draw[very thick] (-1.25,-0.5) -- (-1.25,0.5);
\draw[thick, fill=white] (-2.25,-0.5) circle (0.1cm); 
\draw[thick, fill=white] (-1.25,-0.5) circle (0.1cm); 
\end{tikzpicture}
+\frac{d}{d^2+1}\,\,
\begin{tikzpicture}[baseline=(current  bounding  box.center), scale=1]
\draw[very thick] (-2.25,0.5) -- (-2.25,-0.5);
\draw[very thick] (-1.25,-0.5) -- (-1.25,0.5);
\draw[thick, fill=white] (-2.35,-0.6) rectangle (-2.15,-0.4); 
\draw[thick, fill=white] (-1.35,-0.6) rectangle (-1.15,-0.4); 
\end{tikzpicture},\label{eq:RUrel1}\\
\notag\\
&\begin{tikzpicture}[baseline=(current  bounding  box.center), scale=1]
\def\eps{0.5};
\draw[very thick] (-4.25,0.5) -- (-3.25,-0.5);
\draw[very thick] (-4.25,-0.5) -- (-3.25,0.5);
\draw[ thick, fill=gray, rounded corners=2pt] (-4,0.25) rectangle (-3.5,-0.25);
\draw[thick] (-3.75,0.15) -- (-3.6,0.15) -- (-3.6,0); 
\draw[thick, fill=white] (-3.25,-0.5) circle (0.1cm); 
\draw[thick, fill=white] (-4.35,-0.6) rectangle (-4.15,-0.4); 
\end{tikzpicture}
=
\frac{d}{d^2+1}\,\,
\begin{tikzpicture}[baseline=(current  bounding  box.center), scale=1]
\draw[very thick] (-2.25,0.5) -- (-2.25,-0.5);
\draw[very thick] (-1.25,-0.5) -- (-1.25,0.5);
\draw[thick, fill=white] (-2.25,-0.5) circle (0.1cm); 
\draw[thick, fill=white] (-1.25,-0.5) circle (0.1cm); 
\end{tikzpicture}
+\frac{d}{d^2+1}\,\,
\begin{tikzpicture}[baseline=(current  bounding  box.center), scale=1]
\draw[very thick] (-2.25,0.5) -- (-2.25,-0.5);
\draw[very thick] (-1.25,-0.5) -- (-1.25,0.5);
\draw[thick, fill=white] (-2.35,-0.6) rectangle (-2.15,-0.4); 
\draw[thick, fill=white] (-1.35,-0.6) rectangle (-1.15,-0.4); 
\end{tikzpicture}\,,
\label{eq:RUrel2}
\end{align}
which can be verified by a direct calculation.

These identities can be used to write down some recursive equations for $\mathbb{E}_{\rm Haar}[ Z_1(x,y)]$ and $\mathbb{E}_{\rm Haar}  Z_2(x,y)]$ that uniquely specify them. Let us consider, for instance, the averaged partition function $\mathbb{E}_{\rm Haar}\left[ Z_1(x,y)\right]$, where $Z_1(x,y)$ is defined in Eq.~\eqref{eq:SAC'}:  using Eq.~\eqref{eq:RUrel2} for the bottom right corner, it is immediate to derive the following equation, expressed in graphical notation
\begin{widetext}
\be
\mathbb{E}_{\rm Haar}\left[ Z_1(x,y)\right]=\frac{d}{d^2+1} 
\begin{tikzpicture}[baseline=(current  bounding  box.center), scale=0.55]
\def\eps{-0.5};
\foreach \i in {0,...,3}
\foreach \j in {1,...,-3}{
\draw[very thick] (-4.5+1.5*\j,1.5*\i) -- (-3+1.5*\j,1.5*\i);
\draw[very thick] (-3.75+1.5*\j,1.5*\i-.75) -- (-3.75+1.5*\j,1.5*\i+.75);
\draw[ thick, fill=gray, rounded corners=2pt, rotate around ={45: (-3.75+1.5*\j,1.5*\i)}] (-4+1.5*\j,0.25+1.5*\i) rectangle (-3.5+1.5*\j,-0.25+1.5*\i);
\draw[thick, rotate around ={45: (-3.75+1.5*\j,1.5*\i)}] (-3.75+1.5*\j,.15+1.5*\i) -- (-3.6+1.5*\j,.15+1.5*\i) -- (-3.6+1.5*\j,1.5*\i);
}
\foreach \i in {1,...,3}
\foreach \j in {2}{
\draw[very thick] (-4.5+1.5*\j,1.5*\i) -- (-3+1.5*\j,1.5*\i);
\draw[very thick] (-3.75+1.5*\j,1.5*\i-.75) -- (-3.75+1.5*\j,1.5*\i+.75);
\draw[ thick, fill=gray, rounded corners=2pt, rotate around ={45: (-3.75+1.5*\j,1.5*\i)}] (-4+1.5*\j,0.25+1.5*\i) rectangle (-3.5+1.5*\j,-0.25+1.5*\i);
\draw[thick, rotate around ={45: (-3.75+1.5*\j,1.5*\i)}] (-3.75+1.5*\j,.15+1.5*\i) -- (-3.6+1.5*\j,.15+1.5*\i) -- (-3.6+1.5*\j,1.5*\i);
}
\foreach \i in {1,...,3}{
\draw[thick, fill=white] (0.1,1.5*\i) circle (0.1cm);
}
\foreach \i in {0,...,3}{
\draw[thick, fill=white] (-9-0.1,1.5*\i-0.1) rectangle (-9+0.1,1.5*\i+0.1);
}
\foreach \i in {-3,...,2}{
\draw[thick, fill=white] (-3.75+1.5*\i,5.25) circle (0.1cm);
}
\foreach \i in {-3,...,1}{
\draw[thick, fill=white] (-3.75+1.5*\i-0.1,-0.75-0.1) rectangle (-3.75+1.5*\i+0.1,-0.75+0.1);
}
\draw[thick, fill=white] (-3.75+1.5*2-0.1, 0.75-0.1) rectangle (-3.75+1.5*2+0.1, 0.75+0.1);
\draw[thick, fill=white] (-3.75+2.25-0.1, 0-0.1) rectangle (-3.75+2.25+0.1, 0+0.1);
\Text[x=-8.25,y=3]{}
\end{tikzpicture}
+\frac{d}{d^2+1} 
\begin{tikzpicture}[baseline=(current  bounding  box.center), scale=0.55]
\def\eps{-0.5};
\foreach \i in {0,...,3}
\foreach \j in {1,...,-3}{
\draw[very thick] (-4.5+1.5*\j,1.5*\i) -- (-3+1.5*\j,1.5*\i);
\draw[very thick] (-3.75+1.5*\j,1.5*\i-.75) -- (-3.75+1.5*\j,1.5*\i+.75);
\draw[ thick, fill=gray, rounded corners=2pt, rotate around ={45: (-3.75+1.5*\j,1.5*\i)}] (-4+1.5*\j,0.25+1.5*\i) rectangle (-3.5+1.5*\j,-0.25+1.5*\i);
\draw[thick, rotate around ={45: (-3.75+1.5*\j,1.5*\i)}] (-3.75+1.5*\j,.15+1.5*\i) -- (-3.6+1.5*\j,.15+1.5*\i) -- (-3.6+1.5*\j,1.5*\i);
}
\foreach \i in {1,...,3}
\foreach \j in {2}{
\draw[very thick] (-4.5+1.5*\j,1.5*\i) -- (-3+1.5*\j,1.5*\i);
\draw[very thick] (-3.75+1.5*\j,1.5*\i-.75) -- (-3.75+1.5*\j,1.5*\i+.75);
\draw[ thick, fill=gray, rounded corners=2pt, rotate around ={45: (-3.75+1.5*\j,1.5*\i)}] (-4+1.5*\j,0.25+1.5*\i) rectangle (-3.5+1.5*\j,-0.25+1.5*\i);
\draw[thick, rotate around ={45: (-3.75+1.5*\j,1.5*\i)}] (-3.75+1.5*\j,.15+1.5*\i) -- (-3.6+1.5*\j,.15+1.5*\i) -- (-3.6+1.5*\j,1.5*\i);
}
\foreach \i in {1,...,3}{
\draw[thick, fill=white] (0.1,1.5*\i) circle (0.1cm);
}
\foreach \i in {0,...,3}{
\draw[thick, fill=white] (-9-0.1,1.5*\i-0.1) rectangle (-9+0.1,1.5*\i+0.1);
}
\foreach \i in {-3,...,2}{
\draw[thick, fill=white] (-3.75+1.5*\i,5.25) circle (0.1cm);
}
\foreach \i in {-3,...,1}{
\draw[thick, fill=white] (-3.75+1.5*\i-0.1,-0.75-0.1) rectangle (-3.75+1.5*\i+0.1,-0.75+0.1);
}
\draw[thick, fill=white] (-3.65+2.9, 0.75) circle (0.1cm);
\draw[thick, fill=white] (-3.75+2.25, 0) circle (0.1cm);
\Text[x=-8.25,y=3]{}
\end{tikzpicture}\,.
\label{eq:graphical_recurrence}
\ee
\end{widetext}
Now, making repeated use of the second identity in Eq.~\eqref{eq:unitary1}, we can ``pull'' to the left the rightmost white square in the last line of the first term in Eq.~\eqref{eq:graphical_recurrence}. By doing this, the diagram is cast into a rectangular lattice. Analogously, we can use the first identity in Eq.~\eqref{eq:unitary1}, to get rid of the rightmost column in the second term of Eq.~\eqref{eq:graphical_recurrence}. Putting all together, we obtain 
\begin{align}
\mathbb{E}\left[ Z_1(x,y) \right]=&\frac{d}{d^2+1}  \mathbb{E}\left[ Z_1(x-1,y)\right]\notag\\
&+\frac{d}{d^2+1} \mathbb{E}\left[ Z_1(x,y-1)\right]\,,
\label{eq:RUrecurrence1}
\end{align}
where we omitted the subscript specifying the type of average and in the final step we used $\braket{\mcirc|\mcirc}=\braket{\msquare|\msquare}=1$.  Eq.~\eqref{eq:RUrecurrence1} is a recurrence relation for the averaged partition function, and completely specifies it once the initial ``boundary'' conditions are assigned, namely once the functions $\mathbb{E}\left[ Z_1(x,0)\right]$ and $\mathbb{E}\left[ Z_1(0,y)\right]$ are known. We stress that similar recurrence relations appeared elsewhere in the recent literature (see, e.g., the recurrence equation for the R\'enyi-$2$ entropy for the state entanglement in Ref.~\cite{RandomCircuitsEnt}).

 The boundary conditions for \eqref{eq:RUrecurrence1} can be determined exactly, by repeating the same graphical derivation for $\mathbb{E}\left[ Z_1(x,1)\right]$ and $\mathbb{E}\left[ Z_1(1,y)\right]$. This is straightforward: in the first case it leads to the equation 
\be
\mathbb{E}\left[ Z_1(x,1) \right]=\frac{d}{d^2+1}  \mathbb{E}\left[ Z_1(x-1,1)\right]+\frac{d^{1-x}}{d^2+1}\,.
\ee
We note that this is of the form~\eqref{eq:RUrecurrence1}, provided that we make the identification
\be
\mathbb{E}\left[ Z_1(x,0)\right]=d^{-x}\,.
\label{eq:initial_cond_haar_tr_1}
\ee
Analogously, we can repeat the graphical derivation in Eq.~\eqref{eq:graphical_recurrence} for the averaged partition function $\mathbb{E}\left[ Z_1(1,y)\right]$. Following the same steps this yields
\be
\mathbb{E}\left[Z_1(0,y)\right]=d^{-y}\,.
\label{eq:initial_cond_haar_tr_2}
\ee
Eqs.~\eqref{eq:initial_cond_haar_tr_1},\eqref{eq:initial_cond_haar_tr_2} provide the initial conditions for the recurrence relation~\eqref{eq:RUrecurrence1}, which thus completely determines $\mathbb{E}\left[ Z_1(x,y) \right]$ for all $x,y>1$.

A very similar derivation can be carried out, up to minor modifications, for the averaged partition function $\mathbb{E}\left[ Z_2(x,y) \right]$. In particular, using $\braket{\mcirc|\msquare}=\braket{\msquare|\mcirc}=1/d$, we find the recurrence relation
\begin{align}
\mathbb{E}\left[ Z_2(x,y)\right] =&\frac{1}{d^2+1}  \mathbb{E}\left[ Z_2(x-1,y)\right]\notag\\
&+\frac{1}{d^2+1} \mathbb{E}\left[ Z_2(x,y-1)\right]\,,
\label{eq:RUrecurrence2}
\end{align}
where now the  boundary conditions are
\be 
\mathbb{E}\left[ Z_2(x,0)\right]=\mathbb{E}\left[ Z_2(0,y)\right]=1\,.
\label{eq:initial_cond_haar_tr_3}
\ee
The recurrence relations \eqref{eq:RUrecurrence1}, \eqref{eq:RUrecurrence2},  with boundary conditions \eqref{eq:initial_cond_haar_tr_1}, \eqref{eq:initial_cond_haar_tr_2}, \eqref{eq:initial_cond_haar_tr_3} can be solved by elementary methods (see Appendix~\ref{app:solvingrecurrence}), and their solution reads
\begin{align}
\mathbb{E}\left[ Z_1(x,y) \right]&=\frac{1}{d^{x-y}}- d^{x+y} f(x,y)\,,\label{eq:solZ1}\\
\mathbb{E}\left[ Z_2(x,y) \right]&= \frac{1}{d^{2y}}+ f(x,y)\,,
\label{eq:solZ2}
\end{align}
with
\begin{align}
f(x,y)=&d^{-2x}g(x+y-1,y-1,p)\notag\\
&-d^{-2y}g(x+y-1,y-1,1-p)\,,
\label{eq:f_g_function}
\end{align}
where we introduced the function
\be
g(n,a,s)=\sum_{k=0}^{a}\binom{n}{k}s^k(1-s)^{n-k}\,,
\ee
and defined 
\be
p=\frac{1}{d^2+1}\,.
\label{eq:p_def}
\ee
Plugging the solutions~\eqref{eq:solZ1} and \eqref{eq:solZ2} into Eq.~\eqref{eq:I3TL} we find an exact expression for $I^{(2)}_3(x,t)$, which is analysed in the rest of this section. Specifically, choosing for simplicity $x\in\mathbb Z^+$ (i.e. $x$ positive integer), it reads as 
\begin{align}
\lim_{L\to \infty} e^{ I^{(2)}_3(x,t)} =&1+\left(d^{2t-2x}-d^{2x+2t} \right)f(x_+,x_-)\notag\\
-&d^{4t} f(x_+,x_-)^2\,.
\label{eq:final_exact_i3}
\end{align}
First, we focus on the asymptotic rate at which the negative tripartite information $-I^{(2)}_3(x,t)$  grows at large times (for a fixed value of $x$), which quantifies how fast quantum information is scrambled by the circuit. To this end, we note that for large values of $t$ the second term in Eq.~\eqref{eq:f_g_function} becomes negligible and we can write
\be
\!\! f(x_+,x_-)\simeq \frac{1}{d^{2t+2x}}\!\sum_{k=0}^{t-x-1}\binom{2t-1}{k}p^k (1-p)^{2t-1-k}\,.
\label{eq:f_approximation}
\ee
At large times this function takes the following asymptotic form, as can be proven by a simple application of the Stirling formula
\be
\!\! f(x_+,x_-)\simeq\frac{1}{d^{2t+2x}}-\frac{1}{d^{2t+2x}}2^{2t-1}p^{t-x} (1-p)^{t+x-1}\!.
\label{eq:final_f}
\ee
Plugging Eq.~\eqref{eq:final_f} into Eq.~\eqref{eq:final_exact_i3} and retaining only the leading terms, we finally arrive at
\be
I^{(2)}_3(x,t)= -2t \log\left(\frac{d^2+1}{2d}\right) + O(1)\,.
\label{eq:asymptotic_I2}
\ee
We have tested this formula against numerical evaluation of Eq.~\eqref{eq:final_exact_i3} for different values of $x$ and $d$, finding perfect agreement. This is shown in Fig.~\ref{fig:largeT}, displaying our numerical results for $\Delta I^{(2)}_3(x,t)/t$ where
\be
\Delta I^{(2)}_3(x,t)= \left|I^{(2)}_3(x,t)+2t \log\left[(d^2+1)/2d\right]\right|\,.
\label{eq:delta_I}
\ee

It is interesting to note that the growth rate for the tripartite information coincides with the purity speed $v_P$ computed in Ref.~\cite{Nahum:operatorspreadingRU}. The latter was defined as the rate at which the R\'enyi-$2$ entropy increases after a quench from any product state.  This result is not at all obvious, since R\'enyi mutual information and entropy are two distinct quantities. For instance, for a circuit made of swap gates, cf. Sec.~\ref{sec:TripartiteInfoDU}, the R\'enyi-$2$ entropy can have maximal growth for particular product states made of Bell pairs~\cite{PBCP20}, but the tripartite information remains zero at all times for any bipartition of the output~\cite{HQRY:tripartiteinfo}.

\begin{figure}
	\includegraphics[scale=0.45]{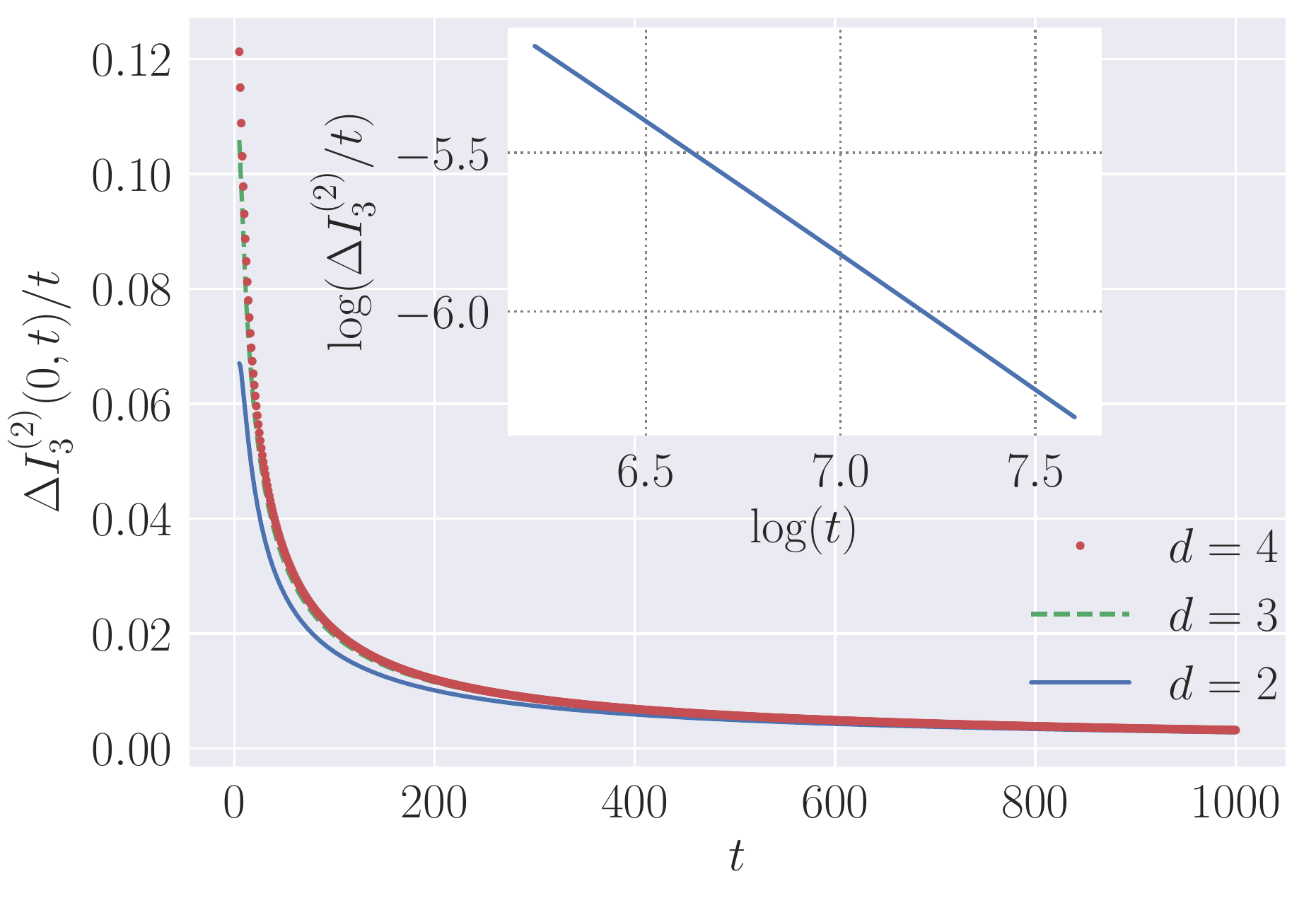}
	\caption{Large-time limit of $I^{(2)}_3(x,t)$. Main panel: the plot shows $\Delta I^{(2)}_3(x,t)/t$ where $\Delta I^{(2)}_3(x,t)$ is defined in Eq.~\eqref{eq:delta_I}, where we chose $x=0$, and increasing local dimension $d$. Inset: log-log plot for the same quantity and $d=2$. }
	\label{fig:largeT}
\end{figure}

Next, we address the variation of the tripartite information as a function of $x$, defined as in Fig.~\ref{fig:tripartite} (we assume again $x\in \mathbb{Z}^{+}$). For a given time $t$, the tripartite information will be zero for $x\gg  t$, since quantum information on the region $A$ has not yet propagated into $D$. On the other hand, for $x\ll  t$ Eq.~\eqref{eq:asymptotic_I2} holds, and the tripartite information becomes negative with large absolute value. We are interested in the intermediate transient regime. Based on physical intuition, we expect $I^{(2)}_3(x,t)$ to develop a front, which propagates at a given velocity and possibly broadens with time, in analogy with the picture established for the operator spreading in Haar-random circuits~\cite{Nahum:operatorspreadingRU,Keyserlingk}.

We can make this intuition precise by focusing on $e^{I^{(2)}_3(x,t)}$, which admits a coarse-grained description at large times. Specifically, in the large-time regime the leading behaviour of $e^{I^{(2)}_3(x,t)}$ is captured by taking a continuous limit approximation in the r.h.s. of Eq.~\eqref{eq:f_approximation}, namely
\be
f(x_+,x_-)\simeq \frac{1}{d^{2t+2x}}\Phi\left(\frac{v_Bt-x}{\sigma(t)}\right)\,,
\label{eq:f_function_hydro}
\ee
where we defined 
\be
\Phi(y)=\frac{1}{\sqrt{2\pi}}\int_{-\infty}^{y}e^{-s^2/2}{\rm d}s\,.
\ee
and introduced 
\be\label{eq:vb_sigma}
v_B=\frac{d^2-1}{d^2+1}\,,\qquad
\sigma(t)=\frac{d\sqrt{2t}}{d^2+1}\,.
\ee
Plugging \eqref{eq:f_function_hydro} into \eqref{eq:final_exact_i3} and considering the leading-order contribution we find   
\be
\lim_{L\to \infty} e^{ I^{(2)}_3(x,t)} \simeq 1-\Phi\left(\frac{v_Bt-x}{\sigma(t)}\right)\,.
\label{eq:I3vsxHaar}
\ee
From this equation we see that $e^{ I^{(2)}_3(x,t)}$ changes from one to zero over a region that moves with ``butterfly velocity'' $v_B$, and which broadens over a region $\sigma(t)$. This is displayed in Fig.~\ref{fig:expI}, where we compare the exact numerical data with the coarse-grained description. 

We stress that $v_B$ and $\sigma(t)$ are exactly the same as the corresponding quantities appearing in the growth of local operators~\cite{Nahum:operatorspreadingRU}, see also Sec.~\ref{sec:otocs_RUC} (note that in our conventions time $t$ and space $x$ are rescaled with respect to Ref.~\cite{Nahum:operatorspreadingRU} by a factor $1/2$, which causes $\sigma(t)$ to have an additional factor $1/\sqrt{2}$). This could be expected based on the close connection between the OTOCs and the tripartite information established in Ref.~\cite{HQRY:tripartiteinfo}.

\begin{figure}
	\includegraphics[scale=0.45]{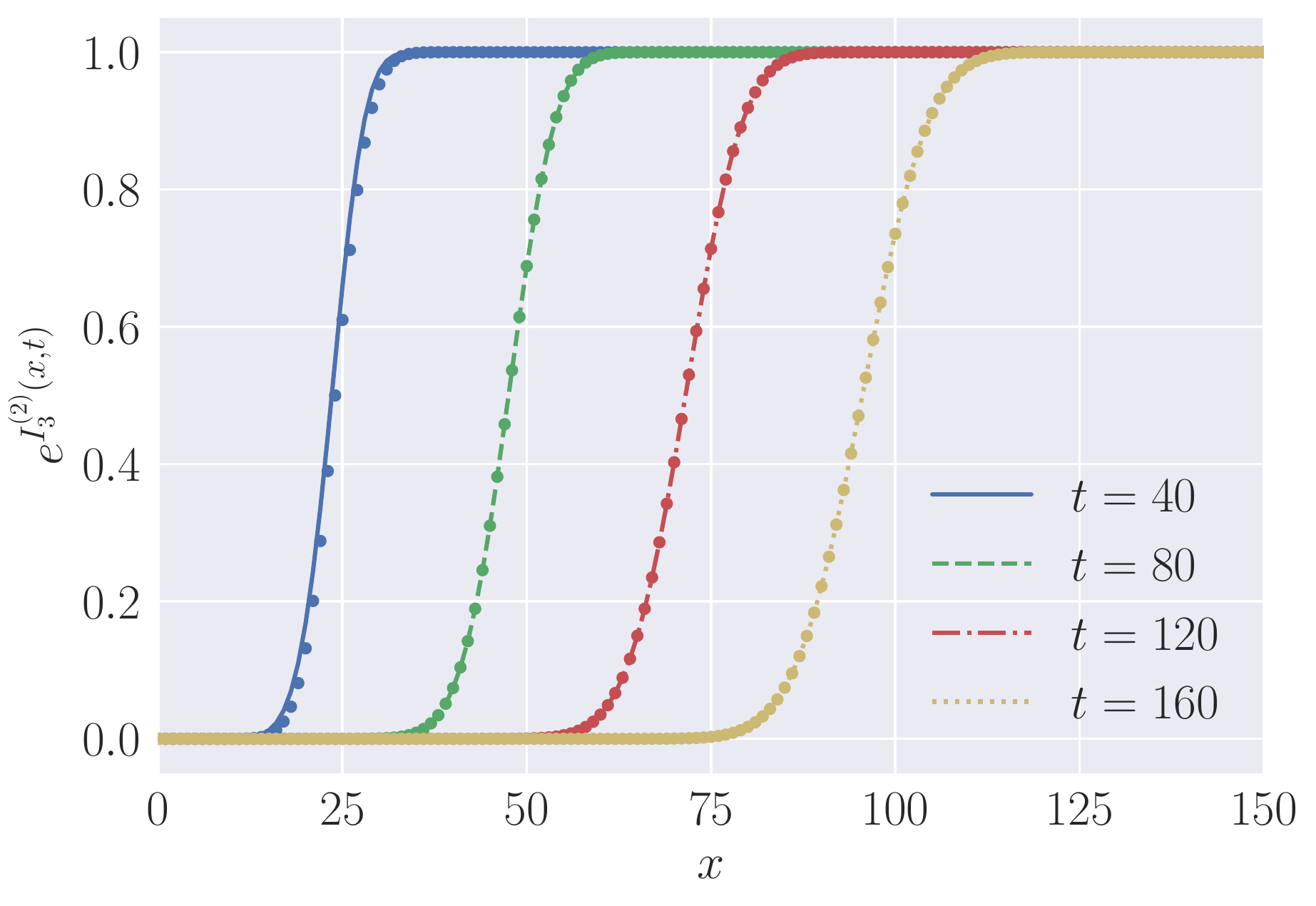}
	\caption{Exponential tripartite information $e^{I_3^{(2)}(x,t)}$ at different times, as a function of $x\in \mathbb{Z}^{+}$. The exponential of the tripartite information develops a front which propagates with butterfly velocity $v_B$, and broadens diffusively in time, cf. the main text. Lines correspond to the exact result~\eqref{eq:final_exact_i3}, while dots are obtained using the analytic function in Eq.~\eqref{eq:f_function_hydro}.}
	\label{fig:expI}
\end{figure}

\subsection{Tripartite Information in Dual Unitary Circuits}
\label{sec:TripartiteInfoDU}

Let us now focus on the tripartite information for dual-unitary quantum circuits. In this case, the folded tensor~\eqref{eq:W} inherits additional properties from the dual-unitarity conditions \eqref{eq:unitary1}--\eqref{eq:unitary2}, which read
\begin{align}
&\begin{tikzpicture}[baseline=(current  bounding  box.center), scale=1]
\def\eps{0.5};
\draw[very thick] (-4.25,0.5) -- (-3.25,-0.5);
\draw[very thick] (-4.25,-0.5) -- (-3.25,0.5);
\draw[ thick, fill=myvioletc, rounded corners=2pt] (-4,0.25) rectangle (-3.5,-0.25);
\draw[thick] (-3.75,0.15) -- (-3.6,0.15) -- (-3.6,0);
\Text[x=-2.75,y=0.0, anchor = center]{$=$}
\draw[thick, fill=white] (-3.25,-0.5) circle (0.1cm); 
\draw[thick, fill=white] (-3.25, 0.5) circle (0.1cm); 
\draw[very thick] (-1.25,0.5) -- (-2.25, 0.5);
\draw[very thick] (-1.25,-0.5) -- (-2.25,-0.5);
\draw[thick, fill=white] (-1.25, 0.5) circle (0.1cm); 
\draw[thick, fill=white] (-1.25,-0.5) circle (0.1cm); 
\Text[x=-1,y=0.0, anchor = center]{,}
\end{tikzpicture}
& &
\begin{tikzpicture}[baseline=(current  bounding  box.center), scale=1]
\def\eps{0.5};
\draw[very thick] (-4.25,0.5) -- (-3.25,-0.5);
\draw[very thick] (-4.25,-0.5) -- (-3.25,0.5);
\draw[ thick, fill=myvioletc, rounded corners=2pt] (-4,0.25) rectangle (-3.5,-0.25);
\draw[thick] (-3.75,0.15) -- (-3.6,0.15) -- (-3.6,0);
\Text[x=-2.75,y=0.0, anchor = center]{$=$}
\draw[thick, fill=white] (-3.35,-0.4) rectangle (-3.15,-0.6); 
\draw[thick, fill=white] (-3.35, 0.4) rectangle (-3.15, 0.6); 
\draw[very thick] (-1.25,0.5) -- (-2.25, 0.5);
\draw[very thick] (-1.25,-0.5) -- (-2.25,-0.5);
\draw[thick, fill=white] (-1.35, 0.6) rectangle (-1.15, 0.4); 
\draw[thick, fill=white] (-1.35,-0.6) rectangle (-1.15, -0.4); 
\Text[x=-1,y=0.0, anchor = center]{,}
\end{tikzpicture}\label{eq:dualunitary1}\\
&\notag\\
&\begin{tikzpicture}[baseline=(current  bounding  box.center), scale=1]
\def\eps{0.5};
\draw[very thick] (-4.25,0.5) -- (-3.25,-0.5);
\draw[very thick] (-4.25,-0.5) -- (-3.25,0.5);
\draw[ thick, fill=myvioletc, rounded corners=2pt] (-4,0.25) rectangle (-3.5,-0.25);
\draw[thick] (-3.75,0.15) -- (-3.6,0.15) -- (-3.6,0);
\Text[x=-2.75,y=0.0, anchor = center]{$=$}
\draw[thick, fill=white] (-4.25,0.5) circle (0.1cm); 
\draw[thick, fill=white] (-4.25,-0.5) circle (0.1cm); 
\draw[very thick] (-2.25,0.5) -- (-1.25,0.5);
\draw[very thick] (-2.25,-0.5) -- (-1.25,-0.5);
\draw[thick, fill=white] (-2.25,-0.5) circle (0.1cm); 
\draw[thick, fill=white] (-2.25,0.5) circle (0.1cm);
\Text[x=-1,y=0.0, anchor = center]{,}
\end{tikzpicture}
& &
\begin{tikzpicture}[baseline=(current  bounding  box.center), scale=1]
\def\eps{0.5};
\draw[very thick] (-4.25,0.5) -- (-3.25,-0.5);
\draw[very thick] (-4.25,-0.5) -- (-3.25,0.5);
\draw[ thick, fill=myvioletc, rounded corners=2pt] (-4,0.25) rectangle (-3.5,-0.25);
\draw[thick] (-3.75,0.15) -- (-3.6,0.15) -- (-3.6,0);
\Text[x=-2.75,y=0.0, anchor = center]{$=$}
\draw[thick, fill=white] (-4.35,0.4) rectangle (-4.15,0.6); 
\draw[thick, fill=white] (-4.35,-0.4) rectangle (-4.15,-0.6); 
\draw[very thick] (-2.25,0.5) -- (-1.25,0.5);
\draw[very thick] (-2.25,-0.5) -- (-1.25,-0.5);
\draw[thick, fill=white] (-2.35,-0.4) rectangle (-2.15,-0.6); 
\draw[thick, fill=white] (-2.35,0.4) rectangle (-2.15,0.6); 
\Text[x=-1,y=0.0, anchor = center]{.}
\end{tikzpicture}
\label{eq:dualunitary2}
\end{align}
These relations allow us to immediately simplify the partition function $Z_1(x,y)$ in Eq.~\eqref{eq:SAC}. Indeed, by multiple use of, say, the second of \eqref{eq:dualunitary2} from the bottom-left corner of Eq.~\eqref{eq:SAD'} we have 
\be
Z_1(x,y)= \braket{\mcirc|\msquare}^{x+y} = d^{-(x+y)}\,, 
\ee
so that, from Eq.~\eqref{eq:I3TL}, we find  
\be
\lim_{L\to\infty} e^{I^{(2)}_3(x,t)}=  \mathbb E[Z_2( t+ \lfloor x \rfloor, t- \lceil x \rceil)]\,.
\label{eq:goal}
\ee
Before embarking in the full calculation for random dual-unitary circuits, let us discuss a few general properties of Eq.~\eqref{eq:goal}, holding also in the non-random case. 

\subsubsection{Non-random case: a conjecture}

First, we note that for the SWAP gate
\begin{equation}
\begin{tikzpicture}[baseline=(current  bounding  box.center), scale=1]
\def\eps{0.5};
\draw[thick] (-2.25,0.5) -- (-1.25,-0.5);
\draw[thick] (-2.25,-0.5) -- (-1.25,0.5);
\Text[x=-2.8,y=0.05]{$U_{\rm SWAP}=$}
\end{tikzpicture},
\end{equation}
we can immediately compute   
\be
Z_2( x_+,x_-)\big |_{\rm SWAP}\!\! =\!\! \braket{\msquare|\msquare}^{x_-} \braket{\mcirc|\mcirc}^{x_+} = 1\,. 
\label{eq:Zswap}
\ee
This means that $I^{(2)}_3(x,t)|_{\rm SWAP}=0$, i.e., the SWAP gate does not scramble, as already noted in Ref.~\cite{HQRY:tripartiteinfo}. 

Second we observe that the partition function $Z_2(x,y)$ can be expressed in terms of row and column transfer matrices as follows
\begin{align}
Z_2(x,y)&=\braket{\underbrace{\mcirc\ldots\mcirc}_x| (T^{\msquare\msquare}_x)^y|\mcirc\ldots\mcirc}\notag\\
&= \braket{\underbrace{\msquare\ldots\msquare}_y|(T^{\mcirc\mcirc}_y)^x|\msquare\ldots\msquare}\,,\label{eq:transfmatZ2}
\end{align}
where we introduced 
\begin{align}
&T^{\msquare \msquare}_x =
\begin{tikzpicture}[baseline=(current  bounding  box.center), scale=0.55]
\def\eps{-0.5};
\foreach \i in {3}{
\foreach \j in {-3,...,2}{
\draw[very thick] (-4.5+1.5*\j,1.5*\i) -- (-3+1.5*\j,1.5*\i);
\draw[very thick] (-3.75+1.5*\j,1.5*\i-.75) -- (-3.75+1.5*\j,1.5*\i+.75);
\draw[ thick, fill=myvioletc, rounded corners=2pt, rotate around ={45: (-3.75+1.5*\j,1.5*\i)}] (-4+1.5*\j,0.25+1.5*\i) rectangle (-3.5+1.5*\j,-0.25+1.5*\i);
\draw[thick, rotate around ={45: (-3.75+1.5*\j,1.5*\i)}] (-3.75+1.5*\j,.15+1.5*\i) -- (-3.6+1.5*\j,.15+1.5*\i) -- (-3.6+1.5*\j,1.5*\i);}
\draw[thick, fill=white] (0,1.5*\i-0.1) rectangle (-.2,1.5*\i+0.1);
\draw[thick, fill=white] (-9,1.5*\i-0.1) rectangle (-8.8,1.5*\i+0.1);
}
\draw [thick, decorate, decoration={brace,amplitude=4pt, raise=4pt},yshift=0pt]
(-8.25,5.3) -- (-.75,5.3) node [black,midway,yshift=0.65cm] {$x$};
\Text[x=-8.25,y=2]{}
\end{tikzpicture}\,,
\label{eq:Tsquaresquare}\\[-25pt]
&T^{\mcirc \mcirc}_y =
 \begin{tikzpicture}[baseline=(current  bounding  box.center), scale=0.55]
\def\eps{-0.5};
\foreach \i in {3}{
\foreach \j in {-3,...,2}{
\draw[very thick] (-4.5+1.5*\j,1.5*\i) -- (-3+1.5*\j,1.5*\i);
\draw[very thick] (-3.75+1.5*\j,1.5*\i-.75) -- (-3.75+1.5*\j,1.5*\i+.75);
\draw[ thick, fill=myvioletc, rounded corners=2pt, rotate around ={45: (-3.75+1.5*\j,1.5*\i)}] (-4+1.5*\j,0.25+1.5*\i) rectangle (-3.5+1.5*\j,-0.25+1.5*\i);
\draw[thick, rotate around ={135: (-3.75+1.5*\j,1.5*\i)}] (-3.75+1.5*\j,.15+1.5*\i) -- (-3.6+1.5*\j,.15+1.5*\i) -- (-3.6+1.5*\j,1.5*\i);}
\draw[thick, fill=white] (-.1,1.5*\i) circle (0.1cm);
\draw[thick, fill=white] (-8.9,1.5*\i) circle (0.1cm);
}
\draw [thick, decorate, decoration={brace,amplitude=4pt, raise=4pt},yshift=0pt]
(-8.25,5.3) -- (-.75,5.3) node [black,midway,yshift=0.65cm] {$y$};
\Text[x=-8.25,y=2]{}
\end{tikzpicture}\,.
\label{eq:Tcirccirc}
\end{align}
These matrices fulfil the following two properties\\ 

\noindent a) They are \emph{contracting}, i.e.\ their eigenvalues lie within the unit circle in the complex plane~\cite{note2}.\\
    
\noindent b) The state $\otimes_i^{x} \ket{\mcirc}_i$ is an eigenvector of $T^{\mcirc \mcirc}_x$ with eigenvalue one, while $\otimes_i^{x} \ket{\msquare}_i$ is an eigenvector of $T^{\msquare \msquare}_x$ with eigenvalue one. This can be shown by repeated use of the graphical identities in Eqs.~\eqref{eq:unitary1}, \eqref{eq:unitary2}.\\

Therefore, if the gate $U$ is such that the matrices $T^{\mcirc\mcirc}_x$ and $T^{\msquare\msquare}_x$ have no other eigenvectors with eigenvalue 1 except for $\otimes_i^{x} \ket{\mcirc}$ and $\otimes_i^{x} \ket{\msquare}$ (the class of gates with this property have been termed ``completely chaotic'' in Ref.~\cite{BKP:OEergodicandmixing}) one finds  
\begin{align}
&\lim_{x\to\infty}Z_2(x,y) = \braket{\msquare|\mcirc}^{2 y}=d^{-2y},\label{eq:limit1}  \\
&\lim_{y\to\infty} Z_2(x,y) = \braket{\msquare|\mcirc}^{2x}=d^{-2x}.\label{eq:limit2}  
\end{align}
Using this result one can formulate the following conjecture for the leading order in time of the partition function  
\begin{align}
Z_2( x_+,x_-) \approx& Z_2( x_+,\infty)+Z_2(\infty, x_-)\notag\\
\approx& d^{-2 x_+}+d^{-2 x_-}\,,
\label{BKP:conjecture}
\end{align}
where we neglected subleading terms. This conjecture for the R\'enyi-$2$ Tripartite Information is analogous to the one put forward in Ref.~\cite{BKP:OEergodicandmixing} for the R\'enyi-$n$ Local Operator Entanglement Entropies, and corresponds to assume that the leading configurations in the partition sum $Z_2(x,y)$ are those having unit weight (which is the maximal one) in the bulk and are suppressed only at the boundary.  

Restricting now for simplicity to $x\in\mathbb{Z}$, the conjecture~\eqref{BKP:conjecture} gives us the following result for the tripartite information 
\be
\lim_{L\to\infty} e^{I^{(2)}_3(x,t)} \overset{t\gg 1}{\approx} 
\begin{cases}
d^{-2t-2x}+d^{-2t+2x} & |x|\leq t\\
1 & |x|>t 
\end{cases}\,.
\label{eq:conjecture}
\ee
First we note that, since $(d^2+1)/(2d)< d$ for $d>1$, Eq.~\eqref{eq:conjecture} predicts that completely chaotic dual-unitary circuits scramble faster than random ones. In particular, they display, asymptotically, the maximal possible growth rate for the negative tripartite information, which is the same of perfect tensors~\cite{HQRY:tripartiteinfo}. We stress, however, that in the latter case such rate is exact also at short times, while for dual-unitary circuits it is expected to generically emerge only asymptotically.

Second, we see that for $x\sim t \gg 1$ we find 
\be
\lim_{L\to\infty} e^{I^{(2)}_3(x,t)} \simeq d^{-2 \max[t-x,0]}\,.
\ee
This expression is again showing that $e^{I^{(2)}_3(x,t)} $ changes from one to zero as we approach the centre of the light cone. However, it features two main differences with respect to \eqref{eq:I3vsxHaar}. First we see that in this case $v_B=1$ (in accordance with recent results for OTOCs~\cite{ClLa20}), and second the region over which the transition happens is now independent of time. In other words, dual unitary circuits show no diffusive broadening of the front. 

In the next sub-section we will \emph{prove} the conjecture in Eq.~\eqref{eq:conjecture} for random dual-unitary circuits with $d=2$.

\subsubsection{Random case: an exact result}
\label{sec:TripartiteInfoRDU}

We consider quantum dual-unitary circuits with local Hilbert space of dimension $d=2$. The elementary gates were introduced in Eq.~\eqref{eq:v_mat_dual}, while random averages are taken as discussed in Sec.~\ref{sec:du_random}. By direct calculation, we find that the effective dimension of the local Hilbert space in the folded picture is reduced, after averaging, from $2^4$ to $2$, analogously to the case of Haar-random unitaries. In particular, the non-trivial subspace is again spanned by the non-orthogonal basis \eqref{eq:nonorthogonalbasis}. Thus, recalling the definition of  $Z_2(x,y)$ given in Eq.~\eqref{eq:SAD'},  we denote $\mathbb{E}_{\rm d.u.}\left[Z_2(x,y)\right]$ by
\be
\!\!\!\!\mathbb{E}\!\left[ Z_2(x,y)\right]= \begin{tikzpicture}[baseline=(current  bounding  box.center), scale=0.55]
\def\eps{-0.5};
\foreach \i in {0,...,3}
\foreach \j in {2,...,-3}{
\draw[very thick] (-4.5+1.5*\j,1.5*\i) -- (-3+1.5*\j,1.5*\i);
\draw[very thick] (-3.75+1.5*\j,1.5*\i-.75) -- (-3.75+1.5*\j,1.5*\i+.75);
\draw[ thick, fill=myorange, rounded corners=2pt, rotate around ={45: (-3.75+1.5*\j,1.5*\i)}] (-4+1.5*\j,0.25+1.5*\i) rectangle (-3.5+1.5*\j,-0.25+1.5*\i);
\draw[thick, rotate around ={45: (-3.75+1.5*\j,1.5*\i)}] (-3.75+1.5*\j,.15+1.5*\i) -- (-3.6+1.5*\j,.15+1.5*\i) -- (-3.6+1.5*\j,1.5*\i);}
\foreach \i in {0,...,3}{
\draw[thick, fill=white] (0,1.5*\i-0.1) rectangle (0.2,1.5*\i+0.1);
\draw[thick, fill=white] (-9,1.5*\i-0.1) rectangle (-8.8,1.5*\i+0.1);
}
\foreach \i in {-3,...,2}{
\draw[thick, fill=white] (-3.75+1.5*\i,-0.75) circle (0.1cm);
\draw[thick, fill=white] (-3.75+1.5*\i,5.25) circle (0.1cm);
}
\draw[thick, fill=white] (-8.25,5.25) circle (0.1cm);
\draw[thick, fill=white] (-8.25,-.75) circle (0.1cm);
\draw [thick, decorate, decoration={brace,amplitude=4pt,mirror,raise=4pt},yshift=0pt]
(0.2,0) -- (0.2,4.5) node [black,midway,xshift=.5cm] {$y$};
\draw [thick, decorate, decoration={brace,amplitude=4pt, raise=4pt},yshift=0pt]
(-8.25,5.3) -- (-.75,5.3) node [black,midway,yshift=0.55cm] {$x$};
\Text[x=-8.25,y=-3]{}
\end{tikzpicture}\!\!,
\label{eq:SAD'averaged}
\ee
where we omitted the subscript ``d.u.'', while we used the notation in Eq.~\eqref{eq:states} for the boundary states. Finally, orange squares represent the folded dual-unitary gates averaged from the single-site Haar invariant probability distribution. 

In the following, it will be more convenient to express the averaged gate in an orthogonal basis. We find that two suitable choices are 
\begin{align}
\mathcal B_1 = \{\ket{\mcirc\mcirc}, \ket{\mcircf\mcirc}, \ket{\mcirc\mcircf}, \ket{\mcircf\mcircf}\}\,, \label{eq:new_b1}\\
\mathcal B_2 = \{\ket{\msquare\msquare}, \ket{\msquaref\msquare}, \ket{\msquare\msquaref}, \ket{\msquaref\msquaref}\}\,,\label{eq:new_b2}
\end{align}
where we introduced 
\be
\ket{\mcircf}=\frac{2 \ket{\msquare}-\ket{\mcirc}}{ \sqrt 3}, \qquad  \ket{\msquaref}=\frac{2 \ket{\mcirc} -\ket{\msquare}}{{\sqrt 3}}\,.
\label{eq:fullcircsquare}
\ee
After a simple calculation, we find that the local gate $\mathbb{E}_{\rm d.u.}\left[ W\right]$ takes the following form in both the basis $\mathcal B_1$ and $\mathcal B_2$  
\begin{equation}
\label{eq:averagedgateDU}
\mathbb{E}_{\rm d.u.}\left[W\right]=
\begin{tikzpicture}[baseline=(current  bounding  box.center), scale=1]
\def\eps{0.5};
\draw[thick] (-2.25,0.5) -- (-1.25,-0.5);
\draw[thick] (-2.25,-0.5) -- (-1.25,0.5);
\draw[ thick, fill=myorange, rounded corners=2pt] (-2,0.25) rectangle (-1.5,-0.25);
\draw[thick] (-1.75,0.15) -- (-1.6,0.15) -- (-1.6,0);
\Text[x=-2,y=-0.7]{}
\end{tikzpicture}
= 
\begin{bmatrix}
1 & 0 & 0 & 0\\
0 & 0 & a & \frac{1-a}{\sqrt{3}}\\
0 & a & 0 &  \frac{1-a}{\sqrt{3}}\\
0 &  \frac{1-a}{\sqrt{3}} &  \frac{1-a}{\sqrt{3}} & \frac{1+2a}{3}
\end{bmatrix}\!,
\end{equation}
where 
\be
a= a(J):= 1 - \frac{2}{3} \cos(2J)^2\,,
\label{eq:a(J)}
\ee
and where $J$ is the free parameter introduced in Eq.~\eqref{eq:v_mat_dual}. Even if the matrix in Eq.~\eqref{eq:averagedgateDU} is not unitary, it is straightforward to see that the properties \eqref{eq:unitary1}--\eqref{eq:unitary2} and \eqref{eq:dualunitary1}--\eqref{eq:dualunitary2} still hold, namely
\begin{align}
&\begin{tikzpicture}[baseline=(current  bounding  box.center), scale=1]
\def\eps{0.5};
\draw[very thick] (-4.25,0.5) -- (-3.25,-0.5);
\draw[very thick] (-4.25,-0.5) -- (-3.25,0.5);
\draw[ thick, fill=myorange, rounded corners=2pt] (-4,0.25) rectangle (-3.5,-0.25);
\draw[thick] (-3.75,0.15) -- (-3.6,0.15) -- (-3.6,0);
\Text[x=-2.75,y=0.0, anchor = center]{$=$}
\draw[thick, fill=white] (-4.25,-0.5) circle (0.1cm); 
\draw[thick, fill=white] (-3.25,-0.5) circle (0.1cm); 
\draw[very thick] (-2.25,0.5) -- (-2.25,-0.5);
\draw[very thick] (-1.25,-0.5) -- (-1.25,0.5);
\draw[thick, fill=white] (-2.25,-0.5) circle (0.1cm); 
\draw[thick, fill=white] (-1.25,-0.5) circle (0.1cm); 
\Text[x=-1,y=0.0, anchor = center]{,}
\end{tikzpicture}
&
&\begin{tikzpicture}[baseline=(current  bounding  box.center), scale=1]
\def\eps{0.5};
\draw[very thick] (-4.25,0.5) -- (-3.25,-0.5);
\draw[very thick] (-4.25,-0.5) -- (-3.25,0.5);
\draw[ thick, fill=myorange, rounded corners=2pt] (-4,0.25) rectangle (-3.5,-0.25);
\draw[thick] (-3.75,0.15) -- (-3.6,0.15) -- (-3.6,0);
\Text[x=-2.75,y=0.0, anchor = center]{$=$}
\draw[thick, fill=white] (-4.35,-0.6) rectangle (-4.15,-0.4); 
\draw[thick, fill=white] (-3.35,-0.6) rectangle (-3.15,-0.4); 
\draw[very thick] (-2.25,0.5) -- (-2.25,-0.5);
\draw[very thick] (-1.25,-0.5) -- (-1.25,0.5);
\draw[thick, fill=white] (-2.35,-0.6) rectangle (-2.15,-0.4); 
\draw[thick, fill=white] (-1.35,-0.6) rectangle (-1.15,-0.4); 
\Text[x=-1,y=0.0, anchor = center]{,}
\end{tikzpicture}\label{eq:aveunitary1}\\
&\notag\\
&\begin{tikzpicture}[baseline=(current  bounding  box.center), scale=1]
\def\eps{0.5};
\draw[very thick] (-4.25,0.5) -- (-3.25,-0.5);
\draw[very thick] (-4.25,-0.5) -- (-3.25,0.5);
\draw[ thick, fill=myorange, rounded corners=2pt] (-4,0.25) rectangle (-3.5,-0.25);
\draw[thick] (-3.75,0.15) -- (-3.6,0.15) -- (-3.6,0);
\Text[x=-2.75,y=0.0, anchor = center]{$=$}
\draw[thick, fill=white] (-4.25,0.5) circle (0.1cm); 
\draw[thick, fill=white] (-3.25,0.5) circle (0.1cm); 
\draw[very thick] (-2.25,0.5) -- (-2.25,-0.5);
\draw[very thick] (-1.25,-0.5) -- (-1.25,0.5);
\draw[thick, fill=white] (-2.25,0.5) circle (0.1cm); 
\draw[thick, fill=white] (-1.25,0.5) circle (0.1cm); 
\Text[x=-1,y=0.0, anchor = center]{,}
\end{tikzpicture}
&
&\begin{tikzpicture}[baseline=(current  bounding  box.center), scale=1]
\def\eps{0.5};
\draw[very thick] (-4.25,0.5) -- (-3.25,-0.5);
\draw[very thick] (-4.25,-0.5) -- (-3.25,0.5);
\draw[ thick, fill=myorange, rounded corners=2pt] (-4,0.25) rectangle (-3.5,-0.25);
\draw[thick] (-3.75,0.15) -- (-3.6,0.15) -- (-3.6,0);
\Text[x=-2.75,y=0.0, anchor = center]{$=$}
\draw[thick, fill=white] (-4.35,0.4) rectangle (-4.15,0.6); 
\draw[thick, fill=white] (-3.35,0.4) rectangle (-3.15,0.6); 
\draw[very thick] (-2.25,0.5) -- (-2.25,-0.5);
\draw[very thick] (-1.25,-0.5) -- (-1.25,0.5);
\draw[thick, fill=white] (-2.35,0.4) rectangle (-2.15,0.6); 
\draw[thick, fill=white] (-1.35,0.4) rectangle (-1.15,0.6); 
\Text[x=-1,y=0.0, anchor = center]{,}
\end{tikzpicture}
\label{eq:aveunitary2}\\
&\notag\\
&\begin{tikzpicture}[baseline=(current  bounding  box.center), scale=1]
\def\eps{0.5};
\draw[very thick] (-4.25,0.5) -- (-3.25,-0.5);
\draw[very thick] (-4.25,-0.5) -- (-3.25,0.5);
\draw[ thick, fill=myorange, rounded corners=2pt] (-4,0.25) rectangle (-3.5,-0.25);
\draw[thick] (-3.75,0.15) -- (-3.6,0.15) -- (-3.6,0);
\Text[x=-2.75,y=0.0, anchor = center]{$=$}
\draw[thick, fill=white] (-3.25,-0.5) circle (0.1cm); 
\draw[thick, fill=white] (-3.25, 0.5) circle (0.1cm); 
\draw[very thick] (-1.25,0.5) -- (-2.25, 0.5);
\draw[very thick] (-1.25,-0.5) -- (-2.25,-0.5);
\draw[thick, fill=white] (-1.25, 0.5) circle (0.1cm); 
\draw[thick, fill=white] (-1.25,-0.5) circle (0.1cm); 
\Text[x=-1,y=0.0, anchor = center]{,}
\end{tikzpicture}
& &
\begin{tikzpicture}[baseline=(current  bounding  box.center), scale=1]
\def\eps{0.5};
\draw[very thick] (-4.25,0.5) -- (-3.25,-0.5);
\draw[very thick] (-4.25,-0.5) -- (-3.25,0.5);
\draw[ thick, fill=myorange, rounded corners=2pt] (-4,0.25) rectangle (-3.5,-0.25);
\draw[thick] (-3.75,0.15) -- (-3.6,0.15) -- (-3.6,0);
\Text[x=-2.75,y=0.0, anchor = center]{$=$}
\draw[thick, fill=white] (-4.25,0.5) circle (0.1cm); 
\draw[thick, fill=white] (-4.25,-0.5) circle (0.1cm); 
\draw[very thick] (-2.25,0.5) -- (-1.25,0.5);
\draw[very thick] (-2.25,-0.5) -- (-1.25,-0.5);
\draw[thick, fill=white] (-2.25,-0.5) circle (0.1cm); 
\draw[thick, fill=white] (-2.25,0.5) circle (0.1cm);
\Text[x=-1,y=0.0, anchor = center]{,}
\end{tikzpicture}\label{eq:avedualunitary1}\\
&\notag\\
&\begin{tikzpicture}[baseline=(current  bounding  box.center), scale=1]
\def\eps{0.5};
\draw[very thick] (-4.25,0.5) -- (-3.25,-0.5);
\draw[very thick] (-4.25,-0.5) -- (-3.25,0.5);
\draw[ thick, fill=myorange, rounded corners=2pt] (-4,0.25) rectangle (-3.5,-0.25);
\draw[thick] (-3.75,0.15) -- (-3.6,0.15) -- (-3.6,0);
\Text[x=-2.75,y=0.0, anchor = center]{$=$}
\draw[thick, fill=white] (-3.35,-0.4) rectangle (-3.15,-0.6); 
\draw[thick, fill=white] (-3.35, 0.4) rectangle (-3.15, 0.6); 
\draw[very thick] (-1.25,0.5) -- (-2.25, 0.5);
\draw[very thick] (-1.25,-0.5) -- (-2.25,-0.5);
\draw[thick, fill=white] (-1.35, 0.6) rectangle (-1.15, 0.4); 
\draw[thick, fill=white] (-1.35,-0.6) rectangle (-1.15, -0.4); 
\Text[x=-1,y=0.0, anchor = center]{,}
\end{tikzpicture}
& &
\begin{tikzpicture}[baseline=(current  bounding  box.center), scale=1]
\def\eps{0.5};
\draw[very thick] (-4.25,0.5) -- (-3.25,-0.5);
\draw[very thick] (-4.25,-0.5) -- (-3.25,0.5);
\draw[ thick, fill=myorange, rounded corners=2pt] (-4,0.25) rectangle (-3.5,-0.25);
\draw[thick] (-3.75,0.15) -- (-3.6,0.15) -- (-3.6,0);
\Text[x=-2.75,y=0.0, anchor = center]{$=$}
\draw[thick, fill=white] (-4.35,0.4) rectangle (-4.15,0.6); 
\draw[thick, fill=white] (-4.35,-0.4) rectangle (-4.15,-0.6); 
\draw[very thick] (-2.25,0.5) -- (-1.25,0.5);
\draw[very thick] (-2.25,-0.5) -- (-1.25,-0.5);
\draw[thick, fill=white] (-2.35,-0.4) rectangle (-2.15,-0.6); 
\draw[thick, fill=white] (-2.35,0.4) rectangle (-2.15,0.6); 
\Text[x=-1,y=0.0, anchor = center]{.}
\end{tikzpicture}
\label{eq:avedualunitary2}
\end{align}

First, note that $a(J)\in[1/3,1]$ and that for $a(J)=1$ the gate becomes the SWAP gate. This means that 
\be
\mathbb{E}\left[Z_2(x_+,x_-)\right]\big |_{a=1}=1\,. 
\label{eq:Z2avSWAP}
\ee
This result tells us that if $a(J)=1$, then the circuit does not scramble, despite taking disorder average. This can be easily understood by looking at Eq.~\eqref{eq:v_mat_dual}. Indeed, it is straightforward to see that in this case the evolution consists in a SWAP circuit followed by a single layer of single-site unitary operators. Clearly, we do not expect for such a circuit to scramble quantum information.

In the following we thus consider $a<1$. Our main result consists in proving that there exist a finite value $a^\ast$, with $0.683013< a^\ast \leq 1$, such that 
\be
\lim_{t\to\infty} 4^t \mathbb{E}\left[ Z_2(x_+,x_-)\right]\big |_{a< a^*} =  4^{-x}+4^{x}\,,
\label{eq:goal3}
\ee
where we assumed for simplicity $x\in\mathbb{Z}$. We will prove this limit by providing strict bounds for the partition function in Eq.~\eqref{eq:SAD'averaged}. This will be done by writing a set of recursive relations fulfilled by $\mathbb{E}\left[ Z_2(x,y)\right]$ and appropriately truncating them.   

We begin by defining the following transfer matrices, that are obtained by taking the single-site Haar average of the operators in Eq.~\eqref{eq:Tsquaresquare} and Eq.~\eqref{eq:Tcirccirc}
\begin{align}
&\bar T^{\msquare \msquare}_x =
\begin{tikzpicture}[baseline=(current  bounding  box.center), scale=0.55]
\def\eps{-0.5};
\foreach \i in {3}{
\foreach \j in {-3,...,2}{
\draw[very thick] (-4.5+1.5*\j,1.5*\i) -- (-3+1.5*\j,1.5*\i);
\draw[very thick] (-3.75+1.5*\j,1.5*\i-.75) -- (-3.75+1.5*\j,1.5*\i+.75);
\draw[ thick, fill=myorange, rounded corners=2pt, rotate around ={45: (-3.75+1.5*\j,1.5*\i)}] (-4+1.5*\j,0.25+1.5*\i) rectangle (-3.5+1.5*\j,-0.25+1.5*\i);
\draw[thick, rotate around ={45: (-3.75+1.5*\j,1.5*\i)}] (-3.75+1.5*\j,.15+1.5*\i) -- (-3.6+1.5*\j,.15+1.5*\i) -- (-3.6+1.5*\j,1.5*\i);}
\draw[thick, fill=white] (0,1.5*\i-0.1) rectangle (-.2,1.5*\i+0.1);
\draw[thick, fill=white] (-9,1.5*\i-0.1) rectangle (-8.8,1.5*\i+0.1);
}
\draw [thick, decorate, decoration={brace,amplitude=4pt, raise=4pt},yshift=0pt]
(-8.25,5.3) -- (-.75,5.3) node [black,midway,yshift=0.65cm] {$x$};
\Text[x=-8.25,y=2]{}
\end{tikzpicture}\,,
\label{eq:Tsquaresquareave}\\[-25pt]
&\bar T^{\mcirc \mcirc}_y =
 \begin{tikzpicture}[baseline=(current  bounding  box.center), scale=0.55]
\def\eps{-0.5};
\foreach \i in {3}{
\foreach \j in {-3,...,2}{
\draw[very thick] (-4.5+1.5*\j,1.5*\i) -- (-3+1.5*\j,1.5*\i);
\draw[very thick] (-3.75+1.5*\j,1.5*\i-.75) -- (-3.75+1.5*\j,1.5*\i+.75);
\draw[ thick, fill=myorange, rounded corners=2pt, rotate around ={45: (-3.75+1.5*\j,1.5*\i)}] (-4+1.5*\j,0.25+1.5*\i) rectangle (-3.5+1.5*\j,-0.25+1.5*\i);
\draw[thick, rotate around ={135: (-3.75+1.5*\j,1.5*\i)}] (-3.75+1.5*\j,.15+1.5*\i) -- (-3.6+1.5*\j,.15+1.5*\i) -- (-3.6+1.5*\j,1.5*\i);}
\draw[thick, fill=white] (-.1,1.5*\i) circle (0.1cm);
\draw[thick, fill=white] (-8.9,1.5*\i) circle (0.1cm);
}
\draw [thick, decorate, decoration={brace,amplitude=4pt, raise=4pt},yshift=0pt]
(-8.25,5.3) -- (-.75,5.3) node [black,midway,yshift=0.65cm] {$y$};
\Text[x=-8.25,y=2]{}
\end{tikzpicture}\,.
\label{eq:Tcirccircave}
\end{align}
Since $\mathbb{E}_{\rm d.u.}\left[W\right]$ has the same form in the basis $\mathcal B_1$ and $\mathcal B_2$ [cf. \eqref{eq:new_b1} and \eqref{eq:new_b2}] and it is real and symmetric, the two matrices are the same up to a basis transformation, namely  
\be
\bar{T}^{\msquare \msquare}_x  = \left[\bigotimes_i^x \bar R^\dag\right] \bar{T}^{\mcirc \mcirc}_x   \left[\bigotimes_i^x \bar R\right],
\ee
where
\be
\bar R = \frac{1}{2}\begin{pmatrix}
1 & {\sqrt 3} \\
{\sqrt 3} & -{1}
\end{pmatrix},
\ee
so from now on we will consider only one of them, say $\bar{T}^{\msquare \msquare}_x $.

To find recursive relations for $\mathbb{E}\!\left[ Z_2(x,y)\right]$ we first note that, taking the average of \eqref{eq:transfmatZ2}, one can view the partition function as the expectation value of $({\bar T}^{\msquare \msquare}_{x})^y$ on the ``boundary state" $\ket{\mcirc \ldots \mcirc}$. Next, we employ the following useful identity (proven in Appendix~\ref{app:proofP2}) 
\begin{property}
\label{prop:P2}
The transfer matrix ${\bar T}^{\msquare \msquare}_{x}$ acts as follows on the boundary state $\ket{\mcirc \ldots \mcirc}$
\be
{\bar T}^{\msquare \msquare}_{x} \ket{\mcirc \ldots \mcirc} = \frac{1}{4} \ket{\mcirc \ldots \mcirc} + \frac{3}{4} \ket{\mcirc' \ldots \mcirc'}
\label{eq:crucialidentity}
\ee
where we introduced 
\be
\!\!\ket{\mcirc'}= a \ket{\mcirc} + \frac{1-a}{\sqrt 3} \ket{\mcircf}=\frac{1}{2} \ket{\msquare} + \frac{\sqrt 3 }{2} \frac{4a-1}{3}  \ket{\msquaref}.
\label{eq:circprimedef}
\ee
\end{property}

\noindent Note that for $a=1/4$ 
\be
\ket{\mcirc'}= \frac{1}{2}\ket{\msquare}
\ee
and Property~\ref{prop:P2} allows one to find an exact expression for $\mathbb{E}\left[ Z_2(x,y)\right]\big |_{\tfrac{1}{4}}$. Indeed, using \eqref{eq:crucialidentity} we immediately find the following recursion equation 
\be
\mathbb{E}\left[Z_2(x,y)]\right]\big |_{a=\tfrac{1}{4}} = \frac{1}{4} \mathbb{E}\left[ Z_2(x,y-1)\right]\big |_{a=\tfrac{1}{4}} + \frac{3}{4^{x+1}}\,,
\ee 
with initial condition $\mathbb{E}\left[ Z_2(x,0)\right]\big |_{a=1/4}= \braket{\mcirc|\mcirc}^x =1$. This relation is directly solved by 
\be
\mathbb{E}\left[ Z_2(x,y)\right]\big |_{\tfrac{1}{4}}= \frac{1}{4^{x}}+\frac{1}{4^{y}}- \frac{1}{4^{x+y}}\,. 
\ee 
The value $a=1/4$, however, is ``unphysical'' because Eq.~\eqref{eq:a(J)} implies $a\in[1/3,1]$ for $J$ real. 

When $a$ is generic, instead, the recurrence equation reads as 
\be
\!\!\! \mathbb{E}\!\left[Z_2(x,y)\right]= \frac{1}{4} \mathbb{E}\!\left[  Z_2(x,y-1) \right]+ \frac{3}{4} \mathbb{E}\!\left[ Z_3(x,y-1)\right]\!,
\label{eq:Z2Z3}
\ee
where we introduced
\be
\!\!\!\!\mathbb{E}\!\left[ Z_3(x,y) \right] \! = \begin{tikzpicture}[baseline=(current  bounding  box.center), scale=0.55]
\def\eps{-0.5};
\foreach \i in {0,...,3}
\foreach \j in {2,...,-3}{
\draw[very thick] (-4.5+1.5*\j,1.5*\i) -- (-3+1.5*\j,1.5*\i);
\draw[very thick] (-3.75+1.5*\j,1.5*\i-.75) -- (-3.75+1.5*\j,1.5*\i+.75);
\draw[ thick, fill=myorange, rounded corners=2pt, rotate around ={45: (-3.75+1.5*\j,1.5*\i)}] (-4+1.5*\j,0.25+1.5*\i) rectangle (-3.5+1.5*\j,-0.25+1.5*\i);
\draw[thick, rotate around ={45: (-3.75+1.5*\j,1.5*\i)}] (-3.75+1.5*\j,.15+1.5*\i) -- (-3.6+1.5*\j,.15+1.5*\i) -- (-3.6+1.5*\j,1.5*\i);}
\foreach \i in {0,...,3}{
\draw[thick, fill=white] (0,1.5*\i-0.1) rectangle (0.2,1.5*\i+0.1);
\draw[thick, fill=white] (-9,1.5*\i-0.1) rectangle (-8.8,1.5*\i+0.1);
}
\foreach \i in {-3,...,2}{
\draw[thick, fill=white] (-3.75+1.5*\i,-0.75) circle (0.1cm);
\Text[x=-3.75+1.5*\i+0.21,y=-0.75-0.01]{$'$};
\draw[thick, fill=white] (-3.75+1.5*\i,5.25) circle (0.1cm);
}
\draw[thick, fill=white] (-8.25,5.25) circle (0.1cm);
\draw[thick, fill=white] (-8.25,-.75) circle (0.1cm);
\draw [thick, decorate, decoration={brace,amplitude=4pt,mirror,raise=4pt},yshift=0pt]
(0.25,0) -- (0.25,4.5) node [black,midway,xshift=.5cm] {$y$};
\draw [thick, decorate, decoration={brace,amplitude=4pt, raise=4pt},yshift=0pt]
(-8.25,5.3) -- (-.75,5.3) node [black,midway,yshift=0.55cm] {$x$};
\Text[x=-8.25,y=-3]{}
\end{tikzpicture}\!\!\!.
\label{eq:Z3}
\ee
Note the symbol $\mcirc'$ refers to the state~\eqref{eq:circprimedef} and that $\mathbb{E}\left[ Z_3(x,0)\right]= \braket{\mcirc|\mcirc'}^x=a^x$. These relations are complemented by 
\be
\!\!\!\!\mathbb{E}\left[ Z_3(x,y)\right]\! = \frac{1}{4} \mathbb{E}\left[ Z_3(x,y-1)\right] \!+\! \frac{3}{4} \mathbb{E}\left[Z_4(x,y-1)\right],
\label{eq:Z3Z4}
\ee
where we introduced
\be
\!\!\!\!\!\mathbb{E}\!\left[ Z_4(x,y) \right] \! = \begin{tikzpicture}[baseline=(current  bounding  box.center), scale=0.55]
\def\eps{-0.5};
\foreach \i in {0,...,3}
\foreach \j in {2,...,-3}{
\draw[very thick] (-4.5+1.5*\j,1.5*\i) -- (-3+1.5*\j,1.5*\i);
\draw[very thick] (-3.75+1.5*\j,1.5*\i-.75) -- (-3.75+1.5*\j,1.5*\i+.75);
\draw[ thick, fill=myorange, rounded corners=2pt, rotate around ={45: (-3.75+1.5*\j,1.5*\i)}] (-4+1.5*\j,0.25+1.5*\i) rectangle (-3.5+1.5*\j,-0.25+1.5*\i);
\draw[thick, rotate around ={45: (-3.75+1.5*\j,1.5*\i)}] (-3.75+1.5*\j,.15+1.5*\i) -- (-3.6+1.5*\j,.15+1.5*\i) -- (-3.6+1.5*\j,1.5*\i);}
\foreach \i in {0,...,3}{
\draw[thick, fill=white] (0,1.5*\i-0.1) rectangle (0.2,1.5*\i+0.1);
\draw[thick, fill=white] (-9,1.5*\i-0.1) rectangle (-8.8,1.5*\i+0.1);
}
\foreach \i in {-3,...,2}{
\draw[thick, fill=white] (-3.75+1.5*\i,-0.75) circle (0.1cm);
\Text[x=-3.75+1.5*\i+0.21,y=-0.75-0.01]{$'$};
\draw[thick, fill=white] (-3.75+1.5*\i,5.25) circle (0.1cm);
\Text[x=-3.75+1.5*\i+0.21,y=5.25+0.01]{$'$};
}
\draw[thick, fill=white] (-8.25,5.25) circle (0.1cm);
\draw[thick, fill=white] (-8.25,-.75) circle (0.1cm);
\draw [thick, decorate, decoration={brace,amplitude=4pt,mirror,raise=4pt},yshift=0pt]
(0.25,0) -- (0.25,4.5) node [black,midway,xshift=.5cm] {$y$};
\draw [thick, decorate, decoration={brace,amplitude=4pt, raise=4pt},yshift=0pt]
(-8.25,5.3) -- (-.75,5.3) node [black,midway,yshift=0.55cm] {$x$};
\Text[x=-8.25,y=-3]{}
\end{tikzpicture}\!\!\!\!,
\label{eq:Z4}
\ee
fulfilling $\mathbb{E}\left[ Z_4(x,0)\right]=\braket{\mcirc' |\mcirc'}^x =(a^2+(a-1)^2/3)^x$. The inductive relation \eqref{eq:Z3Z4} is found by using Property~\ref{prop:P2} for bra states. 

Since the recurrence relations  \eqref{eq:Z2Z3} and \eqref{eq:Z3Z4} do not form a complete set, one would need to find a third independent equation in order to solve them. This, however, is not possible only using the dual-unitary conditions \eqref{eq:aveunitary1}--\eqref{eq:avedualunitary2}. Instead, we will find suitable lower and upper bounds for $\mathbb{E}\left[ Z_4(x,y)\right]$. We will see that for small enough $a$ these bounds give the same leading order in time. 

First of all we note that, as shown in Appendix~\ref{app:solvingrecurrence}, Eqs.~\eqref{eq:Z2Z3}, \eqref{eq:Z3Z4} are formally solved by 
\begin{align}
&\mathbb{E}\left[ Z_2(x,y) \right]=\frac{1}{4^y}+\frac{3}{4^y} \sum_{k=0}^{y-1} 4^k  \mathbb{E}\left[Z_3(x,k)\right]\,, \label{eq:formalsolZ2}\\ 
&\mathbb{E}\left[ Z_3(x,y)\right] =\frac{a^x}{4^y}+ \frac{3}{4^y} \sum_{k=0}^{y-1}  4^k \mathbb{E}\left[ Z_4(x,k)\right]\,. \label{eq:formalsolZ3}
\end{align}
These relations can be used to eliminate $\mathbb{E}\left[Z_3(x,y)\right]$: indeed, plugging the second equation into the first one, we obtain 
\begin{align}
\mathbb{E}\left[ Z_2(x,y)\right]=  \frac{1}{4^{y}} + \frac{3 y a^x}{4^{y}} + \frac{9}{4^{y}}  \sum_{k=0}^{y-1} \sum_{h=0}^{k-1}  {4^{h}}  \mathbb{E} \left[Z_4(x,h)\right]\,.
\end{align}
Now, we note that $\mathbb{E}\left[ Z_4(x,y)\right]$ can be written as an expectation value of $(\bar T^{\msquare \msquare}_x)^y$ as follows 
\be
\mathbb{E}\left[ Z_4(x,y) \right]=\braket{\mcirc'\ldots\mcirc'|(\bar T^{\msquare \msquare}_x)^y|\mcirc'\ldots\mcirc'}\,.
\label{eq:Z4EV}
\ee
This object can be bounded by finding the first largest eigenvalues of $T^{\msquare \msquare}_x$ (and the relative eigenvectors). This is achieved by using the following property, which is proven in Appendix~\ref{app:proofP1}. 

\begin{property}
\label{prop:P1}
The matrix $\bar{T}^{\msquare \msquare}_x $ is positive definite and has the following spectral decomposition 
\begin{align}
&\!\!\bar{T}^{\msquare \msquare}_x \!\!= \!P^{\msquare}_0 \!+ \!a\! \sum_{k=1}^x  P^{\msquare}_k \!+\!\! \left[ a^2\!+\!\frac{(a-1)^2}{3}\right] \!\sum_{k=2}^{x}  Q^{\msquare}_k \!+ \!\!R^{\msquare}_x, \label{eq:specdec1}
\end{align}
where we defined 
\begin{align}
P^{\msquare}_k &:= \ket{\underbrace{\msquare\ldots\msquare\msquaref}_{k}\msquare\ldots\msquare}\!\!\bra{\underbrace{\msquare\ldots\msquare\msquaref}_{k}\msquare\ldots\msquare},\\
Q^{\msquare}_k &:= \ket{\underbrace{\msquare\ldots\msquare\msquaref\msquaref}_{k}\msquare\ldots\msquare}\!\!\bra{\underbrace{\msquare\ldots\msquare\msquaref\msquaref}_{k}\msquare\ldots\msquare},
\end{align}
and the ``reminder'' $R^{\msquare}_x$ has operator norm 
\be
\!\!|R^{\msquare}_x| \leq \left[ a^2+\frac{(a-1)^2}{3}\right].
\ee
\end{property}
Note that for $1/4 < a < 1$
\be
a > a^2+\frac{(a-1)^2}{3}.
\ee 
so that Property~\ref{prop:P1} gives the $2x$ largest eigenvalues of $\bar{T}^{\msquare \msquare}_x $ and the corresponding eigenvectors.
    
By means of the above property we see that performing the replacement 
\be
\bar{T}^{\msquare \msquare}_x \mapsto P^{\msquare}_0 + a \sum_{k=1}^x  P^{\msquare}_k 
\ee
in Eq.\eqref{eq:Z4EV} we obtain a lower bound for $\mathbb{E}\left[ Z_4(x,y) \right]$. This is true because $\bar{T}^{\msquare \msquare}_x$ is positive definite and yields 
\be
\mathbb{E}\left[Z_4(x,y) \right]\geq \frac{1}{4^x} + \frac{3 a^y x}{4^x}\,. 
\ee
An upper bound can instead be found by replacing $R^{\msquare}_x$ with $[a^2+{(a-1)^2}/{3}] \1$ in \eqref{eq:specdec1} and leads to 
\be
\mathbb{E}\left[Z_4(x,y)\right] \leq \frac{1}{4^x} + \frac{3 a^y x}{4^x}  + \left[a^2+\frac{(a-1)^2}{3}\right]^y \braket{\Psi' |\Psi'}\,,
\ee
where we defined 
\be
\ket{\Psi'}= \left[\1- P^{\msquare}_0 - \sum_{k=1}^x  P^{\msquare}_k \right] \ket{\mcirc' \ldots \mcirc'}\,.
\ee
We note that
\begin{align}
\braket{\Psi' |\Psi'}&=\left( \left[a^2+\frac{(a-1)^2}{3}\right]^{x}-\frac{1}{4^{x}}- \frac{(4 a-1)^2}{ 3\cdot 4^{x}} \right)\notag\\
&\sim \left[a^2+\frac{(a-1)^2}{3}\right]^{x}\,,
\end{align}
Finally, retaining only the leading terms, we arrive at the final result 
\begin{align}
\mathbb{E}\left[Z_2(x_+,x_-)\right] \geq& \frac{1}{4^{x_+}} + \frac{1}{4^{x_-}}\,,\\
\mathbb{E}\left[ Z_2(x_+,x_-)\right]\leq& \frac{1}{4^{x_+}} + \frac{1}{4^{x_-}}\notag\\
+&\frac{81}{(4a-1)^4}\!\! \left[a^2\!+\!\frac{(a-1)^2}{3}\right]^{x_+ + x_-}\,.
\end{align}
Therefore Eq.~\eqref{eq:goal3} holds as long as 
\be
\left[a^2+\frac{(1-a)^2}{3}\right]^2 \leq \frac{1}{4}\,,
\ee
which, as anticipated, gives $a< 0.683013$. 

As discussed before this result implies that for $a< a^*$, with $0.683013 <a^*\leq 1$, the R\'enyi-$2$ Tripartite Information decreases with the maximal slope. An interesting question concerns the exact value of $a^*$, or, better whether it is equal to one or smaller. Indeed, a value strictly smaller than one would imply a non-trivial gate-dependence on the rate of growth (we remind the reader that the latter vanishes at $a=1$). This kind of behaviour has been recently observed in the Local Operator Entanglement~\cite{BKP:OEergodicandmixing}. In the current case, however, we find evidence that this does not happen, as we explain below. Accordingly, we conjecture that $a^{\ast}=1$.

First we remark that a direct consequence of Property~\ref{prop:P1} is that the average gate \eqref{eq:averagedgateDU} is completely chaotic for any $a\neq1$, namely
\be
\lim_{x_\pm \to\infty }4^{x_\mp }\mathbb{E}\left[Z_2(x_+,x_-)\right]= 1\,, \qquad \forall a<1\,.
\label{eq:fake_limit}
\ee 
This means that in our case Conjecture~\eqref{BKP:conjecture} gives $a^\ast=1$, while the very same conjecture revealed the gate-dependence of the Operator Entanglement's slope in Ref.~\cite{BKP:OEergodicandmixing}.

Second, our numerical experiments support $a^\ast=1$. As a representative example, we report in Fig.~\ref{fig:logZ} numerical results obtained by direct contraction of the tensor network in Eq.~\eqref{eq:SAD'averaged} for $x=0$, corresponding to $x_-=x_+=t$, and increasing values of time. This was done using the iTensor library~\cite{iTensor}. Although we are clearly limited to fairly short times, our data are consistent with $a^\ast=1$.

\begin{figure}
	\includegraphics[scale=0.45]{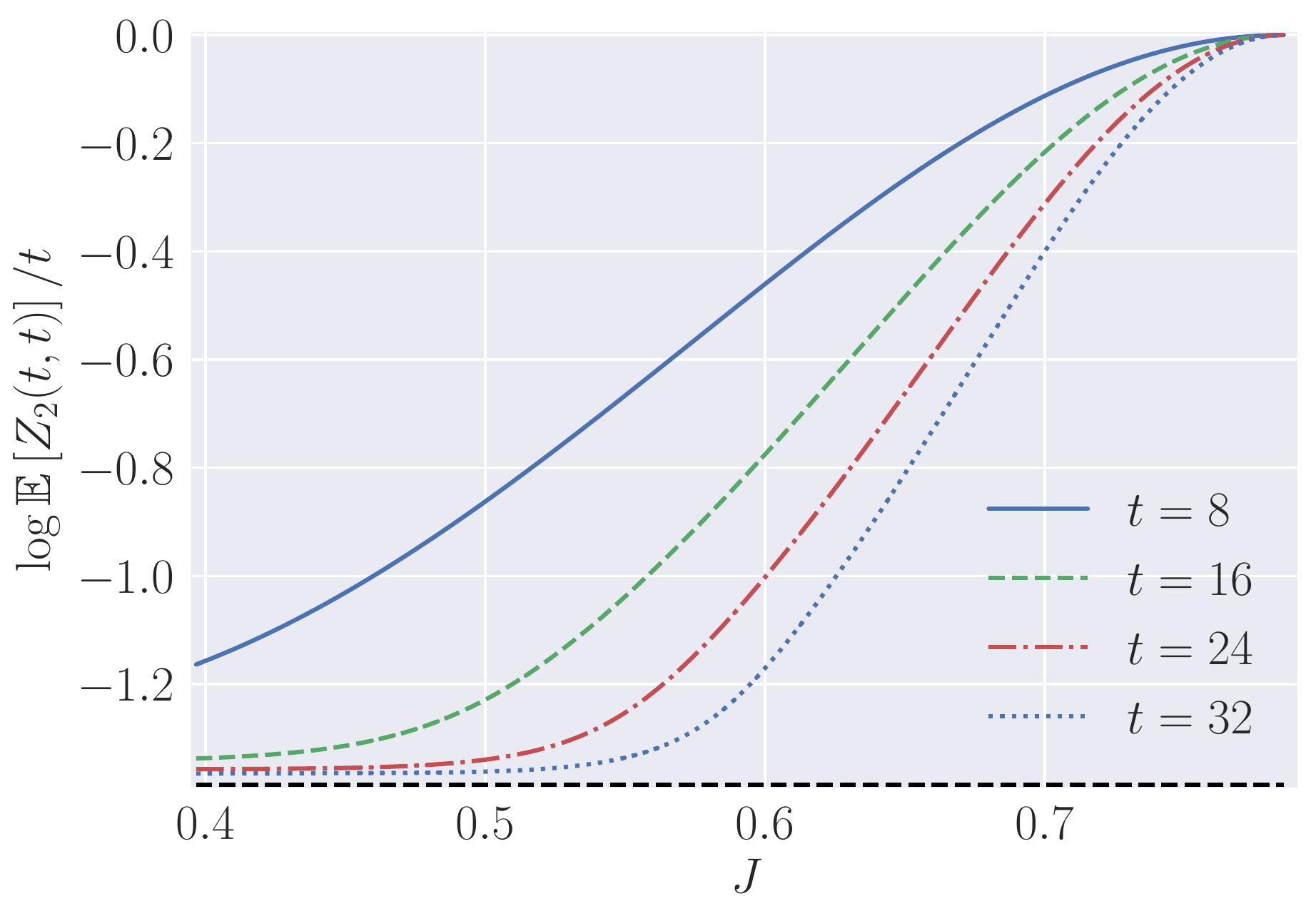}
	\caption{Numerical results for the averaged partition function $\log \mathbb{E}\left[Z_2(t,t)\right]/t$, as a function of $J$, and at different times $t$. The data are obtained by numerical contraction of the tensor network~\eqref{eq:SAD'averaged}, by means of the iTensor library~\cite{iTensor}. The black dashed line corresponds to the maximal rate $-2\log 2$.}
	\label{fig:logZ}
\end{figure}

\section{Out-of-time-order correlators}
\label{sec:OTOC}

The approach developed in the last section is adequate to study a large variety of quantities involving four copies of the evolution operator $\mathbb{U}$. As a further non-trivial example, in this section we revisit the computation of the OTOCs, which has been considered both for Haar-random~\cite{Nahum:operatorspreadingRU,Keyserlingk} and dual-unitary circuits~\cite{CL:OTOCsDU}.

Let us denote by $\{a_x^\alpha\}$ a Hilbert-Schmidt orthonormal basis of local operators at site $x$ cf.~\eqref{eq:squciralphabeta}. The OTOCs are defined as
\be
 O_{\alpha\beta}(\tilde x_+, \tilde x_-)=\frac{1}{d^L}{\rm tr}\left[{a^\alpha_0 \mathbb U^{-t} a^\beta_x \mathbb U^t a^\alpha_0 \mathbb U^{-t} a^\beta_x \mathbb U^t }\right]\,,
\label{eq:OTOC}
\ee
Here, in order to make direct contact with other works in the literature, we use slightly different ``light cone'' coordinates
\be
\tilde x_+=t+1+\lfloor x\rfloor,\qquad \tilde x_-=t-\lfloor x\rfloor\,.
\label{eq:xtildepm}
\ee
We can now apply the folding procedure, as explained in Appendix~\ref{sec:folding}, and represent $O_{\alpha\beta}(x,y) $ in terms of a partition function. In particular, using the usual ``folded" notation, we have
\be
\!\!\!\!\!\!\!\!O_{\alpha\beta}(x,y) \!\!=\!\!d^{x+y}\!\!\!\!\begin{tikzpicture}[baseline=(current  bounding  box.center), scale=0.55]
\def\eps{-0.5};
\foreach \i in {0,...,3}
\foreach \j in {2,...,-3}{
	\draw[very thick] (-4.5+1.5*\j,1.5*\i) -- (-3+1.5*\j,1.5*\i);
	\draw[very thick] (-3.75+1.5*\j,1.5*\i-.75) -- (-3.75+1.5*\j,1.5*\i+.75);
	\draw[ thick, fill=myvioletc, rounded corners=2pt, rotate around ={45: (-3.75+1.5*\j,1.5*\i)}] (-4+1.5*\j,0.25+1.5*\i) rectangle (-3.5+1.5*\j,-0.25+1.5*\i);
	\draw[thick, rotate around ={45: (-3.75+1.5*\j,1.5*\i)}] (-3.75+1.5*\j,.15+1.5*\i) -- (-3.6+1.5*\j,.15+1.5*\i) -- (-3.6+1.5*\j,1.5*\i);}
\foreach \i in {0,...,3}{
	\draw[thick, fill=white] (-0.1,1.5*\i-0.1) rectangle (0.1,1.5*\i+0.1);
	\draw[thick, fill=white] (-8.9,1.5*\i) circle (0.1cm);
}
\foreach \i in {-3,...,2}{
	\draw[thick, fill=white] (-3.85+1.5*\i,-0.85) rectangle (-3.65+1.5*\i,-0.65);
	\draw[thick, fill=white] (-3.75+1.5*\i,5.25) circle (0.1cm);
}
\draw [thick, decorate, decoration={brace,amplitude=4pt, raise=4pt},yshift=0pt]
(-8.25,5.4) -- (-.75,5.4) node [black,midway,yshift=0.55cm] {$x$};
\draw[thick, fill=white] (-.75-0.3,-.75-0.3) rectangle (-.75+0.3,-.75+0.3);
\Text[x=-0.75,y=-.75]{\tiny $\alpha\alpha$}
\draw[thick, fill=white] (-8.25,5.25) circle (0.3);
\Text[x=-8.25,y=5.25]{\tiny $\beta\beta$}
\draw [thick, decorate, decoration={brace,amplitude=4pt,mirror,raise=4pt},yshift=0pt]
(0,0) -- (0,4.5) node [black,midway,xshift=.45cm] {$y$};
\Text[x=-8.25,y=-3]{}
\end{tikzpicture}\!\!\!\!\!\!\,,
\label{eq:OTOCpart}
\ee
where we used the notation~\eqref{eq:states} for the boundary states, while $\text{\small\encircle{$\alpha\beta$}}$ and $\text{\small\boxed{$\alpha\beta$}}$ were defined in Eq.~\eqref{eq:squciralphabeta}. Here, we assumed $x\in\mathbb Z_L$. In the case where $x\in\mathbb Z_L+{1}/{2}$, the graphical representation~\eqref{eq:OTOCpart} should be modified by flipping the states at the top left corner. 

Performing the average of \eqref{eq:OTOC} we see that the operator dependence is drastically simplified. Specifically we find 
\begin{align}
	&\mathbb E[O_{\alpha\beta}(1,0)] = \frac{1}{d}{\rm tr}[{a^\alpha_0 a^\beta_0 a^\alpha_0 a^\beta_0}],\label{eq:O10}\\ 
	&\mathbb E[O_{\alpha\beta}(x\geq1,0)]=\mathbb E[O_{\alpha\beta}(0,y\geq0)]=1,\label{eq:O01}\\
	&\mathbb E[O_{\alpha\beta}(x\geq 1,y\geq 1)]= \mathbb E[O(x,y)],\label{eq:Oxy}
\end{align}
where we defined 
\be
\!\!\!\!\! O(x,y) \!\!=\!\! \frac{d^{x+y}}{d^2-1} \begin{tikzpicture}[baseline=(current  bounding  box.center), scale=0.55]
\def\eps{-0.5};
\foreach \i in {0,...,3}
\foreach \j in {2,...,-3}{
	\draw[very thick] (-4.5+1.5*\j,1.5*\i) -- (-3+1.5*\j,1.5*\i);
	\draw[very thick] (-3.75+1.5*\j,1.5*\i-.75) -- (-3.75+1.5*\j,1.5*\i+.75);
	\draw[ thick, fill=mygreen, rounded corners=2pt, rotate around ={45: (-3.75+1.5*\j,1.5*\i)}] (-4+1.5*\j,0.25+1.5*\i) rectangle (-3.5+1.5*\j,-0.25+1.5*\i);
	\draw[thick, rotate around ={45: (-3.75+1.5*\j,1.5*\i)}] (-3.75+1.5*\j,.15+1.5*\i) -- (-3.6+1.5*\j,.15+1.5*\i) -- (-3.6+1.5*\j,1.5*\i);}
\foreach \i in {0,...,3}{
	\draw[thick, fill=white] (-0.1,1.5*\i-0.1) rectangle (0.1,1.5*\i+0.1);
	\draw[thick, fill=white] (-8.9,1.5*\i) circle (0.1cm);
}
\foreach \i in {-3,...,2}{
	\draw[thick, fill=white] (-3.85+1.5*\i,-0.85) rectangle (-3.65+1.5*\i,-0.65);
	\draw[thick, fill=white] (-3.75+1.5*\i,5.25) circle (0.1cm);
}
\draw [thick, decorate, decoration={brace,amplitude=4pt, raise=4pt},yshift=0pt]
(-8.25,5.4) -- (-.75,5.4) node [black,midway,yshift=0.55cm] {$x$};
\draw[thick, fill=black] (-.75-0.1,-.75-0.1) rectangle (-.75+0.1,-.75+0.1);
\draw[thick, fill=black] (-8.25,5.25) circle (0.1cm);
\draw [thick, decorate, decoration={brace,amplitude=4pt,mirror,raise=4pt},yshift=0pt]
(0,0) -- (0,4.5) node [black,midway,xshift=.45cm] {$y$};
\Text[x=-8.25,y=-3]{}
\end{tikzpicture}\!\!\!\!\!\!.
\label{eq:OTOCave}
\ee
Here we introduced the states $\ket{\mcircf}$ and $\ket{\msquaref}$ by generalising the definition \eqref{eq:fullcircsquare} to any $d$ as follows
\be
\ket{\mcircf}=\frac{d \ket{\msquare}-\ket{\mcirc}}{ \sqrt{d^2-1}}, \qquad  \ket{\msquaref}=\frac{d \ket{\mcirc} -\ket{\msquare}}{{\sqrt{d^2-1}}}\,.
\label{eq:fullcircsquaredgen}
\ee
The partition function~\eqref{eq:OTOCave} can be analysed using the method developed in the previous section. As before, we treat separately the cases of Haar-random circuits and random dual-unitary circuits.

\subsection{OTOCs in Random Unitary Circuits}
\label{sec:otocs_RUC}

In Ref.~\cite{Nahum:operatorspreadingRU,Keyserlingk} the  OTOCs  for  Haar-random circuits were computed by mapping the folded tensor network onto a partition function of a $2D$ Ising model (see also Refs.~\cite{ZN:nonrandommembrane,ZN:statmech}). In this section, we show that one can arrive at the same result by deriving a set of recurrence relations starting directly from the average of the partition function~\eqref{eq:OTOCave}.

First we observe that, using the definition \eqref{eq:fullcircsquaredgen} and the identities \eqref{eq:unitary1}--\eqref{eq:unitary2}, the average of the partition function~\eqref{eq:OTOCave} can be written as
\be
\mathbb E[O(x,y)] \!=\! \frac{d^{x+y+2}}{(d^2-1)^2}W_{x-1}(x,y)\!-\!\frac{(2 d^{2}-1)}{(d^2-1)^2}\,,
\label{eq:otoc_W_relation}
\ee
where we introduced the function $W_{k}(x,y)$ which admits the graphical representation 
\be
\!\!\!\!\!\!W_k(x,y) =\begin{tikzpicture}[baseline=(current  bounding  box.center), scale=0.55]
\def\eps{-0.5};
\foreach \i in {0,...,3}
\foreach \j in {2,...,-3}{
	\draw[very thick] (-4.5+1.5*\j,1.5*\i) -- (-3+1.5*\j,1.5*\i);
	\draw[very thick] (-3.75+1.5*\j,1.5*\i-.75) -- (-3.75+1.5*\j,1.5*\i+.75);
	\draw[ thick, fill=gray, rounded corners=2pt, rotate around ={45: (-3.75+1.5*\j,1.5*\i)}] (-4+1.5*\j,0.25+1.5*\i) rectangle (-3.5+1.5*\j,-0.25+1.5*\i);
	\draw[thick, rotate around ={45: (-3.75+1.5*\j,1.5*\i)}] (-3.75+1.5*\j,.15+1.5*\i) -- (-3.6+1.5*\j,.15+1.5*\i) -- (-3.6+1.5*\j,1.5*\i);}
\foreach \i in {0,...,3}{
	\draw[thick, fill=white] (-0.1,1.5*\i-0.1) rectangle (0.1,1.5*\i+0.1);
	\draw[thick, fill=white] (-8.9,1.5*\i) circle (0.1cm);
}
\foreach \i in {-2,...,2}{
	\draw[thick, fill=white] (-3.75+1.5*\i,-0.75) circle (0.1cm);
	\draw[thick, fill=white] (-3.75+1.5*\i,5.25) circle (0.1cm);
}
\draw [thick, decorate, decoration={brace,amplitude=4pt, raise=4pt},yshift=0pt]
(-8.25,5.3) -- (-.75,5.3) node [black,midway,yshift=0.55cm] {$x$};
\draw[thick, fill=white] (-8.25-0.1,5.25-0.1) rectangle (-8.25+0.1,5.25+0.1);
\draw[thick, fill=white] (-8.25,-.75) circle (0.1cm);
\draw[thick, fill=white] (-3.75-1.5-0.1,-0.75-0.1) rectangle (-3.75-1.5+0.1,-0.75+0.1);
\draw[thick, fill=white] (-3.75-0.1,-0.75-0.1) rectangle (-3.75+0.1,-0.75+0.1);
\draw[thick, fill=white] (-3.75-3-0.1,-0.75-0.1) rectangle (-3.75-3+0.1,-0.75+0.1);
\draw[thick, fill=white] (-3.75-4.5-0.1,-0.75-0.1) rectangle (-3.75-4.5+0.1,-0.75+0.1);
\draw [thick, decorate, decoration={brace,amplitude=4pt,mirror,raise=4pt},yshift=0pt]
(-8.25,-0.7) -- (-3.75,-.7) node [black,midway,yshift=-.45cm] {$k$};
\draw [thick, decorate, decoration={brace,amplitude=4pt,raise=4pt,mirror},yshift=0pt]
(0,0) -- (0,4.5) node [black,midway,xshift=.45cm] {$y$};
\Text[x=-8.25,y=-3]{}
\end{tikzpicture}\!\!\!\!\!\!\,,
\label{eq:OTOCHaarave}
\ee
where, once again, we denoted the Haar-averaged folded gate by a grey square.

Now, using Eqs.~\eqref{eq:RUrel1} and \eqref{eq:RUrel2}, and following the graphical derivation of Eq.~\eqref{eq:RUrecurrence1}, it is easy to write down a recurrence relation for the partition functions $\{W_k(x,y) \}_{k=0}^x$. For instance, one can start from the bottom right corner and use \eqref{eq:RUrel1} to simplify the gate. Then the recurrence relations are found by multiple use of the first of \eqref{eq:unitary1}, to ``pull up" the squares, and of the second of \eqref{eq:unitary1}, to ``pull left" the circles. The final result reads as
\begin{align}
	W_k(x,y)=& \frac{1}{d^2+1} W_{k}(x-1,y)+W_{0}(x,y-1) \left[\frac{d}{d^2+1}\right]^{k+1}\notag\\
	&+\frac{1}{d}  \sum_{r=0}^{k-1}  W_{k-r}(x,y-1) \left[\frac{d}{d^2+1}\right]^{r+2}\!\!\!\!,
	\label{eq:OTOCequations}
\end{align}
with boundary conditions $W_k(x,0)=1/d^{|k-1|}$ and $W_k(k,y)=1/d^{|y+k-1|}$~\cite{note3}. 

This recurrence relation is more involved than the one for the tripartite information because it features $x$ functions instead of just one~\cite{notecomplexity}. Still, it can still be solved exactly.  A particularly elegant solution is obtained by means of the so-called \emph{kernel method}~\cite{Prodinger2004,Bous05,Bostan2010,PemantleBoook2013}. Specifically, this method allows one to write down directly the generating function for the coefficients $W_{x-1}(x,y)$ (directly related to the OTOCs via Eq.~\eqref{eq:otoc_W_relation}). The details of the calculation are relegated to Appendix~\ref{sec:solution_recurrence}, while here we only report the final result. Defining
\be
\mathcal{G}(z,w)=\sum_{m,n=0}^{\infty}W_m(m+1,n)z^{m}w^{n}\,.
\label{eq:g_function}
\ee
we find $\mathcal{G}(z,w)=\widetilde{\mathcal{G}}(z/d,w/d)$, where
\begin{widetext}
	\begin{align}
		\widetilde{\mathcal{G}}(z,w)&=
		\frac{b z \left(d^2 z-1\right)}{2 a d^3 (1-z) (-p w-q z+1)}-\frac{b^2 z \sqrt{p^2 q^2 (z-w)^2-2 p q (z+w)+1}}{2 d^3 (-p w-q z+1) (-p z-q w+1)}\nonumber\\
		&+\frac{b q z
			\left(z-d^2\right)}{2 d^3 (1-z) (-p z-q w+1)}+\frac{d}{(1-z) (1-w)}-2ab\,,
		\label{eq:generating_function}
	\end{align}
\end{widetext}	
with $a=d^2+1$, $ b=d^2-1$, while $p$ was defined in Eq.~\eqref{eq:p_def} and $q=1-p$. 

The expression for the OTOCs can be recovered by deriving $\mathcal{G}(z,w)$ with respect to $z$ and $w$ (respectively $x-1$ and $y$ times) and plugging the result into \eqref{eq:otoc_W_relation}. The final expression agrees with that reported in Eq. (73) of Ref.~\cite{Nahum:operatorspreadingRU} and, in particular, takes the following asymptotic form for $t\gg1$
\be
\!\! O_{\alpha\beta}(\tilde x_+, \tilde x_-) \approx 1-\frac{1}{4}\Phi\left( \frac{x-v_Bt}{\sigma(t)}\right)\Phi\left(\frac{x+v_Bt}{\sigma(t)}\right)\!,
\ee
where $v_B$ and $\sigma(t)$ were defined in Eq.~\eqref{eq:vb_sigma}.


\subsection{OTOCs in Random Dual-Unitary Circuits}
\label{sec:otocs_DU}

Let us now move to consider the calculation of OTOCs in dual-unitary circuits. This problem has been recently considered in Ref.~\cite{CL:OTOCsDU} where the authors presented exact results for a non-interacting dual-unitary circuit (the self-dual kicked Ising model in zero longitudinal field) and computed the asymptotic limits $ \tilde x_\pm\to\infty$ in the class of completely chaotic dual-unitary circuits, where $\tilde{x}_\pm$ is defined in Eq.~\eqref{eq:xtildepm}. The latter result allows one to obtain the asymptotic behaviour of OTOCs at a finite distance from the light-cone edge. Here we show that, by using our approach, we can go beyond these results in the case of random dual-unitary circuits. Indeed, as we now see, the introduction of single-site averages introduces key simplifications. 

First, Property~\ref{prop:P1} provides a rigorous proof that all random dual-unitary gates \eqref{eq:averagedgateDU} are completely chaotic as long as $a\neq 1$~\cite{noteOTOC}. This means that we explicitly determined a class of circuits for which the findings of Ref.~\cite{CL:OTOCsDU} are exact. In particular, since taking the limits $\tilde x_\pm\to \infty$ corresponds to projecting onto the eigenspace associated to the maximal eigenvalue (which is $1/2$ by Property~\ref{prop:P1}) we have    
\begin{widetext}
	\be
	\lim_{\tilde x_-\to \infty}\bar O(\tilde x_+,\tilde x_-) = \frac{2^{\tilde x_+}}{3}\braket{\mcircf\underbrace{\mcirc\ldots\mcirc}_{\tilde x_+-1}|\sum_{k=0}^{\tilde x_+} P^{\msquare}_k |\msquare\ldots\msquare\msquaref} = 
	\begin{cases}
		\displaystyle\frac{2}{3}\braket{\mcircf|\msquaref}=-\frac{1}{3} & \tilde x_+=1\\
		\\
		\displaystyle \frac{2^{\tilde x_+}}{3}\braket{\mcircf\overbrace{\mcirc\cdots\mcirc}^{\tilde x_+-1}|\overbrace{\msquare\cdots\msquare}^{\tilde x_+-1}\msquaref}=0 \qquad & \tilde x_+ >1
	\end{cases}
	\label{eq:OTOClimit1}
	\ee
	and 
	\begin{align}
		\!\!\!\!\!\lim_{\tilde x_+\to \infty}\bar O(\tilde x_+,\tilde x_-) =&  \frac{4}{9} \sum_{j=1}^{\tilde x_-} \left(4\braket{\mcirc|(\bar{\mathcal M}^{\msquare})^{j}|\msquaref}-\braket{\mcirc|(\bar{\mathcal M}^{\msquare})^{j-1}|\msquaref} \right) \left(\braket{\mcircf|(\bar{\mathcal M}^{\mcirc})^{\tilde x_--j}|\msquare}- \braket{\mcircf|(\bar{\mathcal M}^{\mcirc})^{\tilde x_--j+1}|\msquare}\right)\notag\\
		&+\frac{4}{3} \braket{\mcirc|\msquaref} \braket{\mcircf|(\bar{\mathcal M}^{\mcirc})^{\tilde x_-}|\msquare}
		=  (4a-1)(1-a)  \frac{\tilde x_-  a^{\tilde x_-}}{3a} + a^{\tilde x_-}\,.
		\label{eq:OTOClimit2}
	\end{align}
Here we introduced the 1-qubit maps 
\be
	\bar{\mathcal M}^{\msquare}  =
	\begin{tikzpicture}[baseline=(current  bounding  box.center), scale=0.65]
	\def\eps{-0.5};
	\foreach \i in {3}{
		\foreach \j in {-3}{
			\draw[very thick] (-4.5+1.5*\j,1.5*\i) -- (-3+1.5*\j,1.5*\i);
			\draw[very thick] (-3.75+1.5*\j,1.5*\i-.75) -- (-3.75+1.5*\j,1.5*\i+.75);
			\draw[ thick, fill=myorange, rounded corners=2pt, rotate around ={45: (-3.75+1.5*\j,1.5*\i)}] (-4+1.5*\j,0.25+1.5*\i) rectangle (-3.5+1.5*\j,-0.25+1.5*\i);
			\draw[thick, rotate around ={45: (-3.75+1.5*\j,1.5*\i)}] (-3.75+1.5*\j,.15+1.5*\i) -- (-3.6+1.5*\j,.15+1.5*\i) -- (-3.6+1.5*\j,1.5*\i);}
		\draw[thick, fill=white] (-7.6-0.1,1.5*\i-0.1) rectangle (-7.6+0.1,1.5*\i+0.1);
		\draw[thick, fill=white] (-9,1.5*\i-0.1) rectangle (-8.8,1.5*\i+0.1);
	}
	\Text[x=-8.25,y=3.5]{}
	\end{tikzpicture}= \ket{\msquare}\!\!\bra{\msquare} + a \ket{\msquaref}\!\!\bra{\msquaref}\,,  
	\qquad\qquad 
	\bar{\mathcal M}^{\mcirc} =
	\begin{tikzpicture}[baseline=(current  bounding  box.center), scale=0.65]
	\def\eps{-0.5};
	\foreach \i in {3}{
		\foreach \j in {-3}{
			\draw[very thick] (-4.5+1.5*\j,1.5*\i) -- (-3+1.5*\j,1.5*\i);
			\draw[very thick] (-3.75+1.5*\j,1.5*\i-.75) -- (-3.75+1.5*\j,1.5*\i+.75);
			\draw[ thick, fill=myorange, rounded corners=2pt, rotate around ={135: (-3.75+1.5*\j,1.5*\i)}] (-4+1.5*\j,0.25+1.5*\i) rectangle (-3.5+1.5*\j,-0.25+1.5*\i);
			\draw[thick, rotate around ={45: (-3.75+1.5*\j,1.5*\i)}] (-3.75+1.5*\j,.15+1.5*\i) -- (-3.6+1.5*\j,.15+1.5*\i) -- (-3.6+1.5*\j,1.5*\i);}
		\draw[thick, fill=white] (-7.6,1.5*\i) circle (0.1cm);
		\draw[thick, fill=white] (-8.9,1.5*\i) circle (0.1cm);
	}
	\Text[x=-8.25,y=3.5]{}
	\end{tikzpicture}= \ket{\mcirc}\!\!\bra{\mcirc} + a \ket{\mcircf}\!\!\bra{\mcircf}\,.
\ee
\end{widetext}

As discussed in Ref.~\cite{{CL:OTOCsDU}} these maps can be expressed in terms of those  determining the correlation functions of one-site observables~\cite{BKP:dualunitary, PBCP20}. The fact that these ``correlation maps" determine the light-cone behaviour of scrambling measures appear to be quite general, indeed the same has been observed for the operator entanglement~\cite{BKP:OEergodicandmixing}.

The results~\eqref{eq:OTOClimit1} and \eqref{eq:OTOClimit2} are interesting but somewhat limited as they only give information on the ``edges of the light cone" of the out-of-time-ordered correlators. In other words they only describe the case when the OTOC is evaluated at $|x|\approx v_{\rm max} t$ (the symbol $\approx$ denotes equal up to corrections scaling like $t^\eta$ with $\eta<1$). Using our method we can go beyond these results and find the leading behaviour of the OTOCs for  $x\approx \xi t$, for all ``rays" $\xi\in[\xi^*(a),1]$ with 
\be
\!\!\!\xi \!\geq\! \xi^*_a\equiv \frac{\log[a/\mathcal A(a)\mathcal B(a)]}{\log[a]\!-\! {\rm sgn}[a\!-\!\mathcal A(a)\mathcal B(a)] \!\log[\mathcal A(a)/\mathcal B(a)]},
\ee
and  
\be
\mathcal A(a):=1-\frac{4}{3}(a-1)^2, \quad \mathcal B(a):= 2a^2+\frac{2}{3}(a-1)^2\,. 
\label{eq:mathcAB}
\ee
The function $\xi^*_a$ is monotonically increasing in $a$ and its boundary values are $\xi^*_{1/3}\approx-0.532$ and $\xi^*_{1}=1$ (see Fig.~\ref{fig:xi_plot}).  

The idea is again to write some recursive relations for the average of the partition function in \eqref{eq:OTOCave} and truncate them to obtain useful bounds. Specifically, following the calculations detailed in Appendix~\ref{app:OTOCsDU} arrive at 
\begin{align}
	\mathbb E\left[O(x,y)\right]  =& \left[{a^{y}} +(4a-1)(1-a)  \frac{y {a^{y}}}{3a}\right] (1-\delta_{x,1}) \notag\\ 
	&-\frac{1}{3}\delta_{x,1} + r(x,y)
	\label{eq:OTOCbound}
\end{align}
where the ``reminder" $r(x,y)$ fulfils the following bound    
\be
|r(x,y)| \leq C  \mathcal A(a)^{{\rm max}(x,y)} \mathcal B(a)^{{\rm min}(x,y)},
\ee
where $C$ is an appropriate constant. 

Equation \eqref{eq:OTOCbound} implies that the leading contribution to the OTOC \eqref{eq:OTOC} for $t\gg 1$ and fixed $\xi=x/t\in[\xi^*_a,1]$ is given by
\be
\!\!\!\mathbb E\!\left[O(\tilde x_+,\tilde x_-)\right]   \!\approx\!
a^{t(1-\xi)}\!\!\left[1\!+\!t (4a\!-\!1)(1\!-\!a)  \frac{(1-\xi)}{3a}\right]\!\!.
\label{eq:asyOTOCDU}
\ee
This is because for $\xi \geq \xi^*_a$
\be
a^{1-\xi}> \mathcal A(a)\mathcal B(a) \left(\frac{\mathcal A(a)}{\mathcal B(a)}\right)^{|\xi|}, 
\ee
and, therefore, the reminder gives a sub-leading contribution to \eqref{eq:OTOCbound}. This relation also implies that 
\be
\!\!\!\mathbb E\!\left[O(\tilde x_+,\tilde x_-)\right]   \!\!<\!\!
a^{t(1-|\xi|)}\!\!\left[1\!+\!t (4a\!-\!1)(1\!-\!a)  \frac{(1\!-\!|\xi|)}{3a}\right]\!\!,
\ee
for $\xi\in[-1,-\xi^*_a[$. Finally we note that, since $\mathcal A(a)\leq1$ for all $a\in[1/3,1]$, the reminder vanishes in the limit $x_-\to \infty$ and the the expression \eqref{eq:OTOCbound} reproduces \eqref{eq:OTOClimit1}.

\begin{figure}
	\includegraphics[width=8.cm]{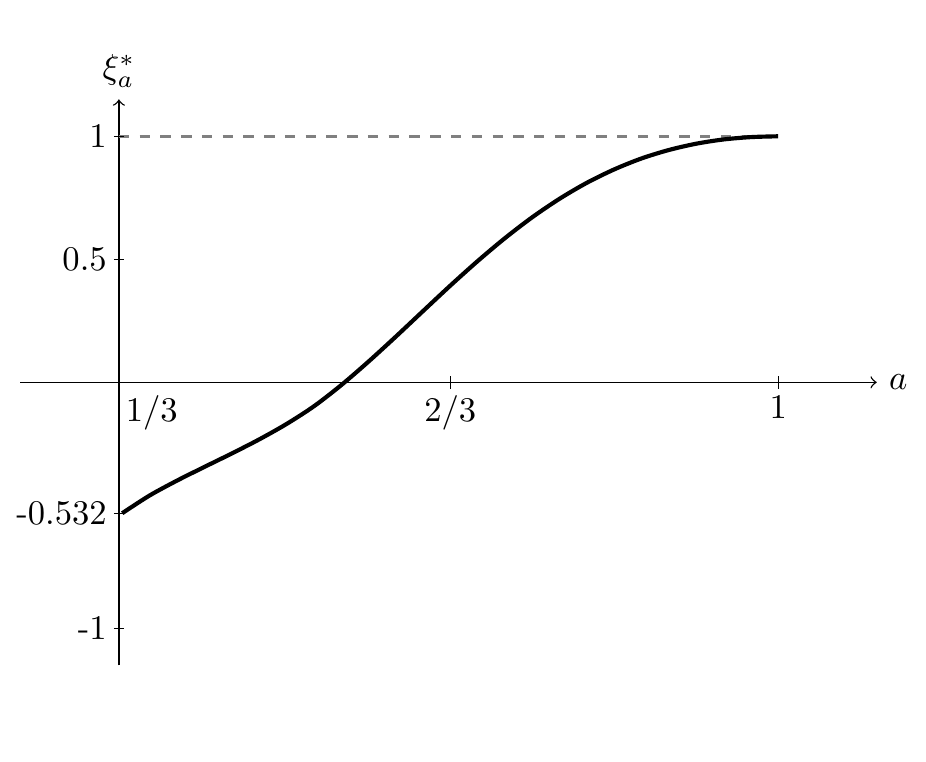}
	\caption{Plot of the transition ray $\xi^*_a$, after which the OTOC follows the asymptotic form \eqref{eq:asyOTOCDU}, as a function of $a$.}
	\label{fig:xi_plot}
\end{figure}

\section{Conclusions}
\label{sec:conclusions}

We have studied the scrambling of quantum information in random unitary circuits, focusing on two cases: $(i)$ the local gates are Haar random and $(ii)$ the local gates are dual-unitary and randomly sampled from a single-site Haar-invariant measure. 

We characterised the scrambling through the R\'enyi-$2$ tripartite information introduced in Ref.~\cite{HQRY:tripartiteinfo}, which can be thought of as a minimal chaos indicator. When non-zero, its (negative) slope signals dynamical chaos, while it is constructed with the minimal number ($n=4$) of replicated copies of the time-evolution operator (two evolving forward and two backward). We computed this quantity exactly for both Haar random circuits and random dual-unitary circuits and proved rigorously that there exists a ``maximally chaotic" subset of gates for which the R\'enyi-$2$ tripartite information grows at the maximal speed, which is strictly larger than one observed in the Haar-random case.   

To find these results we employed a standard mapping onto a folded tensor network that we then studied by means of simple recurrence relations. As such, this approach can be viewed as an alternative formulation (at least in the Haar-random case) of the one developed in Refs.~\cite{ZN:statmech,ZN:nonrandommembrane}, which is based on a mapping onto a classical spin system. 

Our approach is adequate to study also other quantities involving four copies of the time-evolution operator. As an example, here we also considered OTOCs. We recovered the exact results of  Ref.~\cite{Nahum:operatorspreadingRU,Keyserlingk} for the Haar random case, and presented an exact expression for random dual-unitary circuits, extending recent findings of Ref.~\cite{ClLa20}. The same approach can be used to study growth of the R\'enyi-$2$ entropy of finite connected regions after a quench, and gives a promising starting point to tackle the calculation of the same quantity for disjoint intervals. This is particularly interesting in connection to the mutual information of disjoint regions, which was recently studied in the context of quantum scrambling~\cite{BBCC11,AsBe14,ABGH15,LeMo15,AlCa19,AlCa19_II,MoAC20}.

Finally, it is natural to wonder whether the perspective adopted in this work can give some useful insights into the calculation of quantities involving $n>4$ copies of the time evolution operators, giving access to the average of higher R\'enyi entropies and to the averaged operator space entanglement. The findings of Ref.~\cite{ZN:statmech}, however, reveal that this problem is far from straightforward.

\section{Acknowledgments}
\label{sec:acknowledgments}
We thank T. Prosen and P. Kos for useful discussions, and in particular A. Nahum for a careful reading of the manuscript and valuable comments. LP acknowledges support from the Alexander von Humboldt foundation.  BB acknowledges support by the EU Horizon 2020 program through the ERC Advanced Grant OMNES No. 694544, and by the Slovenian Research Agency (ARRS) under the Programme P1-0402. 

\appendix

\begin{widetext}

\section{The $2D$ folded tensor network}
\label{sec:map_2d_pf}

In this appendix we describe how to obtain a diagrammatic representation of Eq.~\eqref{eq:I3} in terms of the ``folded" gate~\eqref{eq:W}. We illustrate the procedure in the case of $S^{(2)}_{AD}$ as the one for $S^{(2)}_{AC}$ is completely analogous. Representing $\rho_{AD}$ [cf. Eq.~\eqref{eq:rhoAD}] diagrammatically, we find  
\be
[\rho_{AD}]_{\boldsymbol s'_A \boldsymbol s'_D}^{\boldsymbol s_A \boldsymbol s_D} = \frac{1}{d^{2L}} 
\begin{tikzpicture}[baseline=(current  bounding  box.center), scale=0.45]
\foreach \i in {-4,...,-1}{
\draw[thick] (-\i-4.5,0) -- (-\i-4.5,2.5);
}
\foreach \i in {1,...,3}{
\draw[thick, rounded corners=1pt]  (\i-9.5,-6) -- (\i-9.5,-6-0.2*\i) -- (-0.2*\i-10.5,-6-0.2*\i) -- (-0.2*\i-10.5,8.5+0.2*\i) -- (\i-9.5,8.5+0.2*\i) -- (\i-9.5,8.5) ;
}
\foreach \kk in {0}{
\def\eps{13*\kk};

\foreach \ii[evaluate=\ii as \i using \ii-0.5] in {1,...,5}
{
\draw[thick] (-.5-2*\i,-1+\eps) -- (0.5-2*\i,\eps);
\draw[thick] (-0.5-2*\i,\eps) -- (0.5-2*\i,-1+\eps);
\draw[thick, fill=myblue, rounded corners=2pt] (-0.25-2*\i,-0.25+\eps) rectangle (.25-2*\i,-0.75+\eps);
\draw[thick] (-2*\i,-0.35+\eps) -- (.15-2*\i,-.35+\eps) -- (.15-2*\i,-0.5+\eps);
}
\foreach \ii[evaluate=\ii as \i using \ii-0.5] in {1,...,5}
{
\draw[thick] (.5-2*\i,-6+\eps) -- (1-2*\i,-5.5+\eps);
\draw[thick] (1.5-2*\i,-6+\eps) -- (1-2*\i,-5.5+\eps);
}
\foreach \jj[evaluate=\jj as \j using 2*(ceil(\jj/2)-\jj/2)] in {0,...,3}
\foreach \i in {1,...,5}
{
\draw[thick] (.5-2*\i-1*\j+1,-2-1*\jj+\eps) -- (1-2*\i-1*\j+1,-1.5-\jj+\eps);
\draw[thick] (1-2*\i-1*\j+1,-1.5-1*\jj+\eps) -- (1.5-2*\i-1*\j+1,-2-\jj+\eps);
}
\foreach \jj[evaluate=\jj as \j using 2*(ceil(\jj/2)-\jj/2)] in {0,...,4}
\foreach \i in {1,...,5}
{
\draw[thick] (.5-2*\i-1*\j+1,-1-1*\jj+\eps) -- (1-2*\i-1*\j+1,-1.5-\jj+\eps);
\draw[thick] (1-2*\i-1*\j+1,-1.5-1*\jj+\eps) -- (1.5-2*\i-1*\j+1,-1-\jj+\eps);
\draw[thick, fill=myblue, rounded corners=2pt] (0.75-2*\i-1*\j+1,-1.75-\jj+\eps) rectangle (1.25-2*\i-1*\j+1,-1.25-\jj+\eps);
\draw[thick] (1-2*\i-1*\j+1,-1.35-1*\jj+\eps) -- (1.15-2*\i-1*\j+1,-1.35-1*\jj+\eps) -- (1.15-2*\i-1*\j+1,-1.5-1*\jj+\eps);
}
}
\foreach \kk in {0}{
\def\eps{13*\kk+8.5};
\foreach \i in {0,...,4}
{
\draw[thick] (-.5-2*\i,-1+\eps) -- (0.5-2*\i,\eps);
\draw[thick] (-0.5-2*\i,\eps) -- (0.5-2*\i,-1+\eps);
\draw[thick, fill=myred, rounded corners=2pt] (-0.25-2*\i,-0.25+\eps) rectangle (.25-2*\i,-0.75+\eps);
\draw[thick] (-2*\i,-0.35+\eps) -- (.15-2*\i,-.35+\eps) -- (.15-2*\i,-0.5+\eps);
}
\foreach \i in {1,...,5}
{
\draw[thick] (.5-2*\i,-6+\eps) -- (1-2*\i,-5.5+\eps);
\draw[thick] (1.5-2*\i,-6+\eps) -- (1-2*\i,-5.5+\eps);
}
\foreach \jj[evaluate=\jj as \j using -2*(ceil(\jj/2)-\jj/2)] in {0,...,3}
\foreach \i in {1,...,5}
{
\draw[thick] (.5-2*\i-1*\j,-2-1*\jj+\eps) -- (1-2*\i-1*\j,-1.5-\jj+\eps);
\draw[thick] (1-2*\i-1*\j,-1.5-1*\jj+\eps) -- (1.5-2*\i-1*\j,-2-\jj+\eps);
}
\foreach \jj[evaluate=\jj as \j using -2*(ceil(\jj/2)-\jj/2)] in {0,...,4}
\foreach \i in {1,...,5}
{
\draw[thick] (.5-2*\i-1*\j,-1-1*\jj+\eps) -- (1-2*\i-1*\j,-1.5-\jj+\eps);
\draw[thick] (1-2*\i-1*\j,-1.5-1*\jj+\eps) -- (1.5-2*\i-1*\j,-1-\jj+\eps);
\draw[thick, fill=myred, rounded corners=2pt] (0.75-2*\i-1*\j,-1.75-\jj+\eps) rectangle (1.25-2*\i-1*\j,-1.25-\jj+\eps);
\draw[thick] (1-2*\i-1*\j,-1.35-1*\jj+\eps) -- (1.15-2*\i-1*\j,-1.35-1*\jj+\eps) -- (1.15-2*\i-1*\j,-1.5-1*\jj+\eps);
}
}
\Text[x=-9.6,y=2]{\scriptsize $s_{A,1}$}
\Text[x=-7.1,y=2]{$\cdots$}
\Text[x=-4.45,y=2]{\scriptsize$s_{A,\ell_A}$}
\Text[x=-9.6,y=0.65]{\scriptsize$s'_{A,1}$}
\Text[x=-7.1,y=0.65]{$\cdots$}
\Text[x=-4.45,y=0.65]{\scriptsize$s'_{A,\ell_A}$}
\Text[x=-5.25,y=8.9]{\scriptsize$s_{D,1}$}
\Text[x=-2.75,y=8.9]{$\cdots$}
\Text[x=1,y=8.9]{\scriptsize$s_{D,\ell_D}$}
\Text[x=-5.25,y=-6.65]{\scriptsize$s'_{D,1}$}
\Text[x=-2.75,y=-6.65]{$\cdots$}
\Text[x=1,y=-6.65]{\scriptsize$s'_{D,\ell_D}$}
\label{eq:rho_AD}
\end{tikzpicture},
\ee
where we considered open boundary conditions. From Eq.~\eqref{eq:rhoAD}, it is straightforward to obtain the graphical representation  for the purity ${\rm tr}\left[\rho_{AD}^2\right]$, which is reported in Fig.~\ref{fig:trrho2gen}.
\begin{figure}
	\centering
\begin{tikzpicture}[baseline=(current  bounding  box.center), scale=0.5]
\scalebox{1.2}{\Text[x=-16,y=8]{${\rm tr}[\rho_{AD}^2] = \frac{1}{d^{4L}} $}};
\draw[thick] (.5,-7.25) -- (2,-7.25) -- (2,26.75) -- (.5,26.75);
\draw[thick] (-13+0.2,-7.25) -- (-16+2,-7.25) -- (-16+2,26.75) -- (-13+0.2,26.75);
\foreach \i in {-6,...,-1}{
\draw[thick]  (.5+\i,-7.25) -- (.5+\i,-7.25+0.2*\i) -- (2-0.2*\i,-7.25+0.2*\i) -- (2-0.2*\i,26.75-0.2*\i) -- (.5+\i,26.75-0.2*\i) -- (.5+\i,26.75);}
\foreach \i in {-5,...,-1}{
\draw[thick]  (-13+0.2-0.2*\i,-7.25) -- (-13+0.2-0.2*\i,-7.25+0.2*\i) -- (-16+2+0.2*\i,-7.25+0.2*\i) -- (-16+2+0.2*\i,26.75-0.2*\i) -- (-13+0.2-0.2*\i,26.75-0.2*\i) -- (-13+0.2-0.2*\i,26.75);
}
\foreach \zz in {0,17}
{
\foreach \i in {-4,...,-1}{
\draw[thick] (-\i-4.5,0+\zz) -- (-\i-4.5,2.5+\zz);
}
\foreach \i in {1,...,3}{
\draw[thick, rounded corners=1pt]  (\i-9.5,-6+\zz) -- (\i-9.5,-6-0.2*\i+\zz) -- (-0.2*\i-10.5,-6-0.2*\i+\zz) -- (-0.2*\i-10.5,8.5+0.2*\i+\zz) -- (\i-9.5,8.5+0.2*\i+\zz) -- (\i-9.5,8.5+\zz) ;
}
\foreach \i in {1,...,5}{
\draw[thick, rounded corners=1pt]  (\i-9.5,-6+\zz+8.5) -- (\i-9.5,-6-0.2*\i+\zz+8.5) -- (-10.4,-6-0.2*\i+\zz+8.5);
\draw[thick, rounded corners=1pt]  (-11.6,-6-0.2*\i+\zz+8.5) -- (-11.8-0.2*\i,-6-0.2*\i+\zz+8.5) -- (-11.8-0.2*\i,-6+\zz+15.75);
\draw[thick, rounded corners=1pt]  (-10.4,8.5+0.2*\i+\zz-8.5) -- (\i-9.5,8.5+0.2*\i+\zz-8.5) -- (\i-9.5,8.5+\zz-8.5) ;
\draw[thick, rounded corners=1pt]  (-11.6,8.5+0.2*\i+\zz-8.5) -- (-11.8-0.2*\i,8.5+0.2*\i+\zz-8.5) -- (-11.8-0.2*\i,8.5+\zz-15.75);
}
\draw[thick, rounded corners=1pt]  (-9.5,-6+\zz+8.5) -- (-10.4,-6+\zz+8.5);
\draw[thick, rounded corners=1pt]  (-11.6,-6+\zz+8.5) -- (-11.8,-6+\zz+8.5) -- (-11.8,-6+\zz+15.75);
\draw[thick, rounded corners=1pt]  (-10.4,8.5+\zz-8.5) -- (-9.5,8.5+\zz-8.5) ;
\draw[thick, rounded corners=1pt]  (-11.6,8.5+\zz-8.5) -- (-11.8,8.5+\zz-8.5) -- (-11.8,8.5+\zz-15.75);
\foreach \i in {-5,...,1}{
\draw[thick] (-\i-4.5,0+\zz+8.5) -- (-\i-4.5,1.25+\zz+8.5);
\draw[thick] (-\i-4.5,1.25+\zz-8.5) -- (-\i-4.5,2.5+\zz-8.5);
}
}
\foreach \kk in {0,1}{
\def\eps{17*\kk};

\foreach \ii[evaluate=\ii as \i using \ii-0.5] in {1,...,5}
{
\draw[thick] (-.5-2*\i,-1+\eps) -- (0.5-2*\i,\eps);
\draw[thick] (-0.5-2*\i,\eps) -- (0.5-2*\i,-1+\eps);
\draw[thick, fill=myblue, rounded corners=2pt] (-0.25-2*\i,-0.25+\eps) rectangle (.25-2*\i,-0.75+\eps);
\draw[thick] (-2*\i,-0.35+\eps) -- (.15-2*\i,-.35+\eps) -- (.15-2*\i,-0.5+\eps);
}
\foreach \ii[evaluate=\ii as \i using \ii-0.5] in {1,...,5}
{
\draw[thick] (.5-2*\i,-6+\eps) -- (1-2*\i,-5.5+\eps);
\draw[thick] (1.5-2*\i,-6+\eps) -- (1-2*\i,-5.5+\eps);
}
\foreach \jj[evaluate=\jj as \j using 2*(ceil(\jj/2)-\jj/2)] in {0,...,3}
\foreach \i in {1,...,5}
{
\draw[thick] (.5-2*\i-1*\j+1,-2-1*\jj+\eps) -- (1-2*\i-1*\j+1,-1.5-\jj+\eps);
\draw[thick] (1-2*\i-1*\j+1,-1.5-1*\jj+\eps) -- (1.5-2*\i-1*\j+1,-2-\jj+\eps);
}
\foreach \jj[evaluate=\jj as \j using 2*(ceil(\jj/2)-\jj/2)] in {0,...,4}
\foreach \i in {1,...,5}
{
\draw[thick] (.5-2*\i-1*\j+1,-1-1*\jj+\eps) -- (1-2*\i-1*\j+1,-1.5-\jj+\eps);
\draw[thick] (1-2*\i-1*\j+1,-1.5-1*\jj+\eps) -- (1.5-2*\i-1*\j+1,-1-\jj+\eps);
\draw[thick, fill=myblue, rounded corners=2pt] (0.75-2*\i-1*\j+1,-1.75-\jj+\eps) rectangle (1.25-2*\i-1*\j+1,-1.25-\jj+\eps);
\draw[thick] (1-2*\i-1*\j+1,-1.35-1*\jj+\eps) -- (1.15-2*\i-1*\j+1,-1.35-1*\jj+\eps) -- (1.15-2*\i-1*\j+1,-1.5-1*\jj+\eps);
}
}
\foreach \kk in {0,1}{
\def\eps{17*\kk+8.5};
\foreach \i in {0,...,4}
{
\draw[thick] (-.5-2*\i,-1+\eps) -- (0.5-2*\i,\eps);
\draw[thick] (-0.5-2*\i,\eps) -- (0.5-2*\i,-1+\eps);
\draw[thick, fill=myred, rounded corners=2pt] (-0.25-2*\i,-0.25+\eps) rectangle (.25-2*\i,-0.75+\eps);
\draw[thick] (-2*\i,-0.35+\eps) -- (.15-2*\i,-.35+\eps) -- (.15-2*\i,-0.5+\eps);
}
\foreach \i in {1,...,5}
{
\draw[thick] (.5-2*\i,-6+\eps) -- (1-2*\i,-5.5+\eps);
\draw[thick] (1.5-2*\i,-6+\eps) -- (1-2*\i,-5.5+\eps);
}
\foreach \jj[evaluate=\jj as \j using -2*(ceil(\jj/2)-\jj/2)] in {0,...,3}
\foreach \i in {1,...,5}
{
\draw[thick] (.5-2*\i-1*\j,-2-1*\jj+\eps) -- (1-2*\i-1*\j,-1.5-\jj+\eps);
\draw[thick] (1-2*\i-1*\j,-1.5-1*\jj+\eps) -- (1.5-2*\i-1*\j,-2-\jj+\eps);
}
\foreach \jj[evaluate=\jj as \j using -2*(ceil(\jj/2)-\jj/2)] in {0,...,4}
\foreach \i in {1,...,5}
{
\draw[thick] (.5-2*\i-1*\j,-1-1*\jj+\eps) -- (1-2*\i-1*\j,-1.5-\jj+\eps);
\draw[thick] (1-2*\i-1*\j,-1.5-1*\jj+\eps) -- (1.5-2*\i-1*\j,-1-\jj+\eps);
\draw[thick, fill=myred, rounded corners=2pt] (0.75-2*\i-1*\j,-1.75-\jj+\eps) rectangle (1.25-2*\i-1*\j,-1.25-\jj+\eps);
\draw[thick] (1-2*\i-1*\j,-1.35-1*\jj+\eps) -- (1.15-2*\i-1*\j,-1.35-1*\jj+\eps) -- (1.15-2*\i-1*\j,-1.5-1*\jj+\eps);
}
}
\end{tikzpicture}\,.
\caption{Pictorial representation of the purity  ${\rm tr}\left[\rho^{2}_{AD}\right]$, where $\rho_{AD}$ was defined in Eq.~\eqref{eq:rho_AD}.}
\label{fig:trrho2gen}
\end{figure}
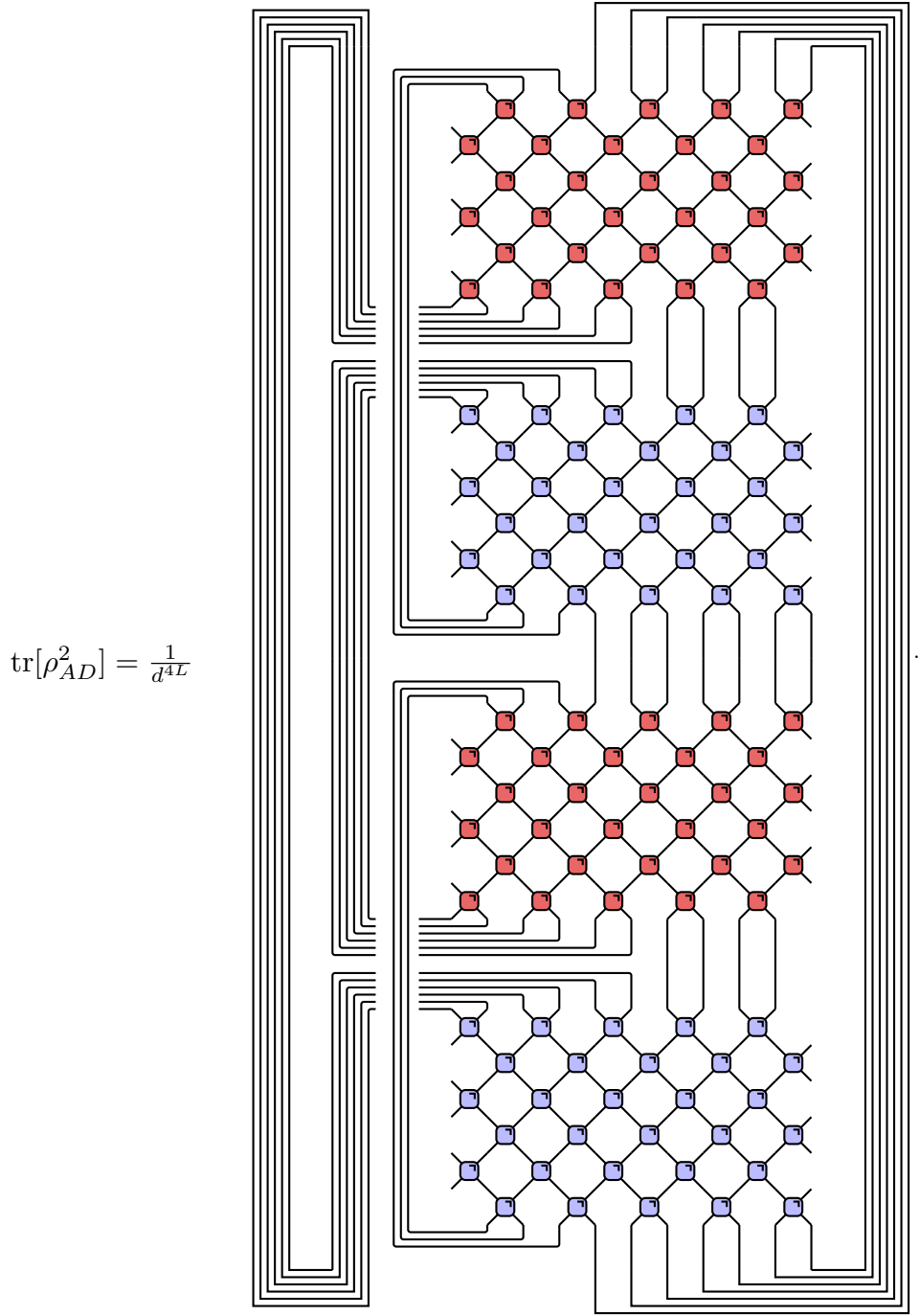
We can proceed by folding the rectangles of gates representing $\mathbb U^t$ (red gates) and $\mathbb U^{-t}$ (blue gates) according to the following procedure. First number them from 1 (bottom-most) to 4 (top-most). Then fold each rectangle underneath the previous one. The resulting figure features rectangles piled up in the order $1,2,3,4$. At this point, using the definitions~\eqref{eq:W}--\eqref{eq:circle} we find  
\be
{\rm tr}[\rho_{AD}^2] = 
\begin{tikzpicture}[baseline=(current  bounding  box.center), scale=0.5]
\def\eps{8.5};

\foreach \i in {0,...,4}
{
\draw[thick] (-.5-2*\i,-1+\eps) -- (0.5-2*\i,\eps);
\draw[thick] (-0.5-2*\i,\eps) -- (0.5-2*\i,-1+\eps);
\draw[thick, fill=mygreen, rounded corners=1pt] (-0.25-2*\i,-0.25+\eps) rectangle (.25-2*\i,-0.75+\eps);
\draw[thick] (-2*\i,-0.35+\eps) -- (.15-2*\i,-.35+\eps) -- (.15-2*\i,-0.5+\eps);
}
\foreach \i in {1,...,5}
{
\draw[thick] (.5-2*\i,-6+\eps) -- (1-2*\i,-5.5+\eps);
\draw[thick] (1.5-2*\i,-6+\eps) -- (1-2*\i,-5.5+\eps);
}
\foreach \jj[evaluate=\jj as \j using -2*(ceil(\jj/2)-\jj/2)] in {0,...,3}
\foreach \i in {1,...,5}
{
\draw[thick] (.5-2*\i-1*\j,-2-1*\jj+\eps) -- (1-2*\i-1*\j,-1.5-\jj+\eps);
\draw[thick] (1-2*\i-1*\j,-1.5-1*\jj+\eps) -- (1.5-2*\i-1*\j,-2-\jj+\eps);
}
\foreach \jj[evaluate=\jj as \j using -2*(ceil(\jj/2)-\jj/2)] in {0,...,4}
\foreach \i in {1,...,5}
{
\draw[thick] (.5-2*\i-1*\j,-1-1*\jj+\eps) -- (1-2*\i-1*\j,-1.5-\jj+\eps);
\draw[thick] (1-2*\i-1*\j,-1.5-1*\jj+\eps) -- (1.5-2*\i-1*\j,-1-\jj+\eps);
\draw[thick, fill=mygreen, rounded corners=1pt] (0.75-2*\i-1*\j,-1.75-\jj+\eps) rectangle (1.25-2*\i-1*\j,-1.25-\jj+\eps);
\draw[thick] (1-2*\i-1*\j,-1.35-1*\jj+\eps) -- (1.15-2*\i-1*\j,-1.35-1*\jj+\eps) -- (1.15-2*\i-1*\j,-1.5-1*\jj+\eps);
}
\foreach \i in {-4,...,2}{
\draw[thick, fill=white] (-3.85+\i+1.25,2.35) rectangle (-3.65+\i+1.25,2.55);
}
\foreach \i in {-7,...,-5}{
\draw[thick, fill=white] (-3.8+\i+1.3,2.45) circle (0.1cm);
}
\foreach \i in {0,...,3}{
\draw[thick, fill=white] (-3.75+\i+1.25,2.45+6.125) circle (0.1cm);
}
\foreach \i in {-6,...,-1}{
\draw[thick, fill=white] (-3.8+\i+1.3-0.1,2.45+6.125-0.1) rectangle (-3.8+\i+1.3+0.1,2.45+6.125+0.1);
}
\end{tikzpicture}\,.
\ee
Considering now the partition~\eqref{eq:relevantpartition1} with $L>|x|+t$ and using the ``unitarity rules" \eqref{eq:unitarity_pic}, we can simplify the above diagram, obtaining   
\be
{\rm tr}[\rho_{AD}^2]=  Z_2(t+ \lfloor x \rfloor,t- \lceil x \rceil) d^{2L-(x_++x_-)}\,,
\label{eq:rhoZ2}
\ee
where $Z_2(x,y)$ is defined in Eq.~\eqref{eq:SAD'}.

\section{Proof of Property~\ref{prop:P2}}
\label{app:proofP2}

The identity \eqref{eq:crucialidentity} can be established by decomposing ${\bar T}_x^{\msquare \msquare}$ as follows 
\be
{\bar T}_x^{\msquare \msquare}= \frac{1}{4}{\bar T}_x^{\mcirc \mcirc}+\frac{\sqrt 3}{4} {\bar T}_x^{\mcirc \mcircf}+\frac{\sqrt 3}{4} {\bar T}_x^{\mcircf \mcirc}+ \frac{3}{4} {\bar T}_x^{\mcircf \mcircf}\,,
\ee
where we defined 
\begin{align}
&\bar T^{\mcirc \mcircf}_x =
\begin{tikzpicture}[baseline=(current  bounding  box.center), scale=0.55]
\def\eps{-0.5};
\foreach \i in {3}{
\foreach \j in {-3,...,2}{
\draw[very thick] (-4.5+1.5*\j,1.5*\i) -- (-3+1.5*\j,1.5*\i);
\draw[very thick] (-3.75+1.5*\j,1.5*\i-.75) -- (-3.75+1.5*\j,1.5*\i+.75);
\draw[ thick, fill=myorange, rounded corners=2pt, rotate around ={45: (-3.75+1.5*\j,1.5*\i)}] (-4+1.5*\j,0.25+1.5*\i) rectangle (-3.5+1.5*\j,-0.25+1.5*\i);
\draw[thick, rotate around ={45: (-3.75+1.5*\j,1.5*\i)}] (-3.75+1.5*\j,.15+1.5*\i) -- (-3.6+1.5*\j,.15+1.5*\i) -- (-3.6+1.5*\j,1.5*\i);}
\draw[thick, fill=black] (0.1,1.5*\i) circle (0.1cm);
\draw[thick, fill=white] (-8.9,1.5*\i) circle (0.1cm);
}
\draw [thick, decorate, decoration={brace,amplitude=4pt, raise=4pt},yshift=0pt]
(-8.25,5.3) -- (-.75,5.3) node [black,midway,yshift=0.65cm] {$x$};
\Text[x=-8.25,y=2]{}
\end{tikzpicture}\,,
\qquad
\bar T^{\mcircf \mcirc}_x =
\begin{tikzpicture}[baseline=(current  bounding  box.center), scale=0.55]
\def\eps{-0.5};
\foreach \i in {3}{
\foreach \j in {-3,...,2}{
\draw[very thick] (-4.5+1.5*\j,1.5*\i) -- (-3+1.5*\j,1.5*\i);
\draw[very thick] (-3.75+1.5*\j,1.5*\i-.75) -- (-3.75+1.5*\j,1.5*\i+.75);
\draw[ thick, fill=myorange, rounded corners=2pt, rotate around ={45: (-3.75+1.5*\j,1.5*\i)}] (-4+1.5*\j,0.25+1.5*\i) rectangle (-3.5+1.5*\j,-0.25+1.5*\i);
\draw[thick, rotate around ={45: (-3.75+1.5*\j,1.5*\i)}] (-3.75+1.5*\j,.15+1.5*\i) -- (-3.6+1.5*\j,.15+1.5*\i) -- (-3.6+1.5*\j,1.5*\i);}
\draw[thick, fill=white] (0.1,1.5*\i) circle (0.1cm);
\draw[thick, fill=black] (-8.9,1.5*\i) circle (0.1cm);
}
\draw [thick, decorate, decoration={brace,amplitude=4pt, raise=4pt},yshift=0pt]
(-8.25,5.3) -- (-.75,5.3) node [black,midway,yshift=0.65cm] {$x$};
\Text[x=-8.25,y=2]{}
\end{tikzpicture}\,,
\label{eq:Tcirclecircleffave}
\end{align}
and analogously
\be
\bar T^{\mcircf \mcircf}_x =
\begin{tikzpicture}[baseline=(current  bounding  box.center), scale=0.55]
\def\eps{-0.5};
\foreach \i in {3}{
\foreach \j in {-3,...,2}{
\draw[very thick] (-4.5+1.5*\j,1.5*\i) -- (-3+1.5*\j,1.5*\i);
\draw[very thick] (-3.75+1.5*\j,1.5*\i-.75) -- (-3.75+1.5*\j,1.5*\i+.75);
\draw[ thick, fill=myorange, rounded corners=2pt, rotate around ={45: (-3.75+1.5*\j,1.5*\i)}] (-4+1.5*\j,0.25+1.5*\i) rectangle (-3.5+1.5*\j,-0.25+1.5*\i);
\draw[thick, rotate around ={45: (-3.75+1.5*\j,1.5*\i)}] (-3.75+1.5*\j,.15+1.5*\i) -- (-3.6+1.5*\j,.15+1.5*\i) -- (-3.6+1.5*\j,1.5*\i);}
\draw[thick, fill=black] (0.1,1.5*\i) circle (0.1cm);
\draw[thick, fill=black] (-8.9,1.5*\i) circle (0.1cm);
}
\draw [thick, decorate, decoration={brace,amplitude=4pt, raise=4pt},yshift=0pt]
(-8.25,5.3) -- (-.75,5.3) node [black,midway,yshift=0.65cm] {$x$};
\Text[x=-8.25,y=2]{}
\end{tikzpicture}\,.
\label{eq:Tcirclefcircleffave}
\ee
Using the graphical rules \eqref{eq:aveunitary1}--\eqref{eq:avedualunitary2} it is then straightforward to verify that 
\be
{\bar T}_x^{\mcirc \mcirc} \ket{\mcirc \ldots \mcirc}= \ket{\mcirc \ldots \mcirc},\qquad {\bar T}_x^{\mcirc \mcircf} \ket{\mcirc \ldots \mcirc}={\bar T}_x^{\mcircf \mcirc} \ket{\mcirc \ldots \mcirc}=0\,. 
\ee
Finally to evaluate ${\bar T}_x^{\mcircf \mcircf} \ket{\mcirc \ldots \mcirc}$ we proceed iteratively. Inserting the identity in the form $\1=\ket{\mcirc}\!\!\bra{\mcirc} + \ket{\mcircf}\!\!\bra{\mcircf}$ in the first ``auxiliary'' leg we have 
\be
{\bar T}_x^{\mcircf \mcircf} \ket{\mcirc \ldots \mcirc} =
 \begin{tikzpicture}[baseline=(current  bounding  box.center), scale=0.55]
\def\eps{-0.5};
\foreach \i in {3}{
\foreach \j in {-2,...,4}{
\draw[very thick] (-4.5+1.5*\j,1.5*\i) -- (-3+1.5*\j,1.5*\i);
\draw[very thick] (-3.75+1.5*\j,1.5*\i-.75) -- (-3.75+1.5*\j,1.5*\i+.75);
\draw[ thick, fill=myorange, rounded corners=2pt, rotate around ={45: (-3.75+1.5*\j,1.5*\i)}] (-4+1.5*\j,0.25+1.5*\i) rectangle (-3.5+1.5*\j,-0.25+1.5*\i);
\draw[thick, rotate around ={45: (-3.75+1.5*\j,1.5*\i)}] (-3.75+1.5*\j,.15+1.5*\i) -- (-3.6+1.5*\j,.15+1.5*\i) -- (-3.6+1.5*\j,1.5*\i);
\draw[thick, fill=white] (-3.75+1.5*\j,1.5*\i-.7) circle (.1cm);
}
\draw[thick, fill=black] (2.9,1.5*\i) circle (.1cm);
}

\foreach \i in {3}{
\foreach \j in {-3.25}{
\draw[very thick] (-4.5+1.5*\j,1.5*\i) -- (-3+1.5*\j,1.5*\i);
\draw[very thick] (-3.75+1.5*\j,1.5*\i-.75) -- (-3.75+1.5*\j,1.5*\i+.75);
\draw[ thick, fill=myorange, rounded corners=2pt, rotate around ={45: (-3.75+1.5*\j,1.5*\i)}] (-4+1.5*\j,0.25+1.5*\i) rectangle (-3.5+1.5*\j,-0.25+1.5*\i);
\draw[thick, rotate around ={45: (-3.75+1.5*\j,1.5*\i)}] (-3.75+1.5*\j,.15+1.5*\i) -- (-3.6+1.5*\j,.15+1.5*\i) -- (-3.6+1.5*\j,1.5*\i);
\draw[thick, fill=black] (-4.5+1.5*\j,1.5*\i) circle (.1cm);
\draw[thick, fill=black] (-3+1.5*\j,1.5*\i) circle (.1cm);
\draw[thick, fill=black] (-3+1.5*\j+0.4,1.5*\i) circle (.1cm);
\draw[thick, fill=white] (-3.75+1.5*\j,1.5*\i-.7) circle (.1cm);
}
}
\Text[x=0,y=3.5]{}
\end{tikzpicture}\,,
\ee
where we used that the contribution of $\ket{\mcirc}\!\!\bra{\mcirc}$ is zero by \eqref{eq:aveunitary2}. Telescoping we find 
\be
{\bar T}_x^{\mcircf \mcircf} \ket{\mcirc \ldots \mcirc} =
 \begin{tikzpicture}[baseline=(current  bounding  box.center), scale=0.55]
\def\eps{-0.5};
\foreach \i in {3}{
\foreach \j in {-3.25, -2, -.75, .5, 1.75, 3, 4.25, 5.5}{
\draw[very thick] (-4.5+1.5*\j,1.5*\i) -- (-3+1.5*\j,1.5*\i);
\draw[very thick] (-3.75+1.5*\j,1.5*\i-.75) -- (-3.75+1.5*\j,1.5*\i+.75);
\draw[ thick, fill=myorange, rounded corners=2pt, rotate around ={45: (-3.75+1.5*\j,1.5*\i)}] (-4+1.5*\j,0.25+1.5*\i) rectangle (-3.5+1.5*\j,-0.25+1.5*\i);
\draw[thick, rotate around ={45: (-3.75+1.5*\j,1.5*\i)}] (-3.75+1.5*\j,.15+1.5*\i) -- (-3.6+1.5*\j,.15+1.5*\i) -- (-3.6+1.5*\j,1.5*\i);
\draw[thick, fill=black] (-4.5+1.5*\j,1.5*\i) circle (0.1cm);
\draw[thick, fill=black] (-3+1.5*\j,1.5*\i) circle (0.1cm);
\draw[thick, fill=white] (-3.75+1.5*\j,1.5*\i-.7) circle (0.1cm);
}
}
\Text[x=0,y=3.5]{}
\end{tikzpicture}\,,
\ee
The identity \eqref{eq:crucialidentity} follows by evaluating explicitly 
\be
\begin{tikzpicture}[baseline=(current  bounding  box.center), scale=0.55]
\def\eps{-0.5};
\foreach \i in {3}{
	\foreach \j in {-3.25}{
		\draw[very thick] (-4.5+1.5*\j,1.5*\i) -- (-3+1.5*\j,1.5*\i);
		\draw[very thick] (-3.75+1.5*\j,1.5*\i-.75) -- (-3.75+1.5*\j,1.5*\i+.75);
		\draw[ thick, fill=myorange, rounded corners=2pt, rotate around ={45: (-3.75+1.5*\j,1.5*\i)}] (-4+1.5*\j,0.25+1.5*\i) rectangle (-3.5+1.5*\j,-0.25+1.5*\i);
		\draw[thick, rotate around ={45: (-3.75+1.5*\j,1.5*\i)}] (-3.75+1.5*\j,.15+1.5*\i) -- (-3.6+1.5*\j,.15+1.5*\i) -- (-3.6+1.5*\j,1.5*\i);
		\draw[thick, fill=black] (-4.5+1.5*\j,1.5*\i) circle (0.1cm);
		\draw[thick, fill=black] (-3+1.5*\j,1.5*\i) circle (0.1cm);
		\draw[thick, fill=white] (-3.75+1.5*\j,1.5*\i-.7) circle (0.1cm);
	}
}
\Text[x=-8,y=+3.6]{}
\end{tikzpicture}
= a \ket{\mcirc} + \frac{1-a}{\sqrt 3} \ket{\mcircf}\,.
\ee

\section{Proof of Property~\ref{prop:P1}}
\label{app:proofP1}

To prove Property~\ref{prop:P1} it is useful to introduce the transfer matrix
\begin{align}
&\bar T^{\mcirc \msquare}_x =
\begin{tikzpicture}[baseline=(current  bounding  box.center), scale=0.55]
\def\eps{-0.5};
\foreach \i in {3}{
\foreach \j in {-3,...,2}{
\draw[very thick] (-4.5+1.5*\j,1.5*\i) -- (-3+1.5*\j,1.5*\i);
\draw[very thick] (-3.75+1.5*\j,1.5*\i-.75) -- (-3.75+1.5*\j,1.5*\i+.75);
\draw[ thick, fill=myorange, rounded corners=2pt, rotate around ={45: (-3.75+1.5*\j,1.5*\i)}] (-4+1.5*\j,0.25+1.5*\i) rectangle (-3.5+1.5*\j,-0.25+1.5*\i);
\draw[thick, rotate around ={45: (-3.75+1.5*\j,1.5*\i)}] (-3.75+1.5*\j,.15+1.5*\i) -- (-3.6+1.5*\j,.15+1.5*\i) -- (-3.6+1.5*\j,1.5*\i);}
\draw[thick, fill=white] (0,1.5*\i-0.1) rectangle (-.2,1.5*\i+0.1);
\draw[thick, fill=white] (-8.9,1.5*\i) circle (0.1cm);
}
\draw [thick, decorate, decoration={brace,amplitude=4pt, raise=4pt},yshift=0pt]
(-8.25,5.3) -- (-.75,5.3) node [black,midway,yshift=0.65cm] {$x$};
\Text[x=-8.25,y=2]{}
\end{tikzpicture}\,,
\label{eq:Tcirclesquareave}
\end{align}
and make use of the following inductive relations for $\bar{T}^{\msquare \msquare}_x$ and $\bar{T}^{\mcirc \msquare}_x$ 
\begin{align}
&\bar{T}^{\msquare \msquare}_x  =   \ket{\msquare}\!\!\bra{\msquare} \otimes \bar{T}^{\msquare \msquare}_{x-1} + \ket{\msquaref}\!\!\bra{\msquaref} \otimes \left(\frac{4a-1}{3}{\bar T}^{\msquare \msquare}_{x-1}+\frac{2(1-a)}{3}{\bar T}^{\mcirc \msquare}_{x-1}\right)\,,\label{eq:induction1}\\
&\bar{T}^{\mcirc \msquare}_x  =  \ket{\mcirc}\!\!\bra{\mcirc} \otimes \bar{T}^{\mcirc \msquare}_{x-1}+  \ket{\mcircf}\!\!\bra{\mcircf} \otimes \left(\frac{2(1-a)}{3}{\bar T}^{\msquare \msquare}_{x-1}+\frac{4a-1}{3}{\bar T}^{\mcirc \msquare}_{x-1}\right)\,,\label{eq:induction2}
\end{align}
which are obtained by writing down explicitly the sum over the basis vectors corresponding to the first ``auxiliary'' leg from the right. In particular, we chose the basis \eqref{eq:B1basis} for \eqref{eq:induction1} and the basis \eqref{eq:B2basis} for \eqref{eq:induction2}. 

Equipped with  \eqref{eq:induction1} and  \eqref{eq:induction2} we now proceed to prove the statement of Property~\ref{prop:P1} by induction in $x$. Computing the matrices for $x=1$ and $x=2$ we have 
\begin{align}
\bar{T}^{\msquare \msquare}_1 &=  \ket{\msquare}\!\!\bra{\msquare} +  a \ket{\msquaref}\!\!\bra{\msquaref}\,,\\
\bar{T}^{\mcirc \msquare}_1 &= \frac{1}{2} \ket{\msquare}\!\!\bra{\msquare} + \frac{1}{2} \ket{\msquaref}\!\!\bra{\msquaref}\,,\\
\bar{T}^{\msquare \msquare}_2  &=  \ket{\msquare\msquare}\!\!\bra{\msquare\msquare} +  a \ket{\msquare\msquaref}\!\!\bra{\msquare\msquaref}+a \ket{\msquaref\msquare}\!\!\bra{\msquaref\msquare}+\left(a^2+\frac{(a-1)^2}{3}\right) \ket{\msquaref\msquaref}\!\!\bra{\msquaref\msquaref} \,, \\ 
\bar{T}^{\mcirc \msquare}_2  &= \frac{1}{2} \ket{\msquare\msquare}\!\!\bra{\msquare\msquare} + \frac{1}{2} \ket{\mcircf\msquare}\!\!\bra{\mcircf\msquare}+\frac{1}{2} \ket{\mcirc\mcircf}\!\!\bra{\mcirc\mcircf} + \left(\frac{1}{2}-\frac{2}{3}(a-1)^2\right) \ket{\mcircf\msquaref}\!\!\bra{\mcircf\msquaref}\,.
\end{align}
We see that the property holds for $x=2$. To conclude we show that if it holds for $x-1$ it holds for $x$. 

First we note that, due to the block structure of \eqref{eq:induction1} and \eqref{eq:induction2}, to prove that $\bar{T}^{\msquare \msquare}_x$  and $\bar{T}^{\mcirc \msquare}_x$  are positive definite it is sufficient to prove that the property holds for each separate block. This follows directly from the inductive hypothesis (${\bar T}^{\msquare \msquare}_{x-1}$ and ${\bar T}^{\mcirc \msquare}_{x-1}$ positive definite) and $(4a-1)/3,2(1-a)/3>0$ (the same holds also if one of the two coefficients vanishes as long as the other is positive). 

Let us now find the $2x$ largest eigenvalues of $\bar{T}^{\msquare \msquare}_x $ (the case of $\bar{T}^{\mcirc \msquare}_x $ is totally analogous). Since $\ket{\msquare\ldots\msquare}_{x-1}$ (we added the subscript to stress that it is defined on $x-1$ sites) is a common eigenvector of $\bar{T}^{\msquare \msquare}_{x-1}$ and ${\bar T}^{\mcirc \msquare}_{x-1}$ we can write $\bar{T}^{\msquare \msquare}_x $ as the direct sum of three blocks 
\begin{align}
\bar{T}^{\msquare \msquare}_x  =&\ket{\msquare}\!\!\bra{\msquare} \otimes {T^{\msquare \msquare}_{x-1} }  + a \ket{\msquaref \msquare\ldots\msquare}\bra{\msquaref \msquare\ldots\msquare}+ \ket{\msquaref}\!\!\bra{\msquaref} \otimes \left[\frac{4a-1}{3} {\bar T}^{\msquare \msquare\,\prime}_{x-1}+\frac{2(1-a)}{3}{\bar T}^{\mcirc \msquare\,\prime}_{x-1}\right]
\end{align}
where defined 
\begin{align}
{\bar T}^{\msquare \msquare\,\prime}_{x} :={\bar T}^{\msquare \msquare}_{x}- \ket{\msquare\ldots\msquare}_{x}\bra{\msquare\ldots\msquare}\,,
\qquad
{\bar T}^{\mcirc \msquare\,\prime}_{x} :={\bar T}^{\mcirc \msquare}_{x}- \ket{\msquare\ldots\msquare}_{x}\bra{\msquare\ldots\msquare}\,,
\label{eq:Tcircsquareprime}
\end{align}
and used
\be
\left[\frac{4a-1}{3}{\bar T}^{\msquare \msquare}_{x-1}+\frac{2(1-a)}{3}{\bar T}^{\mcirc \msquare}_{x-1}\right]\ket{\msquare\ldots\msquare}_{x-1}=a \ket{\msquare\ldots\msquare}_{x-1}\,.
\ee 
By inductive hypothesis we have that the $2x-1$ largest eigenvalues of the first two blocks are 
\be
\bigl\{1,\underbrace{a,\ldots,a}_x,\underbrace{a^2+\frac{(a-1)^2}{3},\ldots,a^2+\frac{(a-1)^2}{3}}_{x-2}\bigr\}.
\ee
To conclude we now bound the largest eigenvalue of the third block. Since the matrices are positive definite we have 
\be
\braket{\psi|\frac{4a-1}{3}{\bar T}^{\msquare \msquare\,\prime}_{x-1}+\frac{2(1-a)}{3}{\bar T}^{\mcirc \msquare\,\prime}_{x-1} |\psi}  \leq \frac{(4a-1)a}{3} + \frac{(1-a)}{3} = a^2 + \frac{(1-a)^2}{3} 
\ee 
where we used that by inductive hypothesis the maximal eigenvalues of ${\bar T}^{\msquare \msquare\,\prime}_{x-1}$ and ${\bar T}^{\mcirc \msquare\,\prime}_{x-1}$ are respectively $a$ and $1/2$. Note that for 
\be
\ket{\psi} = \ket{\msquaref \msquare\ldots\msquare}_{x-1}\,,
\ee
the bound is saturated, so there is at least an eigenvalue $a^2 + {(1-a)^2}/{3}$. We then conclude that the $2x$ largest eigenvalues of ${\bar T}^{\msquare \msquare}_{x}$ are
\be
\bigl\{1,\underbrace{a,\ldots,a}_x,\underbrace{a^2+\frac{(a-1)^2}{3},\ldots,a^2+\frac{(a-1)^2}{3}}_{x-1}\bigr\}.
\ee
These eigenvalues correspond to the eigenvectors
\begin{align}
&\{\ket{{\msquare\ldots\msquare}},\overbrace{\ket{\underbrace{\msquare\ldots\msquare}_{x-1}\msquaref},\ket{\underbrace{\msquare\ldots\msquare}_{x-2}\msquaref\msquare},\ldots,\ket{\msquaref\underbrace{\msquare\ldots\msquare}_{x-1}}}^x,\overbrace{\ket{\underbrace{\msquare\ldots\msquare}_{x-2}\msquaref\msquaref},\ket{\underbrace{\msquare\ldots\msquare}_{x-3}\msquaref\msquaref \msquare},\ldots,\ket{\msquaref\msquaref\underbrace{\msquare\ldots\msquare}_{x-2}}}^{x-1}\}\,,
\end{align}
as can be seen by direct application of the graphical rules \eqref{eq:aveunitary1}--\eqref{eq:avedualunitary2}. This concludes the proof.

\section{Solving Recurrence Relations}
\label{app:solvingrecurrence}

Let us consider $Z(x,y)$ fulfilling a generic inhomogeneous linear recurrence relation of the form 
\be
Z(x,y)=a Z(x-1,y)+ b Z(x,y-1)+k(x,y)\,,
\label{eq:generalequation}
\ee
with boundary conditions $Z(x,0)=f(x)$ and $Z(0,y)=g(y)$. Defining the rescaled function
\be
\tilde Z(x,y) := b^{-y} Z(x,y)- b^{-y} g(y)\,,
\ee
we see that it fulfils 
\be
\tilde Z(x,y)- \tilde Z(x,y-1)= a \tilde Z(x-1,y)+k(x,y) b^{-y} +(a-1) b^{-y} g(y)+b^{-y+1} g(y-1)\,,
\ee 
with boundary conditions $\tilde Z(x,0)=f(x)-g(0)$, $\tilde Z(0,y)=0$. Summing over $y$ we find 
\be
\tilde Z(x,y)= a \sum_{k=1}^y \tilde Z(x-1,k)+ G(x,y)\,,
\label{eq:Zsummed}
\ee
where we defined 
\be
G(x,y) := \sum_{k=1}^y k(x,k) b^{-k} + (a-1) \sum_{k=1}^y b^{-k} g(k)+ \sum_{k=1}^y b^{-k+1} g(k-1) + f(x)-g(0)\,.
\ee
The relation \eqref{eq:Zsummed} is solved by 
\be
\tilde Z(x,y)= \sum_{j=0}^{x-1} a^j G^{(j)}(x-j,y)
\ee
where we defined 
\be
G^{(j)}(x,y)= \overbrace{\sum_{k_1=1}^{y}\sum_{k_2=1}^{k_1}\ldots\sum_{k_j=1}^{k_{j-1}}}^j G(x,k_j)\,.
\ee
Putting all together we find 
\be
Z(x,y)= g(y) + b^y \sum_{j=0}^{x-1} a^j G^{(j)}(x-j,y)\,.
\label{eq:generalsolution}
\ee
Substituting the specific functions $f(x)$, $g(y)$, and  $k(x,y)$, this relation provides a solution to all recurrence equations appearing in the main text. This is  reported explicitly in the following subsections with the exception of Eq.~\eqref{eq:OTOCequations}. Indeed, even if one could apply the general solution \eqref{eq:generalsolution} also to Eq.~\eqref{eq:OTOCequations} this would not directly lead to a closed-form solution. This is because for any given integer $r$, the driving $k(x,y)$ depends on all $W_{r^\prime}(x,y)$, with $r^\prime <r$. In fact, in this case, the solution is more easily achieved by means of the more sophisticated kernel method described in Appendix~\ref{sec:solution_recurrence}.

\subsection{Solution of Equation~\eqref{eq:RUrecurrence1}}

To find the solution to \eqref{eq:RUrecurrence1} we make the following replacement  
\be
a,b\mapsto \frac{d}{d^2+1},\qquad f(x)\mapsto d^{-x},\qquad g(x)\mapsto d^{-x},\qquad k(x,y)\mapsto 0.
\ee
This gives 
\be
G(x,y)= (d-1) \left(\frac{d^2+1}{d^2}\right)^y + d^{-x}-d,
\ee
from which we find 
\be
G^{(j)}(x,y)= (d-1)\left(\frac{d^2+1}{d^2}\right)^y (d^2+1)^j + (d^{-x}-d) \binom{y+j-1}{j} - (d-1)\sum_{k=0}^{j-1} (d^2+1)^{j-k} \binom{y+k-1}{k} \,.
\ee
This finally yields  
\begin{align}
 Z(x,y) &= d^{-y} + \left(\frac{d}{d^2+1}\right)^y \sum_{j=0}^{x-1}  \left(\frac{d}{d^2+1}\right)^j G^{(j)}(x-j,y)= \frac{1}{d^{x-y}}- d^{x+y} f(x,y) 
\end{align}
where $f(x,y)$ is defined in \eqref{eq:f_g_function}. The last step is achieved by using standard identities among binomial coefficients. 

\subsection{Solution of Equation~\eqref{eq:RUrecurrence2}}

With the replacements 
\be
a,b\mapsto \frac{1}{d^2+1},\qquad f(x)\mapsto 1,\qquad g(x)\mapsto 1,\qquad k(x,y)\mapsto 0,
\ee
\eqref{eq:generalsolution} gives the solution to Eq.~\eqref{eq:RUrecurrence2}. In particular, this case we find 
\be
G(x,y)= \left(1-\frac{1}{d^2}\right) \left(1-(d^2+1)^y\right)\,,
\ee
and 
\begin{align}
G^{(j)}(x,y)&= \left(1-\frac{1}{d^2}\right) \binom{y+j-1}{j} - \left(1-\frac{1}{d^2}\right) \left({d^2+1}\right)^y \left(\frac{d^2+1}{d^{2}}\right)^j \notag\\
&+  \left(1-\frac{1}{d^2}\right) \sum_{k=0}^{j-1} \left(\frac{d^2+1}{d^{2}}\right)^{j-k}  \binom{y+k-1}{k} \,.
\end{align}
Substituting back in \eqref{eq:generalsolution} we finally arrive at $Z(x,y) =d^{-2y}+ f(x,y)$, where $f(x,y)$ is defined in \eqref{eq:f_g_function}.
 
\subsection{Solution of Equations~\eqref{eq:Z2Z3} and  \eqref{eq:Z3Z4}} 
\label{sec:formalsol1}
 
The general equation \eqref{eq:generalequation} specialises to \eqref{eq:Z2Z3} and \eqref{eq:Z3Z4} with the replacements 
\be
a\mapsto 0,\qquad  b\mapsto \frac{1}{4},\qquad f(x)\mapsto 1,\qquad g(x)\mapsto 1,\qquad k(x,y)\mapsto \frac{3}{4} \bar Z_3(x,y-1),
\ee
and 
\be
a\mapsto 0,\qquad  b\mapsto \frac{1}{4},\qquad f(x)\mapsto 1,\qquad g(x)\mapsto a^x,\qquad k(x,y)\mapsto \frac{3}{4} \bar Z_4(x,y-1),
\ee
respectively. Since $a=0$ the value of $g(x)$ is not needed, indeed in this case  the general solution \eqref{eq:generalsolution} reads as
\be
Z(x,y)=b^y f(x) + b^y  \sum_{k=1}^y k(x,k) b^{-k}\,.
\ee
Plugging the above values we find \eqref{eq:formalsolZ2} and \eqref{eq:formalsolZ3}.

\subsection{Solution of Equations~\eqref{eq:OO2} and  \eqref{eq:O2O3}} 
\label{sec:formalsol2} 
 
The general equation \eqref{eq:generalequation} specialises to \eqref{eq:OO2} and \eqref{eq:O2O3} with the replacements 
\be
a\mapsto 0,\qquad b\mapsto a, \qquad f(x)\mapsto 1-\frac{4}{3}\delta_{x,1},\qquad k(x,y)\mapsto {(1-a)} \mathbb{E}\left[ O_3(x,y-1)\right],
\ee
and 
\be
a\mapsto 0,\qquad b\mapsto a,\qquad f(x)\mapsto \frac{4a-1}{3},\qquad k(x,y)\mapsto {(1-a)}  \mathbb{E}\left[ O_3(x,y-1)\right],
\ee
respectively. As explained in App.~\ref{sec:formalsol1}, since $a=0$ we do not need to specify $g(x)$. Plugging into the general solution \eqref{eq:generalsolution} we find 
\begin{align}
&\mathbb{E}\left[ O(x,y)  \right]=\left (1-\frac{4}{3}\delta_{x,1}\right)a^y + a^y \left(\frac{1-a}{a}\right) \sum_{k=0}^{y-1} a^{-k}\mathbb{E}\left[ O_2(x,k)  \right], \label{eq:formalsolO}\\ 
&\mathbb{E}\left[ O_2(x,y)  \right] =\frac{4a-1}{3} a^y +  a^y \left(\frac{1-a}{a}\right) \sum_{k=0}^{y-1} a^{-k} \mathbb{E}\left[ O_3(x,k)\right]. \label{eq:formalsolO2}
\end{align}
Combining them we obtain \eqref{eq:formalOO3}. 

\section{Solution to the recurrence equation~\eqref{eq:OTOCequations}}
\label{sec:solution_recurrence}

In this appendix we detail a strategy to find a closed form solution to the recurrence relation~\eqref{eq:OTOCequations}. This strategy is based on the application of the so-called kernel method, which is a very powerful and elegant approach to solve certain discrete recurrence relations. We refer the reader to Refs.~\cite{Prodinger2004,Bous05,Bostan2010,PemantleBoook2013} for further details and applications of the method. 

We start by considering Eq.~\eqref{eq:OTOCequations} for $k-1$ (and $k>1)$. Multiplying the latter by $d/(d^2+1)$, and subtracting it from Eq.~\eqref{eq:OTOCequations} for $k$, we obtain
\be
W_k(m,n)=\frac{1}{d^2+1}W_{k}(m-1,n)+\frac{1}{d} W_{k}(m,n-1)\left[\frac{d}{d+1}\right]^2+\frac{d}{d^2+1}W_{k-1}(m,n)-\frac{d}{(d^2+1)^2}W_{k-1}(m-1,n)\,.
\label{eq:partial}
\ee
This equation is defined for $m,n\geq 1$ and $2\leq k \leq m-1$. On the other hand, setting $k=1$ in ~\eqref{eq:OTOCequations} we obtain
\be
W_1(m,n)=\frac{1}{d^2+1} W_{1}(m-1,n)+ \frac{1}{d}  W_{1}(m,n-1) \left[\frac{d}{d^2+1}\right]^{2}+W_{0}(m,n-1) \left[\frac{d}{d^2+1}\right]^{2}\,.
\label{eq:k_1}
\ee
Now, since $W_0(m,n)$ satisfies $W_{0}(m,n)=\frac{1}{(d^2+1)}W_{0}(m-1,n)+W_{0}(m,n-1) \left[\frac{d}{d^2+1}\right]$, we see from Eq.~\eqref{eq:k_1} that~\eqref{eq:partial} actually  holds also for $k=1$.  Next, define
\be
Z_k(m,n)=\frac{(d^{2}+1)^{2k+m+2n}}{d^{k+n}}W_{k}(m+k,n)\,.
\label{eq:z_function}
\ee
This function satisfies the simple recurrence relation
\begin{align}
Z_k(m,n)=&Z_{k}(m-1,n)+ Z_{k}(m,n-1)+Z_{k-1}(m+1,n)-Z_{k-1}(m,n)\,,
\label{eq:simple_recurrence}
\end{align}
for $m,n\geq 1$, and $1\leq k \leq m-1$, with boundary conditions
\begin{align}
Z_k(m,0)&=\frac{(d^2+1)^{2k+m}}{d^{k+|k-1|}}\,,\qquad Z_{k}(0,n)=\frac{(d^2+1)^{2k+2n}}{d^{k+n+|k+n-1|}}\,.\label{eq:initial_cond}
\end{align}
Note that we have also $Z_0(m,n)=Z_0(m-1,n)+(d^2+1)Z_0(m,n-1)$. 

Finally, define the generating function
\begin{align}
\mathcal{F}(x,y,z)=&\sum_{m,n,k=0}^{\infty}Z_{k}(m,n)x^{m}y^{n}z^{k}\,.
\end{align}
Our strategy is to write down a functional equation for $\mathcal{F}(x,y,z)$. To this end, wee first evaluate the following power series, whose computation is straightforward based on the knowledge of the initial conditions~\eqref{eq:initial_cond}
\begin{align}
\alpha(x)=&\sum_{m=0}^{\infty}Z_{0}(m,0)x^{m}=\frac{1}{d}\frac{1}{1-ax}\,,\qquad \beta(y)=\sum_{n=0}^{\infty}Z_{0}(0,n)y^{n}=\frac{a^2 \left(d^2-1\right) y+d^2}{d^3-a^2 d y}\,,\\
\gamma(z)=&\sum_{k=0}^{\infty}Z_{k}(0,0)z^{k}=\frac{a^2 d z}{d^2-a^2 z}+\frac{1}{d}\,,
\qquad \mathcal{A}(x,z)=\sum_{m,k=0}^{\infty}Z_{k}(m,0)x^{m}z^{k}=\frac{a^2 z-d^2 \left(a^2 z+1\right)}{d (a x-1) \left(d^2-a^2 z\right)}\,,\\
\mathcal{B}(y,z)=&\sum_{m,n=0}^{\infty}Z_{k}(0,n)y^{n}z^{k}=\frac{d^5}{\left(d^2-a^2 y\right) \left(d^2-a^2 z\right)}-d+\frac{1}{d}\,.
\end{align}
where we introduced $a=d^2+1$. In  the following, we will also need to compute
\be
\mathcal{C}(x,y)=\sum_{m,n=0}^{\infty}Z_{0}(m,n)x^{m}y^{n}\,.
\ee
This can be obtained by proving the simple identity
\be
(1-x-ay)\mathcal{C}(x,y)=(1-ay)\beta(y)+(1-x)\alpha(x)-Z_{0}(0,0)\,,
\label{eq:simple_id_1}
\ee
leading to
\be
\mathcal{C}(x,y)=\frac{(1-ay)\beta(y)+(1-x)\alpha(x)-1/d}{1-x-ay}\,.
\ee
Eq.~\eqref{eq:simple_id_1} is obtained by writing the r.h.s. as a sum of power series, and regrouping the terms that are multiplied by the same monomials.

In the same way, one can straightforwardly verify the identities
\begin{align}
\mathcal{F}(x,y,z)=& \sum_{m,n,k=1}^{\infty} Z_{k}(m,n)x^{m}y^{n}z^k+Z_{0}(0,0)
-\alpha(x)-\beta(y)-\gamma(z)+\mathcal{A}(x,z)+\mathcal{B}(y,z)+\mathcal{C}(x,y)\,, \label{eq:id_1}\\
x\mathcal{F}(x,y,z)=& \sum_{m,n,k=1}^{\infty} Z_{k}(m-1,n)x^{m}y^{n}z^k-x\alpha(x)+x\left[\mathcal{A}(x,z)+\mathcal{C}(x,y)\right]\label{eq:id_2}\\
y\mathcal{F}(x,y,z)=& \sum_{m,n,k=1}^{\infty} Z_{k}(m,n-1)x^{m}y^{n}z^k-y\beta(y)
+y\left[\mathcal{B}(y,z)+\mathcal{C}(x,y)\right]\label{eq:id_3}\\
z\mathcal{F}(x,y,z)=& \sum_{m,n,k=1}^{\infty} Z_{k-1}(m,n)x^{m}y^{n}z^k-z\gamma(z)
+z\left[\mathcal{A}(x,z)+\mathcal{B}(y,z)\right]\label{eq:id_4}\\
\frac{z}{x}\mathcal{F}(x,y,z)=& \sum_{m,n,k=1}^{\infty} Z_{k-1}(m+1,n)x^{m}y^{n}z^k-az\gamma(z)
+\frac{z}{x}\left[\mathcal{A}(x,z)+\mathcal{B}(y,z)-\gamma(z)\right]+z\mathcal{M}(y,z)\,,\label{eq:id_5}
\end{align}
where we have introduced the (unknown) function
\be
\mathcal{M}(y,z)=\sum_{n,k=0}^\infty Z_{k}(1,n)y^nz^k\,.
\label{eq:m_function}
\ee

Summing Eqs.~\eqref{eq:id_1}, \eqref{eq:id_5}, subtracting Eqs.~\eqref{eq:id_2}, \eqref{eq:id_3}, \eqref{eq:id_4} and finally using Eq.~\eqref{eq:simple_recurrence}, we get
\be
\left(1-x-y-\frac{z}{x}+z\right)\mathcal{F}(x,y,z)=\mathcal{N}(x,y)-z\mathcal{M}(y,z)\,,
\label{eq:almost_there}
\ee
where $\mathcal{N}(x,y)$ is a function which can be expressed explicitly in terms of $\alpha(x)$, $\beta(y)$, $\gamma(z)$, $\mathcal{A}(x,z)$, $\mathcal{B}(y,z)$ and $\mathcal{C}(x,y)$. Since its form is very unwieldy, we will not report it here. We can now rewrite Eq.~\eqref{eq:almost_there} as
\be
\mathcal{F}(x,y,z)=\frac{x\left[\mathcal{N}(x,y)-z\mathcal{M}(y,z)\right]}{(x-x^2-xy-z+xz)}\,.
\label{eq:final_point}
\ee
Note that we have expressed the generating function $\mathcal{F}(x,y,z)$ in terms of a unknown function $\mathcal{M}(y,z)$. At this point comes the key argument at the core of the kernel method. First, we note that the denominator can be factorized as
\be
(x-x^2-xy-z+xz)=-(x- r_{1}(y,z))(x- r_{2}(y,z))\,,
\ee
where
\be
r_{1}(y,z)=\frac{1}{2} \left(1-y+z -\sqrt{(-y+z+1)^2-4 z}\right)\,,\qquad r_{2}(y,z)=\frac{1}{2} \left(1-y+z +\sqrt{(-y+z+1)^2-4 z}\right)\,.
\ee
The function $r_{1}(y,z)$ defines a curve $\xi(x,y)=(r_{1}(y,z),y,z)\in \mathbb{C}^3$ such that $\lim_{x,y\to 0} \xi(x,y)=(0,0,0)$. Now, assuming that the power series $\mathcal{F}(x,y,z)$ has a finite radius of convergence (this assumption can be verified a posteriori), we deduce that, inside of the convergence region, the numerator of Eq.~\eqref{eq:final_point} must be vanishing when evaluated on the curve $\xi(x,y)$. Indeed, if this were not the case, we would have a point inside of the convergence region where $\mathcal{F}(x,y,z)$ diverges, contradicting our assumption. Then, it must be  $\mathcal{N}(r_{1}(y,z),y)-z\mathcal{M}(y,z)=0$, namely
\be
\mathcal{M}(y,z)=\frac{1}{z} \mathcal{N}(r_{1}(y,z),y)\,.
\label{eq:final_result_M}
\ee
The l.h.s. is now a known function. At this point, we could plug this into the r.h.s. of Eq.~\eqref{eq:final_point} and obtain a final expression for the generating function $\mathcal{F}(x,y,z)$. However, since we are ultimately interested in $\mathcal{G}(x,y)$, introduced in Eq.~\eqref{eq:g_function}, we actually do not need to do this. Indeed, from the definitions~\eqref{eq:z_function} and \eqref{eq:m_function} we have
\be
\mathcal{G}(y,z)=\frac{1}{d^2+1}\mathcal{M}\left(\frac{d}{d^2+1}y,\frac{d}{d^2+1}z\right)\,.
\ee
Using now the explicit expression~\eqref{eq:final_result_M}, and rearranging the terms, we finally obtain $\mathcal{G}(x,y)=\widetilde{G}(x/d,y/d)$ where $\widetilde{G}(x,y)$ is defined in Eq.~\eqref{eq:generating_function}.

\section{Bounds on OTOCs for random dual-unitary circuits}
\label{app:OTOCsDU}

To arrive at \eqref{eq:OTOCbound} we provide two complementary bounds which are conveniently obtained writing the OTOC in terms transfer matrices
\be
\mathbb{E}\left[ O(x,y)\right] =\frac{2^{x+y}}{3}\braket{\mcircf{\mcirc\ldots\mcirc}|(\bar{T}^{\mcirc \msquare}_x )^y |\msquare\ldots\msquare\msquaref} = \frac{2^{x+y}}{3}\braket{\msquare\ldots\msquare|{\bar T}^{\mcirc \msquaref}_{y} (\bar{T}^{\mcirc \msquare}_y )^{x-2} {\bar T}^{\mcircf \msquare}_{y} |{\mcirc\ldots\mcirc}},
\label{eq:OTOCTMapp}
\ee
where the transfer matrix ${\bar T}^{\mcirc \msquare}_{x}$ is defined in \eqref{eq:Tcirclesquareave}
while we defined 
\be
\bar T^{\mcirc \msquaref}_x =
\begin{tikzpicture}[baseline=(current  bounding  box.center), scale=0.55]
\def\eps{-0.5};
\foreach \i in {3}{
\foreach \j in {-3,...,2}{
\draw[very thick] (-4.5+1.5*\j,1.5*\i) -- (-3+1.5*\j,1.5*\i);
\draw[very thick] (-3.75+1.5*\j,1.5*\i-.75) -- (-3.75+1.5*\j,1.5*\i+.75);
\draw[ thick, fill=myorange, rounded corners=2pt, rotate around ={45: (-3.75+1.5*\j,1.5*\i)}] (-4+1.5*\j,0.25+1.5*\i) rectangle (-3.5+1.5*\j,-0.25+1.5*\i);
\draw[thick, rotate around ={135: (-3.75+1.5*\j,1.5*\i)}] (-3.75+1.5*\j,.15+1.5*\i) -- (-3.6+1.5*\j,.15+1.5*\i) -- (-3.6+1.5*\j,1.5*\i);}
\draw[thick, fill=black] (0,1.5*\i-0.1) rectangle (0.2,1.5*\i+0.1);
\draw[thick, fill=white] (-8.9,1.5*\i) circle (0.1cm);
}
\draw [thick, decorate, decoration={brace,amplitude=4pt, raise=4pt},yshift=0pt]
(-8.25,5.3) -- (-.75,5.3) node [black,midway,yshift=0.65cm] {$x$};
\Text[x=-8.25,y=2]{}
\end{tikzpicture}\,,
\qquad 
\bar T^{\mcircf \msquare}_x =
\begin{tikzpicture}[baseline=(current  bounding  box.center), scale=0.55]
\def\eps{-0.5};
\foreach \i in {3}{
\foreach \j in {-3,...,2}{
\draw[very thick] (-4.5+1.5*\j,1.5*\i) -- (-3+1.5*\j,1.5*\i);
\draw[very thick] (-3.75+1.5*\j,1.5*\i-.75) -- (-3.75+1.5*\j,1.5*\i+.75);
\draw[ thick, fill=myorange, rounded corners=2pt, rotate around ={45: (-3.75+1.5*\j,1.5*\i)}] (-4+1.5*\j,0.25+1.5*\i) rectangle (-3.5+1.5*\j,-0.25+1.5*\i);
\draw[thick, rotate around ={135: (-3.75+1.5*\j,1.5*\i)}] (-3.75+1.5*\j,.15+1.5*\i) -- (-3.6+1.5*\j,.15+1.5*\i) -- (-3.6+1.5*\j,1.5*\i);}
\draw[thick, fill=white] (0,1.5*\i-0.1) rectangle (0.2,1.5*\i+0.1);
\draw[thick, fill=black] (-8.9,1.5*\i) circle (0.1cm);
}
\draw [thick, decorate, decoration={brace,amplitude=4pt, raise=4pt},yshift=0pt]
(-8.25,5.3) -- (-.75,5.3) node [black,midway,yshift=0.65cm] {$x$};
\Text[x=-8.25,y=2]{}
\end{tikzpicture}\,.
\label{eq:Tcirclefsquareave}
\ee
Note that, since the gate~\eqref{eq:averagedgateDU} is real and symmetric, the orientation is irrelevant. 

The first bound is found by considering the first transfer-matrix expression in \eqref{eq:OTOCTMapp} and observing that 
\begin{property}
	The transfer matrix ${\bar T}^{\mcirc \msquare}_{x}$ acts as follows on the boundary states $\ket{\mcircf \mcirc \ldots \mcirc}$ and $\bra{\msquare \ldots \msquare \msquaref}$
	\begin{align}
		&\!\!\!\!\!\!{\bar T}^{\mcirc \msquare}_{x} \ket{\msquare \ldots \msquare \msquaref} \!=\! \frac{a}{2} \ket{\msquare \ldots \msquare \msquaref}\!+\! \frac{1-a}{2} \ket{\msquare' \ldots \msquare' \msquaref},
		\label{eq:crucialidentity2}\\
		&\!\!\!\!\!\!\bra{\mcircf \!\mcirc \ldots \mcirc} {\bar T}^{\mcirc \msquare}_{x} \!\!=\! \frac{a}{2}\! \bra{\mcircf \mcirc \ldots \mcirc} \!+\! \frac{1-a}{2} \bra{\mcircf \mcirc' \ldots \mcirc'},
		\label{eq:crucialidentity3}
	\end{align}
	where $\ket{\mcirc'}$ is defined in \eqref{eq:circprimedef} and we introduced 
	\begin{align}
		\ket{\msquare'} = a \ket{\msquare} + \frac{1-a}{\sqrt 3} \ket{\msquaref}=\frac{1}{2} \ket{\mcirc} + \frac{\sqrt 3 }{2} \frac{4a-1}{3}  \ket{\mcircf}\,.\label{eq:squareprimedef}
	\end{align}
\end{property}
This property is directly proven by using the diagrammatic relations \eqref{eq:aveunitary1}--\eqref{eq:avedualunitary2} in analogy with the proof of Property~\ref{prop:P2}. 

Using \eqref{eq:crucialidentity2}, \eqref{eq:crucialidentity3} we find the following recurrence equations for the partition function $\mathbb{E}\left[ O(x,y)\right]$
\begin{align}
	\mathbb{E}\left[ O(x,y)\right] &= {a} \mathbb{E}\left[ O(x,y-1)\right] + {(1-a)}\mathbb{E}\left[ O_2(x,y-1)\right]\,,
	\label{eq:OO2}\\
	\mathbb{E}\left[ O_2(x,y) \right]&= {a} \mathbb{E}\left[ O_2(x,y-1) ]\right]+ {(1-a)} \mathbb{E}\left[ O_3(x,y-1)\right]\,,
	\label{eq:O2O3}
\end{align}
where we introduced
\be
\mathbb{E}\left[ O_2(x,y) \right]\!\!=\!\!\frac{2^{x+y}}{3}\!\!\begin{tikzpicture}[baseline=(current  bounding  box.center), scale=0.55]
\def\eps{-0.5};
\foreach \i in {0,...,3}
\foreach \j in {2,...,-3}{
	\draw[very thick] (-4.5+1.5*\j,1.5*\i) -- (-3+1.5*\j,1.5*\i);
	\draw[very thick] (-3.75+1.5*\j,1.5*\i-.75) -- (-3.75+1.5*\j,1.5*\i+.75);
	\draw[ thick, fill=myorange, rounded corners=2pt, rotate around ={45: (-3.75+1.5*\j,1.5*\i)}] (-4+1.5*\j,0.25+1.5*\i) rectangle (-3.5+1.5*\j,-0.25+1.5*\i);
	\draw[thick, rotate around ={45: (-3.75+1.5*\j,1.5*\i)}] (-3.75+1.5*\j,.15+1.5*\i) -- (-3.6+1.5*\j,.15+1.5*\i) -- (-3.6+1.5*\j,1.5*\i);}
\foreach \i in {0,...,3}{
	\draw[thick, fill=white] (-0.1,1.5*\i-0.1) rectangle (0.1,1.5*\i+0.1);
	\draw[thick, fill=white] (-8.9,1.5*\i) circle (0.1cm);
}
\foreach \i in {-2,...,1}{
	\draw[thick, fill=white] (-3.85+1.5*\i,-0.85) rectangle (-3.65+1.5*\i,-0.65);
	\draw[thick, fill=white] (-3.75+1.5*\i,5.25) circle (0.1cm);
	\Text[x=-3.95+1.5*\i,y=-1]{$'$};
}
\draw[thick, fill=white] (-3.75+1.5*2,5.25) circle (0.1cm);
\draw[thick, fill=white] (-3.85-1.5*3,-0.85) rectangle (-3.65-1.5*3,-0.65);
\Text[x=-3.95-1.5*3,y=-1]{$'$};
\draw [thick, decorate, decoration={brace,amplitude=4pt, raise=4pt},yshift=0pt]
(-8.25,5.4) -- (-.75,5.4) node [black,midway,yshift=0.55cm] {$x$};
\draw[thick, fill=black] (-.75-0.1,-.75-0.1) rectangle (-.75+0.1,-.75+0.1);
\draw[thick, fill=black] (-8.25,5.25) circle (0.1cm);
\draw [thick, decorate, decoration={brace,amplitude=4pt,mirror,raise=4pt},yshift=0pt]
(0,0) -- (0,4.5) node [black,midway,xshift=.45cm] {$y$};
\Text[x=-8.25,y=-3]{}
\end{tikzpicture}\!\!\!\!\!\!,
\quad 
\mathbb{E}\left[ O_3(x,y) \right]\!\!=\!\!\frac{2^{x+y}}{3}\!\!\begin{tikzpicture}[baseline=(current  bounding  box.center), scale=0.55]
\def\eps{-0.5};
\foreach \i in {0,...,3}
\foreach \j in {2,...,-3}{
	\draw[very thick] (-4.5+1.5*\j,1.5*\i) -- (-3+1.5*\j,1.5*\i);
	\draw[very thick] (-3.75+1.5*\j,1.5*\i-.75) -- (-3.75+1.5*\j,1.5*\i+.75);
	\draw[ thick, fill=myorange, rounded corners=2pt, rotate around ={45: (-3.75+1.5*\j,1.5*\i)}] (-4+1.5*\j,0.25+1.5*\i) rectangle (-3.5+1.5*\j,-0.25+1.5*\i);
	\draw[thick, rotate around ={45: (-3.75+1.5*\j,1.5*\i)}] (-3.75+1.5*\j,.15+1.5*\i) -- (-3.6+1.5*\j,.15+1.5*\i) -- (-3.6+1.5*\j,1.5*\i);}
\foreach \i in {0,...,3}{
	\draw[thick, fill=white] (-0.1,1.5*\i-0.1) rectangle (0.1,1.5*\i+0.1);
	\draw[thick, fill=white] (-8.9,1.5*\i) circle (0.1cm);
}
\foreach \i in {-2,...,1}{
	\draw[thick, fill=white] (-3.85+1.5*\i,-0.85) rectangle (-3.65+1.5*\i,-0.65);
	\draw[thick, fill=white] (-3.75+1.5*\i,5.25) circle (0.1cm);
	\Text[x=-3.95+1.5*\i,y=-1]{$'$};
	\Text[x=-3.5+1.5*\i,y=5.25]{$'$};
}
\draw[thick, fill=white] (-3.75+1.5*2,5.25) circle (0.1cm);
\Text[x=-3.5+1.5*2,y=5.25]{$'$};
\draw[thick, fill=white] (-3.85-1.5*3,-0.85) rectangle (-3.65-1.5*3,-0.65);
\Text[x=-3.95-1.5*3,y=-1]{$'$};
\draw [thick, decorate, decoration={brace,amplitude=4pt, raise=4pt},yshift=0pt]
(-8.25,5.4) -- (-.75,5.4) node [black,midway,yshift=0.55cm] {$x$};
\draw[thick, fill=black] (-.75-0.1,-.75-0.1) rectangle (-.75+0.1,-.75+0.1);
\draw[thick, fill=black] (-8.25,5.25) circle (0.1cm);
\draw [thick, decorate, decoration={brace,amplitude=4pt,mirror,raise=4pt},yshift=0pt]
(0,0) -- (0,4.5) node [black,midway,xshift=.45cm] {$y$};
\Text[x=-8.25,y=-3]{}
\end{tikzpicture}\!\!\!\!\!\!,
\label{eq:O3}
\ee
fulfilling 
\be
\mathbb{E}\left[ O_2(x,0)\right]= \frac{2^{x}}{3} \braket{\mcircf|\msquare'} \braket{\mcirc|\msquaref} \braket{\mcirc|\msquare'}^{x-2}=\frac{4a-1}{3}\,,
\ee
and
\be
\mathbb{E}\left[ O_3(x,0)\right]= \frac{2^{x}}{3}  \braket{\mcircf{\mcirc'\ldots\mcirc'}|\msquare'\ldots\msquare'\msquaref}
= \frac{(4a-1)^2}{9} \Bigl[1-\frac{4}{3}(a-1)^2 \Bigr]^{x-2}\,.
\ee
The equations \eqref{eq:OO2} and \eqref{eq:O2O3} can be formally solved and combined (see Appendix~\ref{sec:formalsol2}) obtaining  
\be
\mathbb{E}\left[ O(x,y)  \right]= \left[{a^{y}} +(4a-1)(1-a)  \frac{y {a^{y}}}{3a}\right] (1-\delta_{x,1}) -\frac{1}{3}\delta_{x,1}+ \frac{(1-a)^2}{a^2}   \sum_{k=0}^{y-1} \sum_{h=0}^{k-1}   {\mathbb{E}\left[ O_3'(x,h)\right]} a^{y-h}\,,
\label{eq:formalOO3}
\ee
where we defined 
\be
\mathbb{E}\left[ O'_3(x,y)\right]:= \frac{2^{x+y}}{3}\braket{\mcircf{\mcirc'\ldots\mcirc'}|\Bigl(\bar{T}^{\mcirc \msquare}_x - \frac{1}{2}\sum_{k=0}^{x} P^{\mcirc}_k\Bigr)^y |\msquare'\ldots\msquare'\msquaref} =\mathbb{E}\left[ O_3(x,y)\right]+\frac{1}{3}\delta_{x,1}\,.
\ee
$\mathbb{E}\left[ O_3'(x,y)\right]$ can be bound by using the following property, which is proven following the reasoning of Appendix~\ref{app:proofP1}.
\begin{property}
\label{prop:P5}
The matrix $\bar{T}^{\mcirc \msquare}_x$ is positive definite and has the following spectral decomposition 
\begin{align}
&\!\!\bar{T}^{\mcirc \msquare}_x \!\!= \frac{1}{2} \sum_{k=0}^x  P^{\mcirc}_k + \left[\frac{1}{2}-\frac{2}{3}(a-1)^2\right] \sum_{k=2}^{x}  Q^{\mcirc}_k +  R^{\mcirc}_x,
\end{align}
where we defined 
\begin{align}
P^{\mcirc}_k &:= \ket{\underbrace{\mcirc\ldots\mcirc\mcircf}_{k}\msquare\ldots\msquare}\!\!\bra{\underbrace{\mcirc\ldots\mcirc\mcircf}_{k}\msquare\ldots\msquare},\label{eq:Psquare}\\
Q^{\mcirc}_k &:= \ket{\underbrace{\mcirc\ldots\mcirc\mcircf\msquaref}_{k}\msquare\ldots\msquare}\!\!\bra{\underbrace{\mcirc\ldots\mcirc\mcircf\msquaref}_{k}\msquare\ldots\msquare},
\end{align}
and the ``reminder'' $R^{\mcirc}_x$ has operator norm 
\be
|R^{\mcirc}_x| \leq \!\!\left[\frac{1}{2}-\frac{2}{3}(a-1)^2\right]<\frac{1}{2}\,,\qquad a\in[1/4,1]. 
\ee
\end{property}
Using Property~\ref{prop:P5} we find 
\be
|\mathbb{E}\left[ O_3'(x,y)\right]| \leq \frac{2}{3} \mathcal A(a)^y \mathcal B(a)^{x-1},
\ee
where $\mathcal A(a)$ and $\mathcal B(a)$ are defined in Eq.~\eqref{eq:mathcAB}.  This gives 
\be
\mathbb{E}\left[ O(x,y) \right] = \left[{a^{y}} +(4a-1)(1-a)  \frac{y {a^{y}}}{3a}\right] (1-\delta_{x,1}) -\frac{1}{3}\delta_{x,1}+ r(x,y)\,,\qquad |r(x,y)|< C_1  \mathcal A(a)^y \mathcal B(a)^{x}\,.
\label{eq:OTOCbound1}
\ee

Another bound can be found by considering the r.h.s. of \eqref{eq:OTOCTMapp} and invoking 
\begin{property}
	The transfer matrices ${\bar T}^{\mcirc \msquaref}_{x}$ and ${\bar T}^{\mcircf \msquare}_{x}$ act as follows on the boundary states $\bra{\mcirc \ldots \mcirc}$ and $\ket{\msquare \ldots \msquare}$
	\begin{align}
		\bra{\msquare \ldots \msquare} {\bar T}^{\mcirc \msquaref}_{x}  = \frac{\sqrt 3}{2} \bra{\msquare' \ldots  \msquare'}
		\qquad
		{\bar T}^{\mcircf \msquare}_{x} \ket{\mcirc \ldots \mcirc}  = \frac{\sqrt 3 }{2} \ket{\mcirc' \ldots \mcirc'}\,,\label{eq:P41}
	\end{align}
	where $\ket{\mcirc'}$ and $\ket{\msquare'}$ are respectively defined in \eqref{eq:circprimedef} and \eqref{eq:squareprimedef}.
\end{property} 
Once again, this property is directly proven using \eqref{eq:aveunitary1}--\eqref{eq:avedualunitary2}. The relations \eqref{eq:P41} give
\be
\mathbb{E}\left[ O(x,y)\right]  = \left[{a^{y}} +(4a-1)(1-a)  \frac{y {a^{y}}}{3a}\right] (1-\delta_{x,1}) -\frac{1}{3}\delta_{x,1} + \mathbb{E}\left[ \tilde O(x-2,y)\right],
\ee
where we introduced 
\be
\mathbb{E}\left[ \tilde O(x,y)\right]:= 2^{x+y}\braket{{\msquare'\ldots\msquare'}|\Bigl(\bar{T}^{\mcirc \msquare}_y - \frac{1}{2}\sum_{k=0}^{y} P^{\mcirc}_k\Bigr)^x |\mcirc'\ldots\mcirc'}\,.
\ee
Using Property~\ref{prop:P5} we have 
\be
|\mathbb{E}\left[ \tilde O(x,y)\right]| \leq \frac{2}{3} \mathcal A(a)^x \mathcal B(a)^{y},
\ee
so that 
\be
\mathbb{E}\left[ O(x,y) \right] = \left[{a^{y}} +(4a-1)(1-a)  \frac{y {a^{y}}}{3a}\right] (1-\delta_{x,1}) -\frac{1}{3}\delta_{x,1}+ r(x,y)\,,\qquad |r(x,y)|< C_2  \mathcal A(a)^x \mathcal B(a)^{y}\,.
\label{eq:OTOCbound2}
\ee
Combining the bounds \eqref{eq:OTOCbound1} and \eqref{eq:OTOCbound2} and using $\mathcal B(a)\geq \mathcal A(a)$ for all $a\in[1/3,1]$, we obtain \eqref{eq:OTOCbound}.  

\end{widetext}


\begin{thebibliography}{99}
\bibitem{note1}
Note that this formula does not only apply in the limit $L\to\infty$: it is exact whenever $L-|x|>t$.
\bibitem{note2}
This can be shown by reasoning on the lines of Appendix A of Ref.~\cite{BKP:OEergodicandmixing}.
\bibitem{note3}
The boundary conditions are found by following the strategy explained after \eqref{eq:RUrecurrence1}. One writes the equations for $W_k(x,k+1)$ and $W_k(1,y)$ and matches with the general form \eqref{eq:OTOCequations}.
\bibitem{notecomplexity}
This still represents a great simplification compared to the generic case where one has exponentially many such functions.
\bibitem{noteOTOC}
From the point of view of OTOCs, being completely chaotic means that the transfer matrix $\bar{T}^{\mcirc \msquare}_x$ has maximal eigenvalue $1/2$ and that the projectors $\{P^{\msquare}_k\}_{k=0}^x$ defined in Eq.~\eqref{eq:Psquare} identify its $x+1$ degenerate eigenspace.
\end{thebibliography}
\end{document}